\documentclass[11pt]{article}
\usepackage{axodraw2,pix}
\usepackage{epsfig}
\usepackage{amsfonts}
\usepackage{amsmath,amssymb}
\usepackage{bbm,bm}
\usepackage{mathrsfs} %
\usepackage{slashed} %
\usepackage{listings}
\usepackage{cite}
\allowdisplaybreaks[1]
\usepackage{color}
\definecolor{codeblue}{rgb}{0,0,0.5}
\definecolor{codered}{rgb}{0.6,0,0}
\definecolor{codegreen}{rgb}{0,0.5,0}

\usepackage{caption}
\DeclareCaptionFormat{listingformat}{\colorbox{black}{\color{white}\thelstlisting}}
\graphicspath{{.}{./figures/}}

\usepackage{tikz}
\usetikzlibrary{shapes,arrows,shadows,automata,positioning}
\newdimen\nodeDist
\def\wallgo{{\tt WallGo}}
\def\wallgoMatrix{\wallgo{\tt Matrix}}
\def\wallgoCollision{\wallgo{\tt Collision}}
\def\wallgoUrl{https://github.com/Wall-Go/WallGo}
\def\wallgoMatrixUrl{https://github.com/Wall-Go/WallGoMatrix}
\def\wallgoCollisionUrl{https://github.com/Wall-Go/WallGoCollision}
\def\wallgoDocsUrl{https://wallgo.readthedocs.io}
\def\wallgoPypiUrl{https://pypi.org/project/WallGo}
\def\wallgoCollisionPypiUrl{https://pypi.org/project/WallGoCollision}
\def\wallgoCollisionDocsUrl{https://wallgocollision.readthedocs.io}
\def\wallgoMatrixPacletUrl{https://resources.wolframcloud.com/PacletRepository/resources/WallGo/WallGoMatrix}
\def\WallGoVersion{{\tt 1.0.0}}
\def\WallGoMatrixVersion{{\tt 1.0.0}}
\def\WallGoCollisionVersion{{\tt 1.0.0}}

\def\dralgo{{\tt DRalgo}}

\usepackage[normalem]{ulem} %

\usepackage[
  colorlinks=true,
  linkcolor={red!50!black},
  citecolor={blue!50!black},
  filecolor=black,
  urlcolor=black,
  breaklinks=true
]{hyperref}

\usepackage{breakurl}
\usepackage{pifont}
\newcommand*\colourcheck[1]{%
  \expandafter\newcommand\csname #1check\endcsname{\textcolor{#1}{\ding{52}}}%
}
\colourcheck{blue}
\colourcheck{green}

\newcommand{\te}{\textemdash}

\renewcommand{\vec}[1]{{\bf #1}}

\newcommand{\alphas}{\alpha_\rmi{s}}

\renewcommand{\tr}{{\rm Tr\,}}
\newcommand{\vw}{v_w}
\newcommand{\vJ}{v_\rmii{$J$}}
\newcommand{\cs}{c_s}

\usepackage{amsmath}
\usepackage{xparse}

\newcommand{\bra}[1]{\langle#1\vert} %
\newcommand{\ket}[1]{\lvert#1\rangle} %

\newcommand{\Bra}[1]{\left[#1 \right\vert} %
\newcommand{\Ket}[1]{\left\vert#1 \right]} %

\newcommand{\braketT}[2]{\left\langle #1 \, #2\right\rangle} %
\newcommand{\BraKetT}[2]{\left[ #1 \,  #2\right]} %

\newcommand{\braket}[3]{\left\langle #1 | #2 | #3\right\rangle} %
\newcommand{\Braket}[3]{\left[ #1 | #2 | #3\right\rangle} %
\newcommand{\braKet}[3]{\left\langle #1 | #2 | #3\right]} %
\newcommand{\BraKet}[3]{\left[ #1 | #2 | #3\right]} %

\newcommand{\abs}[1]{\left\vert #1 \right\vert}

\newcommand{\Tc}{T_{\rm c}}
\newcommand{\Tn}{T_{\rm n}}
\newcommand{\Tp}{T_{\rm p}}

\newcommand{\T}{\rmii{$T$}}

\newcommand{\mD}{m_\rmii{D}}

\newcommand{\gY}{g_\rmii{$Y$}}

\def\lsi{\raise0.3ex\hbox{$<$\kern-0.75em\raise-1.1ex\hbox{$\sim$}}}
\def\gsi{\raise0.3ex\hbox{$>$\kern-0.75em\raise-1.1ex\hbox{$\sim$}}}

\newcommand{\lsim}{\mathop{\lsi}}

\renewcommand{\nn}{\nonumber \\}
\renewcommand{\rmi}[1]{{\mbox{\scriptsize #1}}}
\newcommand{\rmii}[1]{{\mbox{\tiny\rm{#1}}}}

\newcommand{\Tint}[1]{{\hbox{$\sum$}\!\!\!\!\!\!\!\int\,}_{\!\!\!\!\raise-0.9ex\hbox{$\scriptstyle{#1}$}}}
\newcommand{\Tinti}[1]{{{\Sigma}\!\!\!\!\raise0.3ex\hbox{$\int$}_\rmii{${#1}$}}}
\newcommand{\Tintip}[1]{{{\Sigma'}\!\!\!\!\!\raise0.3ex\hbox{$\int$}_\rmii{${#1}$}}}

\newcommand{\bsl}[1]{\,\slash\!\!\!\!{#1}\,}
\newcommand{\msl}[1]{\,\slash\!\!\!{#1}\,}
\newcommand{\deltabar}{\raise-0.02em\hbox{$\bar{}$}\hspace*{-0.8mm}{\delta}}

\newcommand{\vev}{vacuum expectation value}
\renewcommand{\vev}{VEV}
\newcommand{\Veff}{V^{\rmi{eff}}}

\newcommand{\Nchi}{M}
\newcommand{\Nrho}{N}

\def\backtick{\char18}
\lstdefinestyle{backtickavailable}{literate={`}{\backtick}1, escapechar=@}

\makeatletter \@addtoreset{equation}{section} \makeatother
\renewcommand{\theequation}{\arabic{section}.\arabic{equation}}
\makeatletter
\renewcommand\section{\@startsection{section}{1}{\z@}%
  {-5.5ex \@plus -1ex \@minus -.2ex}%
  {2.3ex \@plus.2ex}%
  {\normalfont\large\bfseries}}
\renewcommand\subsection{\@startsection{subsection}{2}{\z@}%
  {-3.25ex\@plus -1ex \@minus -.2ex}%
  {1.5ex \@plus .2ex}%
  {\normalfont\normalsize\bfseries}}
\renewcommand\thesection{\@arabic\c@section}
\renewcommand\thesubsection{\thesection.\@arabic\c@subsection}
\renewcommand{\@seccntformat}[1]{%
  \csname the#1\endcsname.\hspace{1.0em}}
\makeatother

\begin{document}

\flushbottom

\begin{titlepage}

  {
\flushright
CERN-TH-2024-174\\
DESY-24-162\\
HIP-2024-21/TH\\
  }

\begin{centering}
\includegraphics[height=3cm]{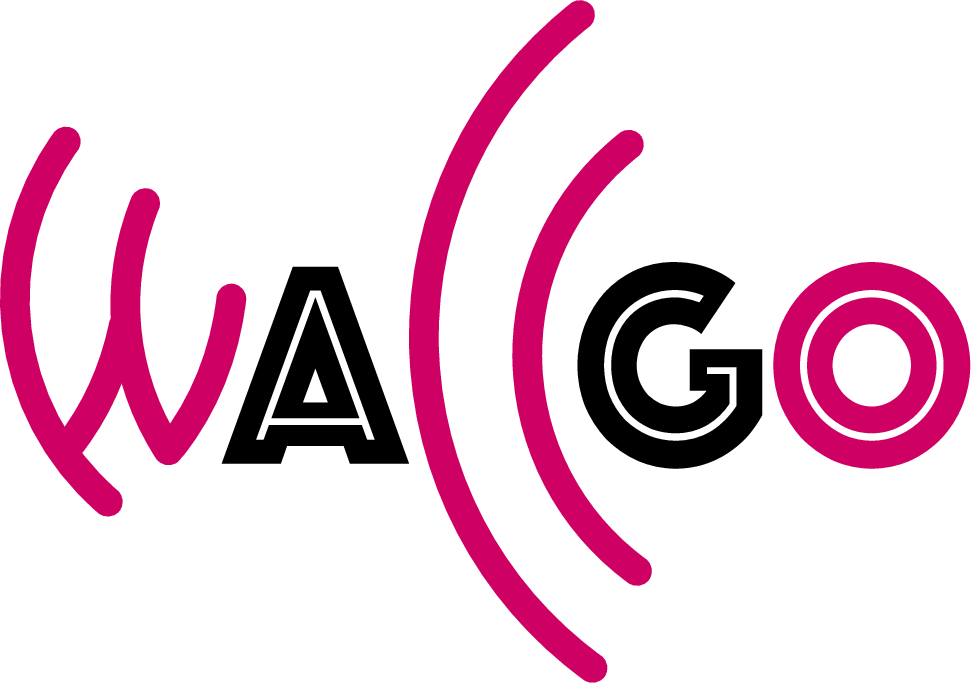}
\vfill

{\Large\bf
    How fast does the \wallgo? \\
  \large\bf
  A package for computing wall velocities in first-order phase transitions
}

\vspace{0.8cm}

\renewcommand{\thefootnote}{\fnsymbol{footnote}}
Andreas Ekstedt$^{\rm a,}$%
\footnotemark[1],
Oliver Gould$^{\rm b,}$%
\footnotemark[2],
Joonas Hirvonen$^{\rm b,\rm c,}$%
\footnotemark[3],
Benoit Laurent$^{\rm d,}$%
\footnotemark[4],\\
Lauri Niemi$^{\rm c,}$%
\footnotemark[5],
Philipp Schicho$^{\rm e,}$%
\footnotemark[6],
Jorinde van de Vis$^{\rm f,}$%
\footnotemark[8]

\vfill

\mbox{\bf Abstract}

\end{centering}

\vspace*{0.3cm}

\noindent
\wallgo{} is an open-source software designed to compute
the bubble wall velocity in first-order cosmological phase transitions.
Additionally, it evaluates
the energy budget available for generating gravitational waves.
The main part of \wallgo{}, built in {\tt Python}, determines the wall velocity by solving the scalar-field(s) equation of motion, the Boltzmann equations and energy-momentum conservation for the fluid velocity and temperature.
\wallgo{} also includes two auxiliary modules:
\wallgoMatrix{}, which computes matrix elements for out-of-equilibrium particles, and
\wallgoCollision{}, which performs higher-dimensional integrals
for Boltzmann collision terms.
Users can implement custom models by
defining an effective potential and
specifying a list of out-of-equilibrium particles and their interactions.

As the first public software to compute the wall velocity including out-of-equilibrium contributions,
\wallgo{}
improves the precision of the computation compared to common assumptions
in earlier computations.
It utilises a spectral method
for the deviation from equilibrium and collision terms that
provides exponential convergence in basis polynomials, and
supports multiple out-of-equilibrium particles,
allowing for Boltzmann mixing terms.
\wallgo{} is tailored for non-runaway wall scenarios where leading-order coupling effects dominate friction.

While this work introduces the software and the underlying theory,
a more detailed documentation can be found in
\url{\wallgoDocsUrl}.

\newpage

\begin{centering}
$^\rmi{a}$%
{\em
  Department of Physics and Astronomy,
  Uppsala University,\\
  P.O. Box 256, SE-751 05 Uppsala, Sweden
}

\vspace{0.3cm}

$^\rmi{b}$%
{\em
  School of Physics and Astronomy, University of Nottingham,\\
  Nottingham NG7 2RD, United Kingdom
}

\vspace{0.3cm}

$^\rmi{c}$%
{\em
  Department of Physics and Helsinki Institute of Physics, University of Helsinki, FI-00014, Finland
}

\vspace{0.3cm}

$^\rmi{d}$%
{\em
  McGill University, Department of Physics, 3600 University St.,
  Montr\'eal, QC H3A2T8 Canada
}

\vspace{0.3cm}

$^\rmi{e}$%
{\em
  D\'epartement de Physique Th\'eorique, Universit\'e de Gen\`eve,\\
  24 quai Ernest Ansermet, CH-1211 Gen\`eve 4, Switzerland
}

\vspace{0.3cm}

$^\rmi{f}$%
{\em
  Theoretical Physics Department, CERN,\\
  1 Esplanade des Particules, CH-1211 Geneva 23, Switzerland
}

\end{centering}

\renewcommand{\thefootnote}{\fnsymbol{footnote}}
\footnotetext[1]{andreas.ekstedt@physics.uu.se}
\footnotetext[2]{oliver.gould@nottingham.ac.uk}
\footnotetext[3]{joonas.hirvonen@nottingham.ac.uk}
\footnotetext[4]{benoit.laurent@mail.mcgill.ca}
\footnotetext[5]{lauri.b.niemi@helsinki.fi}
\footnotetext[6]{philipp.schicho@unige.ch}
\footnotetext[8]{jorinde.van.de.vis@cern.ch}
\clearpage
\end{titlepage}

\renewcommand{\thefootnote}{\arabic{footnote}}
\setcounter{footnote}{0}

{\hypersetup{hidelinks}
\tableofcontents
}
\vfill

\section*{Notation}
\label{sec:notation}

\paragraph*{Indices}
\begin{itemize}
	\item[-] $a, b, c$: particle species
	\item[-] $i, j, k$: polynomial basis elements
	\item[-] $i$: background scalar field components
	\item[-] $I, J, L, M$: spinor indices
	\item[-] $\alpha, \beta, \gamma$: lattice points
	\item[-] $\mu, \nu$: Lorentz indices
	\item[-] $\lambda_1, \lambda_2, \lambda_3$: helicities
\end{itemize}

\clearpage
\section*{Program summary}
{\em License}:
GNU General Public License 3 (GPLv3)
\\%
{\em Operating system:}
Any operating system;
tested
macOS~11--15, Ubuntu 18--24 and Windows 10--11.
\\%
\rule{\textwidth}{0.4pt}
\\%
{\em Program title}: 
\wallgo{}
{\tt [v\WallGoVersion]}
\\%
{\em Program obtainable from}:
\url{\wallgoPypiUrl}
\\%
{\em Documentation link}:
\url{\wallgoDocsUrl}
\\%
{\em Developer's repository link}:
\url{\wallgoUrl}
\\%
{\em Programming languages}:
{\tt Python} 
\\%
{\em Nature of problem}:
Computing terminal velocities of expanding bubble walls for first-order phase transitions, as coupled sets of scalar field, hydrodynamic and Boltzmann equations.
\\%
{\em Solution method}:
The scalar-field equations are solved as a variational problem with an ansatz, and the Boltzmann equations are solved using a spectral decomposition.
\\%
{\em Restrictions}:
{\tt Python} version 3.10 or above.
\\%
\rule{\textwidth}{0.4pt}
\\%
{\em Program title}: 
\wallgoCollision{}
{\tt [v\WallGoCollisionVersion]}
\\%
{\em Program obtainable from}:
\url{\wallgoCollisionPypiUrl}
\\%
{\em Documentation link}:
\url{\wallgoCollisionDocsUrl}
\\%
{\em Developer's repository link}:
\url{\wallgoCollisionUrl}
\\%
{\em Programming languages}:
{\tt C++} and {\tt Python} 
\\%
{\em Nature of problem}:
Computing collision integrals for Boltzmann equations in a spectral expansion.
\\%
{\em Solution method}:
Adaptive Monte-Carlo integration.
\\%
{\em Restrictions}:
{\tt Python} version 3.10 or above. Source builds require a {\tt C++17} compliant compiler and {\tt CMake} 3.18 or newer.
\\%
\rule{\textwidth}{0.4pt}
\\%
{\em Program title}:
\wallgoMatrix{}
{\tt [v\WallGoMatrixVersion]}
\\%
{\em Program obtainable from}:\\
\url{\wallgoMatrixPacletUrl}
\\%
{\em Developer's repository link}:
\url{\wallgoMatrixUrl}
\\%
{\em Programming languages}:
{\tt Mathematica}
\\%
{\em External routines/libraries}:
{\tt DRalgo}~\cite{Ekstedt:2022bff} and {\tt GroupMath}~\cite{Fonseca:2020vke}.
\\%
{\em Nature of problem}:
Computing 2-to-2 scattering matrix elements for arbitrary quantum field theories.
\\%
{\em Solution method}:
Matrix elements are constructed by building coupling tensors for a given model, and contracting these with kinematic factors appropriately.
\\%
{\em Restrictions}:
Tested on {\tt Mathematica} versions 12, 13 and 14.
\clearpage

\section{Introduction}
\label{sec:intro}
Cosmological first-order phase transitions (FOPTs) are exciting possible processes in
the early universe.
They might have played a role in generating the asymmetry between matter
and antimatter (e.g.\ in
electroweak baryogenesis~\cite{Kuzmin:1985mm, Shaposhnikov:1986jp, Shaposhnikov:1987tw, Cohen:1990py, Cohen:1993nk}, or
alternatives~\cite{Katz:2016adq, Baldes:2021vyz, Azatov:2021irb}),
and they could also have sourced gravitational wave (GW) signals that can be observed with the next
generation of GW detectors~\cite{Grojean:2006bp, Caprini:2018mtu,Caprini:2019egz,LISACosmologyWorkingGroup:2022jok}.
It is well-known that the two candidate processes of the Standard Model (SM) of particle physics,
the electroweak phase transition~\cite{Kajantie:1995kf, Kajantie:1996mn, Kajantie:1996qd, Gurtler:1997hr, Csikor:1998eu, Aoki:1999fi} and
the QCD phase transitions~\cite{Aoki:2006we,Bhattacharya:2014ara},
are in fact cross-overs.
There are, however, many possibilities for extending the SM to render these
phase transitions first order
(see e.g.~\cite{Caprini:2019egz} and references therein for a number of examples). Moreover, a significant part of the energy budget of the universe is in dark matter and energy, and phase transitions could have also occurred in dark sectors, with possibly observable GW signatures~\cite{Espinosa:2008kw, Breitbach:2018ddu, Ertas:2021xeh}.

The expansion velocity of bubbles generated during a FOPT, denoted  $\vw$, is
a crucial parameter for predicting GW signals.
Variations in $\vw$ can lead to changes of up to two orders of magnitude in predicted GW amplitudes~\cite{Espinosa:2010hh, Hindmarsh:2017gnf, Caprini:2019egz}, and
the spectral shape of the GW spectrum is also highly sensitive to $\vw$~\cite{Hindmarsh:2016lnk, Hindmarsh:2019phv, Jinno:2020eqg,Jinno:2022mie, Caprini:2024gyk}.
Studies~\cite{Gowling:2021gcy, Giese:2021dnw} suggest that
{\em if} LISA detects a GW signal from a FOPT,
$\vw$ may be the most accurately reconstructible parameter.
Accurate predictions of GW spectra are essential to distinguish signals from FOPTs amid astrophysical sources in data from observatories like LISA~\cite{Alvey:2023npw}. Furthermore, the wall velocity can significantly impact predictions
for the value of the matter-antimatter asymmetry~\cite{DeVries:2018aul, Cline:2020jre, Cline:2021dkf},
which can determine whether a model can successfully explain the observed asymmetry. In alternative mechanisms for baryogenesis, e.g.~\cite{Baldes:2021vyz, Azatov:2021irb, Cataldi:2024pgt}, and dark matter generation~\cite{Azatov:2021ifm, Giudice:2024tcp} during phase transitions, the baryon and dark matter abundances are also expected to depend on $\vw$.

The wall velocity is determined as the speed at which the outward pressure, resulting from the pressure difference between the phases, is balanced by the friction and backreaction from the plasma.
In practice, finding $\vw$ (and the wall width) requires solving the scalar-field(s) equation of motion, the Boltzmann equations for the particles in the plasma, and the energy-momentum conservation equations for the fluid velocity and temperature profile. 
Early work on computing the wall velocity employed a so-called ballistic approximation~\cite{Liu:1992tn, Dine:1992wr}, which assumes that
the mean free path of the particles is much larger than the wall width.
The authors of~\cite{Moore:1995ua, Moore:1995si},
were the first to self-consistently solve
the scalar field and Boltzmann equations,
including leading-log collision terms.
For additional calculations of $\vw$  using similar methods,
see e.g., \cite{Konstandin:2014zta, Kozaczuk:2015owa, DeCurtis:2022hlx, Jiang:2022btc}.

The approach of~\cite{Moore:1995ua, Moore:1995si} describes
deviations from equilibrium in the distribution functions via
the chemical potential $\mu$,
deviation of the temperature $\delta T$, and
fluid velocity $\delta v$.
By taking three moments of the Boltzmann equations, a set of equations for
$\mu, \delta T$ and $\delta v$ is obtained.
However, as noted in~\cite{Laurent:2020gpg, Dorsch:2021nje},
the so-obtained equations contain a singularity around the sound speed.
In~\cite{Dorsch:2021nje},
it was demonstrated that this singularity stems from
the linearisation of the background solution.
To improve the description of the out-of-equilibrium contribution,
the authors introduce a generalised fluid ansatz, and
solve the Boltzmann equations for a larger set of moments.
In contrast, \cite{Laurent:2022jrs} adopts an alternative approach
using a spectral method to describe deviations from equilibrium,
thereby
eliminating the need for a linearisation of the background distribution.
We adopt this latter approach in our work.

Due to the difficulty of computing the wall velocity,
it is often guessed
or treated as a free parameter.
Estimates can be made by assuming
local thermal equilibrium (LTE)~\cite{Balaji:2020yrx, Ai:2021kak, Ai:2023see},
or by considering a large jump in
degrees of freedom~\cite{Sanchez-Garitaonandia:2023zqz}.
The LTE assumption has been shown to provide
a reasonable approximation for
the Standard Model coupled to a singlet~\cite{Laurent:2022jrs},
and should offer an upper bound on $\vw$.
Further improvement in determining $\vw$ requires
accounting for out-of-equilibrium effects.

\wallgo{} provides an
{\bf automated computation of the wall velocity for user-defined models}.
The software package consists of a main part and two auxiliary modules:
\begin{itemize}
  \item
    \wallgo{} finds $\vw$ by solving the equations of motion of the scalar fields, and the Boltzmann equations of the out-of-equilibrium particles.
  \item
    \wallgoMatrix{} computes the $2 \rightarrow 2$ matrix elements for
    user-specified particles and interactions.
  \item
    \wallgoCollision{} computes the collision integrals in the spectral basis.
\end{itemize}
\wallgo{} allows readers to largely reduce the uncertainty related to
the wall velocity in predictions of the GW signal and
the baryon or dark matter abundance.
Additionally,
it will help to identify the largest sources of uncertainty in $\vw$,
offering guidance for
future studies aimed at further minimizing these uncertainties.
A detailed examination of these uncertainty sources will be addressed in future work.

In addition to being the {\em first} publicly available numerical code for solving the wall velocity with out-of-equilibrium effects for user-defined models,
\wallgo{} also provides several options that improve
the computation beyond commonly used approximations in the literature:
\begin{itemize}
  \item
    {\em Matrix elements and collision terms}:
    \wallgo{} computes matrix elements and collision terms for arbitrary particles and interactions, enabling analysis beyond the typical focus on only
    the out-of-equilibrium top quark and strong interactions.
  \item
    {\em User-defined potential}:
    The effective potential is fully customizable,
    allowing calculations beyond leading order.
    Higher-order corrections to the potential,
    shown to significantly affect aspects of phase transitions~\cite{Croon:2020cgk, Gould:2021oba, Gould:2023jbz, Kierkla:2023von, Lewicki:2024xan, Ekstedt:2024etx}, will be explored in future work.
  \item
    {\em Mixing in collision terms and Boltzmann equations}:
    When multiple particles are out of equilibrium, \wallgo{} incorporates mixing in
    the collision terms and Boltzmann equations
    (cf.\ section~\ref{sec:Theorycollision}).
    This is in contrast to the conventional moment expansion where
    interaction rates replace collision terms~\cite{Moore:1995si}.
  \item
    {\em Spectral method for convergence}:
    \wallgo{} uses the spectral method introduced in~\cite{Laurent:2022jrs}, which parameterises the deviation from equilibrium on a basis of polynomials.
    For a sufficiently large basis of polynomials, the solution converges exponentially, which we expect to be
    a relevant improvement over the common three-moment Boltzmann equation method.
    \item
    {\em Computation of the gravitational wave energy budget}:
    \wallgo{} can directly compute the efficiency factor used in the prediction of the GW power spectrum \cite{Hindmarsh:2015qta, Hindmarsh:2017gnf, Caprini:2019egz, Caprini:2024gyk}, 
    by solving the hydrodynamics equations with the fully model-dependent equation of state. This removes the need to map onto a
    simplified equation of state~\cite{Espinosa:2010hh, Giese:2020rtr, Giese:2020znk}.
\end{itemize}

This version of \wallgo{} has been developed to deal with weak to moderately strong phase transitions, where the leading-order pressure from out-of-equilibrium particles and the hydrodynamic backreaction are sufficient to compensate the driving force. For stronger phase transitions, this source of friction might not be sufficient to stop the wall from accelerating. The question whether the wall ``runs away'' in this scenario, or whether it is slowed down by next-to-leading-order contributions to the friction, has been a matter of active discussion in recent literature~\cite{Bodeker:2009qy, Bodeker:2017cim, Hoche:2020ysm, Azatov:2020ufh, Gouttenoire:2021kjv, Ai:2023suz, Long:2024sqg}. We leave the inclusion of these next-to-leading order contributions to the friction for a future version of \wallgo{}.

This paper is organised as follows.
Section~\ref{sec:install} describes how to install \wallgo{} and how to run a first simple model.
In section~\ref{sec:theory}, we detail the physics underlying the computation of the wall velocity.
Section~\ref{sec:code}
overviews the different parts of the software package
and their interactions with each other.
In section~\ref{sec:tests}, we describe convergence tests.
We demonstrate some examples and compare to the literature in
section~\ref{sec:BSM}, and conclude in
section~\ref{sec:conclusion}.

\section{Installation and running}
\label{sec:install}

The main \wallgo{} {\tt Python}
package can be installed directly with pip, by running
the following listing
\makeatletter
\renewcommand\thelstlisting{L.\@arabic\c@lstlisting}
\makeatother
\begin{lstlisting}[language=Bash]
pip install WallGo
\end{lstlisting}
from the command line, and then imported as any other {\tt Python} package.
A first example of how to use \wallgo{} follows in section~\ref{sec:simple}.
A collection of other example models is found at 
\url{\wallgoDocsUrl}.

The \wallgo{} package requires collision integrals as input, loaded from files.
Some pre-computed collision files can be found on the repository, in
the folder {\tt Models}.
Beyond this, installing the \wallgoCollision{} package enables
the computation of new collision integrals, and is also available using pip,
\begin{lstlisting}[language=Bash]
pip install WallGoCollision
\end{lstlisting}
Example usage is also found within the collected example models at \url{\wallgoDocsUrl}.

Finally, the \wallgoCollision{} package requires 2-to-2 scattering matrix elements as input.
The \wallgoMatrix{}
{\tt Mathematica} package allows for computing these for generic models.
To install this package, one must also install
{\tt GroupMath}~\cite{Fonseca:2020vke} on which it depends.
This can be done automatically by setting
\begin{lstlisting}[
  label={lst:wallgomatrix:groupMath},
  language=Mathematica]
WallGo`WallGoMatrix`$InstallGroupMath=True
\end{lstlisting}
in {\tt Mathematica} before loading \wallgoMatrix{}.
The latter can be installed by running the following
\begin{lstlisting}[
  label={lst:wallgomatrix:pacletInstall},
  language=Mathematica]
PacletInstall["WallGo/WallGoMatrix"]
\end{lstlisting}
after which the package can be loaded with
\begin{lstlisting}[style=backtickavailable]
<<WallGo@\backtick@WallGoMatrix@\backtick
\end{lstlisting}
A number of matrix element calculations for example models can be found in the
\href{https://github.com/Wall-Go/WallGoMatrix/tree/main/examples}{{\tt examples}}
folder of the \wallgoMatrix{} GitHub repository.

\subsection{Defining a simple model}
\label{sec:simple}

Defining a model in \wallgo{} requires a few different ingredients:
a scalar potential, a list of the particles in the model together with their properties,
and the matrix elements for interactions between these particles.
The matrix elements are used to compute the collision integrals in \wallgoCollision{}.
The collision integrals are then loaded into \wallgo{} for the wall velocity computation.

We will now describe a simple example which demonstrates how to compute the wall velocity with \wallgo{}.
The relevant file can be found in
{\tt Models/Yukawa/yukawa.py}.
Collision integrals are contained in
{\verb!Models/Yukawa/CollisionOutput_N11!}.
These have been obtained by running the file
{\tt Models/Yukawa/yukawaWithCollisionGeneration.py}.
Generating matrix elements and collisions will be further discussed
in section~\ref{sec:code}.

Concretely,
let us consider a simple model of a real scalar field $\phi$ coupled to
a Dirac fermion $\psi$ via a Yukawa coupling.
Its interaction Lagrangian is given by
\begin{align}
    \mathscr{L} =
	-\frac{1}{2}\partial_\mu \phi \partial^\mu \phi
        - \sigma \phi
        - \frac{m^2}{2}\phi^2
        - \frac{g}{3!} \phi^3
        - \frac{\lambda}{4!} \phi^4
	- i\bar{\psi}\slashed{\partial} \psi
        - m_f \bar{\psi}\psi
        -y \phi \bar{\psi}\psi
        \,.
\end{align}
In this case,
the scalar field may undergo a phase transition,
with the fermion field contributing to the friction for the bubble wall growth.

The first step to implement this model into \wallgo{} is to define a specific model class, inheriting from the base class {\tt WallGo.GenericModel}.
\begin{lstlisting}[language=Python]
import pathlib
import numpy as np
import WallGo
from WallGo import Fields, GenericModel, Particle

class YukawaModel(GenericModel):
    """
    The Yukawa model, inheriting from WallGo.GenericModel.
    """

    def __init__(self) -> None:
        """
        Initialize the Yukawa model.
        """
        self.modelParameters: dict[str, float] = {}

        # Initialize internal effective potential
        self.effectivePotential = EffectivePotentialYukawa(self)

        # Create a list of particles relevant for the Boltzmann equations
        self.defineParticles()

    # ~ GenericModel interface
    @property
    def fieldCount(self) -> int:
        """How many classical background fields"""
        return 1

    def getEffectivePotential(self) -> "EffectivePotentialYukawa":
        return self.effectivePotential

    # ~

    def defineParticles(self) -> None:
        """
        Define the out-of-equilibrium particles for the model.
        """
        self.clearParticles()

        # === left fermion ===
        # Vacuum mass squared
        def psiMsqVacuum(fields: Fields) -> Fields:
            return (
                self.modelParameters["mf"]
                + self.modelParameters["y"] * fields.getField(0)
            ) ** 2

        # Field-derivative of the vacuum mass squared
        def psiMsqDerivative(fields: Fields) -> Fields:
            return (
                2
                * self.modelParameters["y"]
                * (
                    self.modelParameters["mf"]
                    + self.modelParameters["y"] * fields.getField(0)
                )
            )

        psiL = Particle(
            "psiL",
            index=1,
            msqVacuum=psiMsqVacuum,
            msqDerivative=psiMsqDerivative,
            statistics="Fermion",
            totalDOFs=2,
        )
        psiR = Particle(
            "psiR",
            index=2,
            msqVacuum=psiMsqVacuum,
            msqDerivative=psiMsqDerivative,
            statistics="Fermion",
            totalDOFs=2,
        )
        self.addParticle(psiL)
        self.addParticle(psiR)
	\end{lstlisting}

The scalar potential is used both for determining the free energy of homogeneous phases and for the shape and width of the bubble wall.
In principle,
the potentials determining these two phenomena are different, as the former is coarse grained all the way to infinite length scales, while the latter can only consistently be coarse grained on length scales shorter than the bubble wall width~\cite{Langer:1974cpa}.
Nevertheless, at high temperatures and to leading order in powers of the coupling, these two potentials agree.

At high temperatures,
the leading-order effective potential of our simple model is
\begin{equation}
  V^\text{eff}(\phi, T) =
  - \frac{\pi^2}{20} T^4
  + \sigma_\text{eff}\phi
  + \frac{1}{2}m^2_\text{eff}\phi^2
  + \frac{1}{3!}g \phi^3
  + \frac{1}{4!}\lambda \phi^4
  \,,
\end{equation}
where we have defined the effective tadpole coefficient and effective mass as
\begin{align}
  \sigma_\text{eff} &= \sigma + \frac{1}{24}(g + 4y\,m_f)T^2
  \,,
  &
  m^2_\text{eff} &= m^2 + \frac{1}{24}(\lambda + 4y^2)T^2
  \,.
\end{align}
The implementation in \wallgo{} is as follows:
one defines a class, here
{\tt EffectivePotentialYukawa} which inherits from
the base class
{\tt WallGo.EffectivePotential}.
This definition must contain a member function called
{\tt evaluate} which evaluates the potential as a function of
the scalar fields and temperature.

\begin{lstlisting}[language=Python]
class EffectivePotentialYukawa(WallGo.EffectivePotential):
    """
    Effective potential for the Yukawa model.
    """

    def __init__(self, owningModel: YukawaModel) -> None:
        """
        Initialize the EffectivePotentialYukawa.
        """

        super().__init__()

        assert owningModel is not None, "Invalid model passed to Veff"

        self.owner = owningModel
        self.modelParameters = self.owner.modelParameters

    # ~ EffectivePotential interface
    fieldCount = 1
    """How many classical background fields"""

    effectivePotentialError = 1e-15
    """
    Relative accuracy at which the potential can be computed. Here the potential is
    polynomial so we can set it to the machine precision.
    """
    # ~

    def evaluate(self, fields: Fields, temperature: float) -> float | np.ndarray:
        """
        Evaluate the effective potential.
        """
        # getting the field from the list of fields (here just of length 1)
        fields = WallGo.Fields(fields)
        phi = fields.getField(0)

        # the constant term
        f0 = -np.pi**2 / 90 * (1 + 4 * 7 / 8) * temperature**4

        # coefficients of the temperature and field dependent terms
        y = self.modelParameters["y"]
        mf = self.modelParameters["mf"]
        sigmaEff = (
            self.modelParameters["sigma"]
            + 1 / 24 * (self.modelParameters["gamma"] + 4 * y * mf) * temperature**2
        )
        msqEff = (
            self.modelParameters["msq"]
            + 1 / 24 * (self.modelParameters["lam"] + 4 * y**2) * temperature**2
        )

        potentialTotal = (
            f0
            + sigmaEff * phi
            + 1 / 2 * msqEff * phi**2
            + 1 / 6 * self.modelParameters["gamma"] * phi**3
            + 1 / 24 * self.modelParameters["lam"] * phi**4
        )

        return np.array(potentialTotal)
\end{lstlisting}

The
{\tt EffectivePotential} stores the model parameters for use in evaluating the potential. It is possible to override other member functions when defining
{\tt EffectivePotentialYukawa}, such as the initialisation function, or to add additional member functions and variables,
though we have not done so in this simple example.

Once these two classes have been defined, we can now set up the {\tt WallGoManager} and run \wallgo{} to compute the bubble wall speed. What follows now, would be the content of the {\tt main} function of the model file. First, we perform some initialisations.

\begin{lstlisting}[language=Python]
    manager = WallGo.WallGoManager()
    
    # Change the amount of grid points in the spatial coordinates
    # for faster computations
    manager.config.configGrid.spatialGridSize = 20
    # Increase the number of iterations in the wall solving to 
    # ensure convergence
    manager.config.configEOM.maxIterations = 25
    # Decrease error tolerance for phase tracing to ensure stability
    manager.config.configThermodynamics.phaseTracerTol = 1e-8

    pathtoCollisions = pathlib.Path(__file__).resolve().parent / pathlib.Path(
        f"CollisionOutput_N11"
    )
    manager.setPathToCollisionData(pathtoCollisions)

    model = YukawaModel()
    manager.registerModel(model)
\end{lstlisting}

We modify the default value of the spatial grid, the maximum number of iterations in the solution of the equation of motion and the error tolerance of the phase tracer.
We specify the location of the pre-computed collision files.
Now that we have registered the model, we need to provide input parameters, and a number of settings for the computation of the wall velocity.

\begin{lstlisting}[language=Python]
    inputParameters = {
        "sigma": 0.0,
        "msq": 1.0,
        "gamma": -1.2,
        "lam": 0.10,
        "y": 0.55,
        "mf": 0.30,
    }

    model.modelParameters.update(inputParameters)

    manager.setupThermodynamicsHydrodynamics(
        WallGo.PhaseInfo(
            temperature=8.0,  # nucleation temperature
            phaseLocation1=WallGo.Fields([0.4]),
            phaseLocation2=WallGo.Fields([27.0]),
        ),
        WallGo.VeffDerivativeSettings(
            temperatureVariationScale=1.0,
            fieldValueVariationScale=[100.0],
        ),
    )
\end{lstlisting}
In this model, {\tt inputParameters} directly contain the parameters appearing in the potential and the interactions. In a more realistic scenario, the
{\tt inputParameters} would correspond e.g.\ to physical particle masses and interaction strengths.
In the
{\tt WallGo.PhaseInfo} object, we provide the nucleation temperature and the positions of the high-temperature and low-temperature phase. The positions of the phases do not need to be exact.
{\tt temperatureVariationScale} and
{\tt fieldValueVariationScale} are parameters
that are used in the construction of an interpolated free energy and when computing the derivatives of the potential.
A reasonable choice is the difference between the critical and nucleation temperature and the value of the vacuum expectation value (\vev{}), respectively.

Now we can compute the wall velocity. First, we return an estimate of the wall velocity in LTE, computed from hydrodynamics only. Then, the wall velocity is obtained by solving the scalar equation of motion. The first computation ignores the contribution from the out-of-equilibrium fermions, and is therefore very close to {\tt vwLTE}.
Lastly, the wall velocity with out-of-equilibrium contributions is computed.
\begin{lstlisting}[language=Python]
    # ---- Solve wall speed in Local Thermal Equilibrium (LTE) approximation
    vwLTE = manager.wallSpeedLTE()
    print(f"LTE wall speed:    {vwLTE:.6f}")

    solverSettings = WallGo.WallSolverSettings(
        bIncludeOffEquilibrium=False,
        # meanFreePathScale is determined here by the annihilation channels,
        # and scales inversely with y^4 or lam^2. This is why
        # meanFreePathScale has to be so large.
        meanFreePathScale=5000.0,  # In units of 1/Tnucl
        wallThicknessGuess=10.0,  # In units of 1/Tnucl
    )

    results = manager.solveWall(
                    solverSettings
    )

    print(f"Wall velocity without out-of-equilibrium contributions {results.wallVelocity:.6f}")

    solverSettings.bIncludeOffEquilibrium = True

    results = manager.solveWall(
                    solverSettings
    )

    print(f"Wall velocity with out-of-equilibrium contributions {results.wallVelocity:.6f}")
\end{lstlisting}

\section{Theoretical background for the terminal wall velocity of an expanding bubble}
\label{sec:theory}
This section provides an overview of the theoretical background relevant to FOPTs%
\footnote{
  While, in principle, many such phase transitions could occur in the history of our universe, we will always focus on a single transition in this work.
} and
the bubble wall velocity.

\subsection{Effective potential and thermodynamics}
Cosmological phase transitions can occur when the free energy, or effective potential, transitions from a higher to a lower energy state at some temperature. 
In the Standard Model for example, the Higgs field does not break the Electroweak symmetry at high temperatures, but it will transition to a state where the symmetry is broken as the Universe cools.

The temperature at which
two minima of the free energy are degenerate is called the critical temperature, $\Tc$.
If the degenerate minima are separated by a barrier at the critical temperature,
the phase transition will be of first order, and
usually proceeds via the nucleation of bubbles.
Around $\Tc$, the nucleation rate is immensely exponentially suppressed,
and barely any nucleation takes place.
However, the exponential suppression shrinks quickly
as the temperature decreases,
allowing for the phase transition to take place.
For the computation
of the wall velocity, one is usually interested in the temperature
at which non-trivial phase-transition dynamics take place,
such as most of the bubble growth and the collisions of hydrodynamic shock fronts.
A good estimate for this temperature is the percolation temperature, $\Tp$,
at which $\sim 1/e$ fraction of the Universe is in
the metastable phase~\cite{Enqvist:1991xw, Ellis:2018mja}.
Below, we will refer to the temperature of interest as the nucleation temperature, $\Tn$.%
\footnote{%
  Strictly speaking, $\Tn$ and $\Tp$ are not identical, as $\Tn$ corresponds to the temperature
  where the average number of bubbles per Hubble volume is exactly 1. The two quantities are usually very close~\cite{Caprini:2019egz},
  except in models with a large amount of
  supercooling~\cite{Ellis:2018mja, Athron:2023rfq},
  for which \wallgo{} is anyway not applicable.
}
The computation of the temperature of interest is beyond the scope of this work and
it is an input parameter for \wallgo{}.
We refer the reader to~\cite{Wainwright:2011kj, Masoumi:2016wot, Athron:2019nbd, Sato:2019wpo, Guada:2020xnz, Ekstedt:2023sqc}
for numerical tools and to~\cite{Gould:2021ccf, Lofgren:2021ogg, Hirvonen:2021zej, Ekstedt:2022tqk, Hirvonen:2024rfg}
for the theoretical and computational framework for the high-temperature nucleation rate.

Let us now discuss the effective potential in more detail.
We define the scalar fields
as
$\bm\phi = (\phi_1,\phi_2, \cdots)^T$. We will denote the temperature-dependent
effective potential as
$\Veff (\bm\phi, T)$, and it is given by
\begin{equation}
\label{eq:Veff}
  \Veff (\bm\phi, T) =
      V_0(\bm\phi)
    + V_{\rm higher \,order}(\bm\phi, T)
  \,,
\end{equation}
where
$V_0$ denotes the zero-temperature, tree-level potential, and
$V_{\rm higher \,order}$ contains
higher order corrections, e.g.\ the temperature-dependent loop corrections.
We will denote the field values that minimise the potential at $\Tn$ as
$v_\rmii{HT}(\Tn)$ and
$v_\rmii{LT}(\Tn)$ for the
high-temperature and low-temperature phases respectively.

In earlier computations of
the wall velocity~\cite{Moore:1995si,Laurent:2022jrs, Ellis:2022lft},
$V_{\rm higher \, order}$ was approximated by a one-loop effective potential augmented with (in some cases) thermally resumed masses.
In recent years, it has become clear that the one-loop effective potential often does not give accurate predictions
for phase transition parameters such as the nucleation temperature and the phase transition strength~\cite{Croon:2020cgk, Gould:2021oba}, and
higher-loop corrections have to be included. The effective potential determines the critical and nucleation temperature, but also enters in the equation
of motion of the scalar field, as we will see in section~\ref{sec:ScalarEOM}, and therefore directly affects the
value of the wall velocity. As \wallgo{} does not restrict the shape of the potential, we can
now study the effect of these higher loop corrections on the wall velocity for the first time. This will be discussed in a  separate
publication.

We can describe the hot plasma in the early Universe
in terms of the pressure, which is obtained
by evaluating the effective potential at its minima,
\begin{align}
  p_\rmii{HT}(T) &= -\Veff (v_\rmii{HT}(T), T)
	\,,&
  p_\rmii{LT}(T) &= -\Veff (v_\rmii{LT}(T), T)
	\,.
\end{align}
Other thermodynamic quantities such as the enthalpy density $w$, the entropy density $s$, 
the energy density $e$ and
the speed of sound $\cs$ follow from
the relations
\begin{align}
  w(T) &= T\frac{{\rm d}p}{{\rm d}T}
  \,,&
  s(T) &= \frac{w(T)}{T}
  \,,&
  e(T) &= T \frac{{\rm d}p}{{\rm d}T} - p
  \,,&
  \cs^2 &= \frac{{\rm d}p/{\rm d}T}{{\rm d}e/{\rm d}T}
  \,.
\end{align}

\subsection{The scalar equation of motion and the thermal plasma}
\label{sec:ScalarEOM}
The (classical)%
\footnote{%
  Fundamentally all fields participating in the phase transition and the particles are quantum. Nevertheless, over sufficiently large distances and times $t,L\gg T^{-1}$, scalar fields behave classically~\cite{Arnold:1997gh, Arnold:1998cy, Jeon:1995zm}. This follows since low-energy  $E\ll T$ modes are Bose-enhanced, and thus behave classically in accordance with the correspondence principle.
}
equation of motion (EOM) for scalar fields $\phi_i$, coupled to the thermal plasma is:
\begin{equation}
\label{eq:ScalarEOM}
	\partial^2 \phi_i
	+ \frac{\partial \Veff(\bm\phi, T)}{\partial \phi_i}
	+ \sum_a \frac{\partial m_a^2}{\partial \phi_i}
	\int_{\vec{p}} \frac{1}{2E} \delta f^a(p^\mu, \xi)=0
	\,,
\end{equation}
where
$\int_{\vec{p}} = \int\frac{{\rm d}^d p}{(2\pi)^d}$,
$d=3$, and
the sum runs over all particle species in the plasma and $\delta f^a$ denotes the deviation from the equilibrium distribution function.
The equilibrium contributions of all plasma particles have been
absorbed in $V^{\rm eff}$, which equals the potential
of eq.~(\ref{eq:Veff}).

It is useful to choose a coordinate system that follows the expanding bubble. As such our $\delta f^a$ only depends on the momentum $p^\mu$ and the distance from the wall
\begin{equation}
	\xi = - \bar u^\mu_w x_\mu
        \,, 
\end{equation}
where
the 4-velocity
$\bar u^\mu_w$ is
perpendicular to
the 4-velocity
$u^\mu_w$
of the wall, {\em viz.}
\begin{align}
  \bar u_w^\mu &= \gamma_w(\vw,0,0,1)
  \,,&
  u_w^\mu &= \gamma_w (1,0,0,\vw)
  \,.
\end{align}

The last term in eq.~\eqref{eq:ScalarEOM} is the friction caused by out-of-equilibrium particles. Since the friction term is proportional to the field-derivative of the mass, 
in a SM-like plasma, the dominant contribution comes from the top quark, the electroweak gauge bosons and possibly the scalar fields.  

The temperature and fluid-velocity profiles must be solved simultaneously with the scalar-field equations of motion, as the equation of motion depends explicitly on the temperature profile.
To this aim, we obtain two additional equations from energy-momentum, $T^{\mu \nu}$, conservation in the wall frame (here, the wall is assumed to be planar and moving in the $z$-direction)
\begin{align}
\label{eq:EMT}
	T^{30} &=
              w \gamma_{\rm pl}^2 v_{\rm pl}^{ }
            + T^{30}_{\rm out} = c_1
        \,, \nn
        T^{33} &=
            \frac{1}{2}(\partial_z \phi_i)^2
          - \Veff(\bm\phi, T)
          + w \gamma_{\rm pl}^2 v_{\rm pl}^2
          + T^{33}_{\rm out} = c_2
        \,,
\end{align}
where
$v_{\rm pl}$ denotes the local plasma velocity, and
$\gamma_{\rm pl}$ the corresponding Lorentz factor.
The out-of-equilibrium components of the energy-momentum tensor,
$T^{30}_{\rm out}$ and
$T^{33}_{\rm out}$,
are obtained from the following moments of the out-of-equilibrium particle distributions (see~\cite{Laurent:2022jrs} for details)
\begin{align}
\label{eq:Deltas}
    \Delta_{mn}^a(\xi) = \int_{\vec p}\frac{1}{E}E_{\rm pl}^m\,p_{z,{\rm pl}}^n \delta f^a(p^\mu,\xi)
    \,,
\end{align}
where
$E_{\rm pl}$ is the energy and
$p_{z,{\rm pl}}$ the momentum in the $z$ direction
measured in the plasma frame.
The boundary conditions,
$c_1$ and $c_2$
are obtained by solving the hydrodynamic equations for the given
$\Tn$ and $\vw$
(cf.\ e.g.~\cite{LandauLifshitz, Kamionkowski:1993fg, Kurki-Suonio:1995rrv, Espinosa:2010hh}).
These macroscopic hydrodynamic boundary conditions correspond to
$\xi \rightarrow \pm \infty$ for the microscopic description of the bubble wall.
In contrast, when finding the boundary conditions in the macroscopic context of hydrodynamics, they correspond to the temperature and fluid profile immediately behind and in front of the infinitely thin wall.

The previous discussion demonstrated how the $\delta f^a$ need to be known to solve the equation of motion of the $\phi_i$ as well as the temperature and velocity profile. 
Let us assume that we have obtained a solution for $\delta f^a$ and the corresponding $T_{\rm out}^{30}$,
$T_{\rm out}^{33}$.
It is most convenient to solve the equations for $\phi_i$, $T$ and $v_{\rm pl}$
in the rest frame of the bubble wall, where the profiles are simply a function of $\xi = z$.

A common approach is to describe the profiles of the fields that undergo the phase transition by a Tanh-ansatz:
\begin{equation}
	\phi_i(z) =
            v_{\rmii{HT},i}
          + \frac{v_{\rmii{LT},i} - v_{\rmii{HT},i}}{2} \left[1 - \tanh{\left(\frac{z}{L_i} + \delta_i\right)} \right]
        \,,
\end{equation}
where $L_i$ denotes the wall width, $\delta_i$ the offset (the center of the profiles does not need to coincide).
The Tanh-ansatz
does not solve the scalar EOM eq.~(\ref{eq:ScalarEOM}) exactly, but in most known cases the impact of this approximation on the value of the wall velocity is small;
see e.g.~\cite{Moore:1995si,Friedlander:2020tnq}.
\wallgo{} uses the Tanh-ansatz, with the intention to make it optional in
a future version.

There exist several procedures to determine the parameters $L_i$ and $\delta_i$ that give the best solution to the EOM. We chose here to minimise the action that gives rise to
the EOM (\ref{eq:ScalarEOM})
\begin{align}
\label{eq:action}
  S=\int {\rm d}z \left[
      \frac{1}{2}\sum_i(\partial_z \phi_i)^2
      + \Veff\left(\bm\phi,{\bar T}\right)
      - \Veff\left(\bar{\bm\phi},{\bar T}\right)
      + \sum_a m_a^2{\bar \Delta}_{00}^a
    \right]
  \,.
\end{align}
The functions $T$ and $\Delta$ will, in general, depend on the scalar fields.
However, to recover the appropriate EOMs during the minimisation procedure, these need to be independent of $\phi_i$.
Therefore, we replace them by ${\bar T}$ and ${\bar \Delta}_{00}^a$, which are the same functions computed using a fixed scalar field profile ${\bar \phi_i}$.
To find a solution with the correct $T$ and $\Delta_{00}^a$, the minimisation procedure can be repeated iteratively, each time with updated ${\bar T}$ and ${\bar \Delta}_{00}^a$ computed with the previous estimation of $\phi_i$ until convergence is attained.
Note, that the $-\Veff\left(\bar{\bm\phi},{\bar T}\right)$ term does not have an effect on the minimisation as it is just a constant.
It is there to make the integral over $z\in (-\infty,\infty)$ convergent.

One can show that minimising $S$ with respect to $\delta_i$ and $L_i$ is equivalent to solving the moment equations
\begin{align}
  P_i=&\int {\rm d}z \frac{\partial\phi_i}{\partial \delta_i}(\mathrm{EOM})=0
  \,,\\
  G_i=&\int {\rm d}z \frac{\partial\phi_i}{\partial L_i}(\mathrm{EOM})=0
  \,.
\end{align}
The first one can be interpreted as the pressure on the $\phi_i$ wall, as it controls the position of the $\phi_i$ wall with respect to the other walls. The second is the gradient of pressure in the wall, and controls the wall thickness.

From this representation in terms of moment equations, it becomes clear that the minimisation procedure is not sufficient to ensure that the total pressure on the wall $P_{\rm{tot}}=\sum_i P_i$ vanishes, since we need to enforce $\delta_1=0$ to
fix the center of the wall.
There is therefore no corresponding $P_1=0$ equation. However, once the action is minimised with respect to $L_i$ and $\delta_{i>1}$, $P_{\rm{tot}}$ becomes purely a function of $\vw$ and one can finally solve
\begin{align}
P_{\rm tot}(\vw)=0
  \,,
\end{align}
where it is understood that the $L_i$ and $\delta_{i>1}$ are chosen as to minimise $S$.

Finally, even for vanishing $\delta f^a$, the wall still feels a backreaction force from the plasma that gets heated~\cite{Ignatius:1993qn, Konstandin:2010dm, Balaji:2020yrx, Ai:2021kak}. For deflagration and hybrid solutions (wall velocity smaller than the so-called Jouguet velocity $\vJ$), this backreaction force is an increasing function of $\vw$, but for detonation solutions, it decreases with $\vw$, see e.g.~\cite{Ai:2021kak, Laurent:2022jrs}.
 In many cases, this hydrodynamic backreaction effect is already sufficient to obtain a static deflagration or hybrid solution.%
\footnote{%
   Simulations of~\cite{Krajewski:2024gma} suggest that this static solution does not always get reached in a dynamical simulation, and the wall would in fact
   run away without additional friction effects.
  }
For the xSM, it was even shown that the local thermal equilibrium (LTE) approximation gives a reasonable estimate of the wall velocity~\cite{Laurent:2022jrs}.
However, as seen for
the Standard Model with a light Higgs mass in section~\ref{sec:SMLightHiggs} and
the Inert Doublet Model in section~\ref{sec:IDM},
the wall velocity is significantly overestimated in LTE.

\subsection{Boltzmann equations for the plasma particles}
\label{sec:Boltzmann}

As argued before, we need to know the distribution functions of the plasma particles
to solve the scalar-field equation of motion. As the out-of-equilibrium friction is proportional to the derivative of the particle mass, the particles with the greatest shift in their masses are expected to give the dominant contribution to the friction. Nevertheless, light quarks and gluons will still affect the friction indirectly. For example, out-of-equilibrium gluons help top quarks to equilibrate, as the gluon has a much larger cross-section and can act as a catalyst for the quarks.

We will write the distribution function as%
\footnote{%
   Particles such as the Higgs, quarks, and transverse gluons develop gauge-invariant poles at large momenta $\vec{p}^2\sim T^2$; these can be interpreted as thermal masses.
   In defining  $f_{a,{\rm eq}}$,
   we have implicitly included all such thermal masses to one-loop.
   See~\cite{Jeon:1995zm} for a discussion on avoiding double-counting.
}
\begin{align}
  f^a(\vec{p}, \xi) &= f^a_{{\rm eq}}(\vec{p},\xi) +\delta f^a(\vec{p}, \xi)
  \,,&
  f^a_{{\rm eq}} &= \frac{1}{\exp{[p_\mu u^\mu_{\rm pl}(\xi)/T(\xi)] \pm 1}}  \biggr|_{E_a=\vec{p^2}+m_a^2}
  \,,
\end{align}
with
the plus sign for fermions, and
the minus sign for bosons.
For now, we will assume that the temperature $T$ and fluid profile $u^\mu_{\rm pl}$, as well as the scalar field profile are all known (in practice the Boltzmann equation and the scalar field, temperature and fluid equations are solved iteratively).
We will assume that deviations from equilibrium are small
$|\delta f^a| \ll f^a_{\rm eq}$,
which will greatly simplify the computation (the validity of this assumption can be checked in \wallgo{}).
While such nonlinear contributions were also considered in~\cite{DeCurtis:2024hvh},
their effect on the wall velocity was small
for the benchmarks considered there.

The evolution of the distribution function in the wall frame is given by
the Boltzmann equation
\begin{equation}
  \Bigl(
      p^\mu \partial_\mu
    + \frac{1}{2}\vec{\nabla }m_a^2\cdot \nabla_{\vec{p}}
    \Bigr) f^a(\vec{p}, x^\mu) = -\mathcal C_a[\bm f]
    \,,
\end{equation}
where the second term denotes the classical force term that
the particles experience due to the mass change caused by the passing wall.
The collision term $\mathcal C_a$ of particle $a$
describes the interactions between the particles in the plasma and ensures that the distributions
relax to equilibrium far away from the wall.%
\footnote{%
  Far away from the wall, the masses are constant in time, and
  the equilibrium distribution solves the Boltzmann equation exactly.
}
Here, $\bm f$ without a superscript refers to the full set of distribution functions.
The dependence of the collision term on the particle distribution functions makes this system
very difficult to solve in practice.
In turn,
the assumption that the $\delta f^{a}$ are small allows for
linearising the Boltzmann equations which
greatly simplifies the problem.
After the linearisation, the Boltzmann equation in terms of $\xi$ becomes
\begin{equation}\label{eq:BoltzmannXi}
  \Bigl(
    - p_\mu \bar u^\mu_w \partial_\xi
    - \frac{1}{2}\partial_\xi(m_a^2) \bar u^\mu_w \partial_{p^\mu}
    \Bigr) \delta f^a 
    = - \mathcal C_{ab}^{\rm lin}[\delta f^b] + \mathcal S_a
    \,,
\end{equation}
where the source term $\mathcal S_a$ contains the contributions from the equilibrium distribution
\begin{equation}
  \mathcal S_a = \Bigl(
      p_\mu \bar u^\mu_w
    + \frac{1}{2} \partial_\xi(m_a^2) \bar u_w^\mu \partial_{p^\mu}
    \Bigr)f^a_{{\rm eq}}
  \,,
\end{equation}
and the definition of the collision term
$\mathcal C_{ab}^{\rm lin}[\delta f^b]$
can be found in eq.~\eqref{eq:Cactingondeltaf}. 
The linearised Boltzmann equation contains mixing in the collision term.
Hence,
particle $b$ in eq.~\eqref{eq:BoltzmannXi} is not necessarily equal to particle $a$, and
we imply a sum over index $b$.
This mixing effect is usually not considered in computations of $\vw$.

To render the Boltzmann equations in a numerically solvable form,
we follow the procedure developed in~\cite{Laurent:2022jrs}.
This procedure
differs from the seminal work of~\cite{Moore:1995si} in
the description of the distribution functions
which uses the so-called fluid ansatz and
takes three moments of the Boltzmann equations to obtain kinetic equations
for the quantities that parameterise the deviation of equilibrium:
the chemical potential,
the temperature, and
the fluid velocity.%
\footnote{%
  This approach was generalised to a larger number of moments in~\cite{Dorsch:2021nje}.
}
In~\cite{Laurent:2022jrs} however, the distribution functions are expanded on the orthogonal basis of Chebyshev polynomials, $T_i$,
and read
\begin{equation}
\label{eq:deltFlin}
  \delta f^a(\chi, \rho_z , \rho_{\parallel}) =
      \sum_{i=2}^{\Nchi}
      \sum_{j=2}^{\Nrho}
      \sum_{k=1}^{\Nrho-1}
      \delta f^a_{ijk}
      \bar T_i^{ }(\chi)
      \bar T_j^{ }(\rho_z)
      \tilde T_k^{ }(\rho_{\parallel})
      \,,
\end{equation}
where
the compact coordinate
$\chi$ corresponds to the distance from the wall $\xi$,
$\rho_z$ to the momentum perpendicular to the wall $p_z$, and
$\rho_{\parallel}$ to the momentum parallel to the wall $p_\parallel$,
mapped to the interval $[-1,1]$.
See the end of this subsection and
eq.~\eqref{eq:momentumCoordinateEquation} below for details).
The number of basis polynomials
in the spatial direction is $\Nchi$ and
in the momentum direction is $N$.
Utilizing the so-called restricted Chebyshev polynomials
\begin{align}
	\bar T_i(x) &= \begin{cases}
		T_i(x) - T_0(x), & i \text{ even}\,, \\
		T_i(x) - T_1(x), & i \text{ odd}\,,
	\end{cases} \\
	\tilde T_i(x) & = T_i(x) - T_0(x)
        \,,
\end{align}
ensures that
the $\delta f^a$ vanish for
$\xi,p_z \to \pm\infty$ and
$p_\parallel\to\infty$. 
The convergence of the approximation by expanding in polynomials is exponential in the number of basis 
polynomials (this is demonstrated in section~\ref{sec:Convergence}).
Thus, one can approximate the $\delta f^a$ with any desired accuracy.

The virtue of this parameterisation is that the Boltzmann equation
reduces to an algebraic equation of the coefficients
$\delta f^a_{ijk}$,
{\em viz.}
\begin{align}
  \sum_{i,j,k} \bigg\{
    \partial_\xi \chi
    \left[
      \mathcal P_w \partial_\chi - \frac{\gamma_w}{2} \partial_\chi (m^2) (\partial_{p_z} \rho_z)\partial_{\rho_z}
    \right] &
    \bar T_i(\chi) \bar T_j(\rho_z)\tilde T_k(\rho_{\parallel}) \delta f^a_{ijk}
    \nn
    &
    + \bar T_i(\chi)
    \mathcal C_{ab}^{\rm lin} \left[ \bar T_j(\rho_z)\tilde T_k(\rho_{\parallel}) \right] \delta f^b_{ijk}
  \bigg\}
  = \mathcal S_a(\chi, \rho_z, \rho_\parallel)
  \,.\label{eq:BoltzmannMatrix1}
\end{align}

The algebraic equation has $(M-1)(N-1)^2$ coefficients, $\delta f^a_{ijk}$, from the parametrisation in eq.~\eqref{eq:deltFlin}.
They can be uniquely fixed by demanding that the algebraic equation holds on a discrete grid of points, $(\chi^{(\alpha)}, \rho_z^{(\beta)}, \rho_\parallel^{(\gamma)})$:
\begin{align}
  \chi^{(\alpha)} &=-\cos\Bigl(\frac{\pi \alpha}{M}\Bigr)
  \,, &
  \alpha&=1,\cdots,M-1
  \,, \\[1mm]
  \rho_z^{(\beta)} &=-\cos\Bigl(\frac{\pi \beta}{N}\Bigr)
  \,,&
  \beta&=1,\cdots,N-1
  \,,\\[1mm]
  \rho_\parallel^{(\gamma)} &=-\cos\Bigl(\frac{\pi \gamma}{N-1}\Bigr)
  \,,&
  \gamma&=0,\cdots,N-2
  \,.
\end{align}
This precise choice for the grid points ensures the exponential convergence of the spectral method~\cite{boyd2001chebyshev,Laurent:2022jrs}.
See fig.~\ref{fig:ConvergenceLog} below for a numerical example.

With the choice of the grid, $(\chi^{(\alpha)}, \rho_z^{(\beta)}, \rho_\parallel^{(\gamma)})$, the algebraic equation simply becomes a (dense) matrix equation,%
\footnote{%
  The ordering of the indices here is the same as in \wallgo{}.
  Note, that the index ordering in \wallgoCollision{} is transposed, i.e.\
  $\mathcal{C}_{ab}[\alpha,\beta;j,k] \to \mathcal{C}_{ab}[j,k;\alpha,\beta]$.
}
\begin{equation}
	\left(\mathcal{L}[\alpha,\beta,\gamma;i,j,k]\delta_{ab} + \bar T_i(\chi^{(\alpha)})\mathcal{C}_{ab}[\beta,\gamma; j,k] \right) \delta f^b_{ijk} = \mathcal{S}_a[\alpha,\beta,\gamma]
        \,,\label{eq:BoltzmannMatrix2}
\end{equation}
in the indices $\{\alpha, \beta, \gamma \}$ and $\{i,j,k\}$, and where repeated indices are summed.
Here, we have defined the Liouville operator,
the collision operator and the source as
\begin{align}
  \mathcal{L}[\alpha,\beta,\gamma;i,j,k] &\equiv
    \partial_\xi \chi^{(\alpha)}
    \left[
	\mathcal P_w^{(\alpha, \beta, \gamma)} \partial_\chi - \frac{\gamma_w}{2} \partial_\chi (m^2)^{(\alpha)} (\partial_{p_z} \rho_z^{(\beta)})\partial_{\rho_z}
  \right]
  \bar T_i(\chi^{(\alpha)}) \bar T_j(\rho_z^{(\beta)})\tilde T_k(\rho_{\parallel}^{(\gamma)})
  \,, \\[2mm]
  \mathcal{C}_{ab}[\alpha,\beta;j,k] &\equiv
  \mathcal{C}^{\rm lin}_{ab} \left[
      \bar T_j(\rho_z^{(\alpha)})
      \tilde T_k(\rho_{\parallel}^{(\beta)})
      \right]\,,
    \label{eq:collisionTensorGeneral}\\[2mm]
  \mathcal{S}_a[\alpha,\beta,\gamma] &\equiv
    \mathcal{S}_a\bigl(
      \chi^{(\alpha)},
      \rho_z^{(\beta)},
      \rho_\parallel^{(\gamma)}
    \bigr)
    \,.
\end{align}
The derivatives of the basis polynomials arising can be re-expressed
in terms of linear combinations of basis polynomials.

Now that we have the Boltzmann equation in a numerically solvable form,
let us discuss the mapping between compact and physical coordinates in \wallgo{}.
For the momentum directions, these are the same as in~\cite{Laurent:2022jrs},
\begin{align}
\label{eq:momentumCoordinateEquation}
    \rho_z(p_z) &= \tanh\Bigl(\frac{p_z}{2T}\Bigr)
    \,,\nn
    \rho_\parallel(p_\parallel) &= 1-2\exp{\Bigl(-\frac{p_\parallel}{T}\Bigr)}
    \,.
\end{align}
This ensures that the solution vanishes exponentially as $|\mathbf{p}|\to\infty$,
with a decay length of $T$.

In the spatial direction, \wallgo{} uses a more sophisticated mapping
than~\cite{Laurent:2022jrs} to better model
the different scales involved in the solution.
The solution of the Boltzmann equation should decay exponentially with decay lengths $l_\pm$ when $\xi\to\pm\infty$, with $l_-\neq l_+$ in general (we refer to these regions as the \emph{tails}).
Furthermore, the source term is only nonzero within the bubble wall, $|\xi|\lsim L_{\text{wall}}$, where $L_{\text{wall}}$ is the width of the whole wall.
The mapping is therefore constructed in such a way that these three scales are properly resolved.

In practice, the scales $l_\pm$ are estimated from the wall velocity and the parameter {\tt meanFreePathScale},
which must be provided by the user;
see section~\ref{sec:meanFreePathScale} for details on how to choose this parameter properly.
Then, \wallgo{} estimates the parameter
$L_{\tt Grid}\sim L_{\text{wall}}$ from the field profiles $\phi_i(\xi)$.
The mapping also depends on the parameters
{\tt smoothing} and
{\tt ratioPointsWall},
which control how smooth the transition between
the different regions is and the approximate ratio of points used
to resolve the wall in the interval
$\xi\in[-L_{\tt Grid},L_{\tt Grid}]$;
see section~\ref{sec:gridParameters} for more details.

\subsection{Collision terms}
\label{sec:Theorycollision}

In the Boltzmann equation for a given particle species labelled by $a$, the collision term describes the rate at which particles $a$ with given momentum $P_1$ are lost and created due to elastic and inelastic scattering processes. In principle, there exist collision processes which take an initial $a$ particle and $N_\text{in}-1$ other particles into a final state of $N_\text{out}$ particles, i.e.~a process $N_\text{in}\to N_\text{out}$.
However, at leading logarithmic order,
only $2\to 2$ scatterings contribute~\cite{Arnold:2000dr, Arnold:2001ms, Arnold:2002zm};
see section~\ref{sec:Boltzmann:llog}.

The linearised collision terms are linear functionals acting on a set of distribution functions. In principle, these distribution functions carry all the quantum numbers of the corresponding particles, but in cases where certain quantum numbers are irrelevant to the scattering, they can safely be averaged over, e.g.~for helicity in QCD. We label the distribution functions by an index $a,b,c,..$, each denoting a group of degrees of freedom, potentially averaged some quantum numbers.
Then, the linearised collision term appearing in eq.~(\ref{eq:BoltzmannMatrix1})
is given by
\begin{align}
\label{eq:collision-integral}
    \mathcal{C}_{ab}^\mathrm{lin}[\bar T_j(\rho_z) &\tilde T_k(\rho_\parallel)] = \frac{1}{4} \sum_{cde}
    \int_{\vec{p}_{2},\vec{p}_{3},\vec{p}_{4}}
    \frac{1}{2E_2 2E_3 2E_4} (2\pi)^4 \delta^4(P_1 + P_2 - P_3 - P_4)
    \\ & \times
      |M_{ac\rightarrow de}(P_1, P_2 ; P_3, P_4)|^2
      f^a f^c f^d f^e
      \bigl(
          \delta^{ }_{ab} F^c_a
        + \delta^{ }_{cb} F^a_c
        - \delta^{ }_{db} F^e_d
        - \delta^{ }_{eb} F^d_e \bigr)
      \bar T_j(\rho_z) \tilde T_k(\rho_\parallel)
    \,,\notag
\end{align}
where $\rho_z$ and $\rho_\parallel$ are the momenta of the $b$-particle. 
We have also introduced the notation
\begin{equation}
	F^a_b = \frac{e^{E_a/T}}{(f^b)^2}
        \,.
\end{equation}
By denoting
$P_a=(E_a,\vec{p}_a)$,
the particles $\{a,c,d,e\}$ carry four-momenta
$\{P_1,P_2,P_3,P_4\}$
in that order.
$|M_{ac\to de}(P_1, P_2 ; P_3, P_4)|^2 $
are (squared) matrix-elements averaged over the degrees of freedom specified by index $a$ and summed over all other degrees of freedom specified by $c,d,e$,
\begin{align}\label{eq:matrixElementInput}
  |M_{ac\rightarrow de}(P_1, P_2 ; P_3, P_4)|^2 \equiv
    \frac{1}{N_a}
    \sum_{a_i \in a}
    \sum_{c_i \in c}
    \sum_{d_i \in d}
    \sum_{e_i \in e}
    |\mathcal T_{a_i c_i\to d_i e_i}(P_1, P_2 ; P_3, P_4)|^2
    \,,
\end{align}
where $\mathcal T$ is a normal scattering amplitude.
The factor of $\tfrac{1}{N_a}$ arises from the definition of $f^a$ as the occupancy of $a$-particles, averaged over the set of unimportant quantum numbers labelled by $a_i$. The choice of quantum numbers to average over depends on the physical model and situation considered. Considering gluons, for example, here $a_i$ could run over its possible helicities and colours, in which case $N_a=16$. Instead considering a left-handed top-quark, averaging over colour and particle and antiparticle gives $N_a=6$.

The factor of $\tfrac{1}{4}$ on the right hand side of eq.~\eqref{eq:collision-integral} is the product of two factors of $\tfrac{1}{2}$: one from the usual relativistic invariant momentum integration factor $\tfrac{1}{2 E_a}$, and the other which is a symmetry factor if $d$ and $e$ are identical, or a factor which compensates the double-counting in the sum over species if $d$ and $e$ are different~\cite{Arnold:2000dr, Arnold:2002zm}.

The challenge is now to evaluate the 9-dimensional momentum integral for all $n_p^2(\Nrho-1)^4$ components of the collision tensor on the grid;
see eq.~(\ref{eq:collisionTensorGeneral}).
The number of out-of-equilibrium particles is denoted as $n_p$.
Fortunately,
the task becomes easier because half of the components of
$\mathcal C_{ab}^{\rm lin}
[ \bar T_j(\rho_z)\tilde T_k(\rho_{\parallel}) ]
$
are redundant,
\begin{equation}
  \mathcal C_{ab}^{\rm lin}[\bar T_j(-\rho_z) \tilde T_k(\rho_\parallel)] =
  (-1)^j 
  \mathcal C_{ab}^{\rm lin}[\bar T_j(\rho_z) \tilde T_k(\rho_{\parallel})]
  \,,
\end{equation}
which follows from the properties of the Chebyshev polynomials.
Moreover
$\mathcal C_{ab}^{\rm lin} [\bar T_j(0)\tilde T_k(\rho_\parallel)]=0$
for odd $j$, 
or when $\rho_z = \rho_\parallel =0$.
In the absence of vacuum masses,
the momenta can be rescaled as $P_i \to P_i/T$, such that the
temperature only appears as a pre-factor in front of the collision term:
\begin{equation}
  \mathcal C_{ab}^{\rm lin}[\delta f^b] = T^2 \hat{\mathcal C}^{\rm lin}_{ab}[\delta f^b]
  \,.
\end{equation}
We will use this rescaling in \wallgo{} and leave the inclusion of vacuum masses to future work.

Out of the nine integration dimensions, four are trivial due to the momentum-conserving $\delta$-function.
In appendix~\ref{sec:CollisionReduction},
the final form of the collision integral is derived.
The remaining integrals are performed using a Monte Carlo method,
see appendix~\ref{sec:CollisionReduction}.

\subsubsection{The leading-logarithmic approximation}
\label{sec:Boltzmann:llog}

In the integration of collision terms,
kinematic enhancements can increase the parametric size of
certain terms over the naive expectation based on counting powers of couplings.
In particular, the dominant processes involve
$2\to 2$ scatterings of small angle $\theta$, for which the cross-section behaves as
$d\sigma \sim \frac{d \theta}{\theta}$.
Thermal screening softens the divergence at small $\theta$, and leads to a multiplicative $O(|\log g|)$ logarithmic enhancement dependent on
the screening length~\cite{Arnold:2000dr, Arnold:2001ms, Arnold:2002zm}.

State-of-the-art wall velocity computations use the leading-log approximation when calculating the collision rates~\cite{Laurent:2022jrs}.
Such an approximation correctly accounts for all terms which are logarithmically enhanced due to the small angle scatterings, and may truncate or otherwise distort terms which are not so enhanced.
We adopt one particular, common implementation of the leading-logarithmic approximation, wherein masses of the external particles are neglected, and internal lines are regulated by asymptotic thermal masses.%
\footnote{%
  In some of our model files,
  we deviate from this prescription
  to reproduce results obtained in the literature.
}
This approximation was found to lead to an overestimate of the quark diffusion constant by
50\%~\cite{Arnold:2001ms},
consistent with the expected $O(|\log g|^{-1})$ relative errors.

Asymptotic thermal masses result from soft resummation on the lightcone.
Such resummation becomes necessary when
not all momentum components are taken as soft ($P_\mu \sim gT$) but rather
$P^2 \sim (g T)^2$~\cite{Weldon:1982bn, Arnold:2002zm}
and therefore close to the lightcone.
This also happens in a hard regime where $P \sim T$
which is the case for the hard momenta assumed in the matrix elements.
The corresponding resummation near the lightcone gives rise to
a modified dispersion relation
$\omega^2 = k^2 + m_\infty^2$,
where we denote $m_\infty$ as the corresponding asymptotic mass.
For scalars, asymptotic and (static) thermal masses coincide
but are genuinely different for
vector fields and fermions and at LO are
\begin{align}
  m_{\rmii{$V$},\infty}^2 & = \frac{1}{2} \mD^2
  \,,&
  m_{q,\infty}^2 & = 2 m_{q}^2
  \,,
\end{align}
where
$\mD$ is the corresponding Debye mass and
$m_q$ the standard fermionic thermal mass.
See~\cite{Ekstedt:2023anj} for the asymptotic masses of vector fields at
LO and NLO for generic models.
If not stated otherwise,
we employ asymptotic masses throughout our analyses.

Going beyond leading logarithmic accuracy introduces a number of new phenomena and computational difficulties~\cite{Arnold:2001ms, Arnold:2002zm}.
Collision integrals for naively forbidden $1\to 2$ scatterings must be included,
momentum dependence and scalar field background dependence must be resummed in
the particle self-energies, and
gauge and fermion fields experience non-local Landau damping.
Including these effects is beyond the scope of the current work, but we foresee an
investigation of effects beyond leading logarithmic approximation in an upcoming publication.

\section{What happens in the code}
\label{sec:code}

\begin{figure}[t!]
  \centering
  \includegraphics[width=\textwidth]{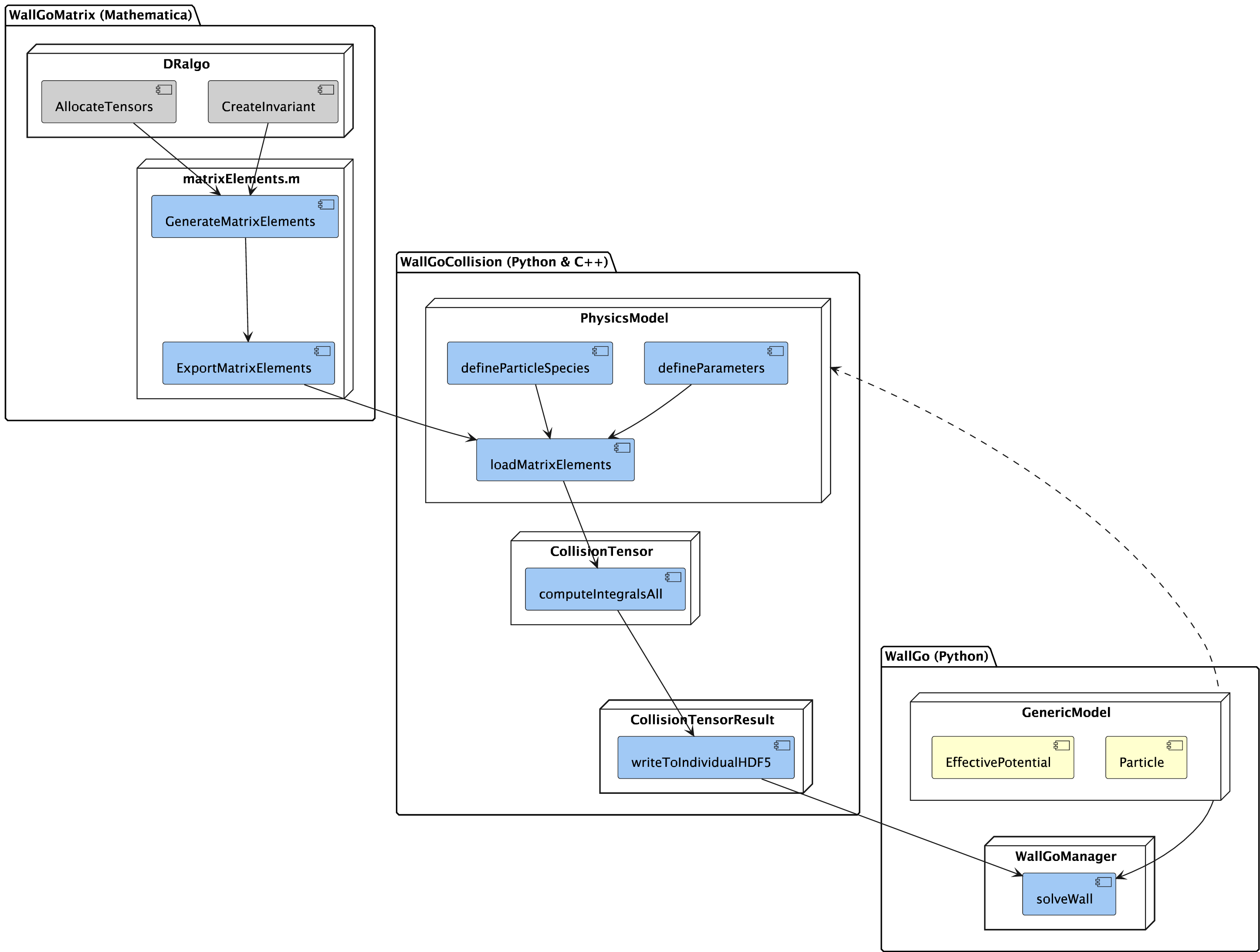}
  \caption{%
    Diagram showing the main parts and functions required to compute the bubble wall velocity for a given particle physics model.
    The main \wallgo{} package solves for the wall velocity with input from the \wallgoCollision{} package, which in turn has input from the \wallgoMatrix{} package.
  }
  \label{fig:CodeDiagram}
\end{figure}

\wallgo{} consists of three parts: 
\begin{itemize}
  \item \wallgo {\tt Matrix}, a {\tt Mathematica} code that determines the matrix elements,
  \item \wallgo {\tt Collision}, a {\tt C++} code (with {\tt Python} wrapper) that computes the collision integrals,
  \item \wallgo{}, the {\tt Python} code that that solves the Boltzmann equations and equations of motion of the scalar 
field(s).
\end{itemize}
Detailed documentation of the code can be found in the online documentation,
\url{\wallgoDocsUrl}.
In this section, we will merely give
a compact overview of the different parts of the code and how they depend on each other.

\subsection{\wallgoMatrix{}}
\label{sec:wallgomatrix}

\wallgoMatrix{} is a {\tt Wolfram Mathematica} tool that generates matrix elements
needed for out-of-equilibrium computations in the \wallgo{} framework.
In \wallgo{}, users can supply their own matrix elements or generate them using \wallgoMatrix{}. The model generation in \wallgoMatrix{} is based on \dralgo{}~\cite{Ekstedt:2022bff}.
The procedure to generate matrix elements is:
\begin{itemize}
  \item[(i)]
    {\bf Loading packages:}
    \wallgoMatrix{} requires
    the files {\tt WallGoMatrix.m},
    {\tt matrixElements.m}, and
    {\tt modelCreation.m} along with its dependency on {\tt GroupMath}.
    Load the package by setting the directory and using the command
    {\tt <<WallGoMatrix'}.

  \item[(ii)]
    {\bf Defining the model:}
    Define gauge groups (e.g., {\tt SU3} and {\tt SU2}) and particle representations. Fermion and scalar representations are specified using their Dynkin coefficients.

  \item[(iii)]
    {\bf Spontaneous symmetry breaking:}
    The tool supports spontaneous symmetry breaking by defining the \vev{} of scalars.
    Use the command {\tt SymmetryBreaking[vev]} to display
    an indexed list of particles according to their symmetry-breaking-induced masses.
    This command is needed if couplings are \vev-dependent or
    if one wants to use the {\tt CreateParticle} command.

  \item[(iv)]
    {\bf Specifying particles:}
    Particles that can be taken in- or out-of-equilibrium are specified by creating their representations and grouping them into distributions
    using e.g.\ the
    {\tt CreateParticle} command.
    For example, left- and right-handed quarks, gluons, and scalar Higgs particles can be defined this way.
    See listing~\ref{lst:no:createparticle}
    or the example
    \href{https://github.com/Wall-Go/WallGoMatrix/blob/main/examples/2scalars.m}{{\tt 2scalars.m}}
    for scenarios that bypass the {\tt CreateParticle} command.
  \item[(v)]
    {\bf Generating matrix elements:}
    Matrix elements for specific particles are generated with the command {\tt ExportMatrixElements}.
    Options for normalisation, truncation at leading logarithmic order, and
    output formats (e.g., {\tt .json}, {\tt .txt}) are available through the {\tt OptionPattern}.
\end{itemize}
The corresponding main functions of \wallgoMatrix{} are documented in
appendix~\ref{sec:Matrix}.
Appendix~\ref{sec:GenMat} details the internal computation of
the matrix elements.
Here, we exemplify a typical matrix element file which can be used to compute the collisions\te
for QCD given in the example
{\tt qcd.m}.
In this scenario,
only the top quark is out-of-equilibrium, and only strong interactions are included in
the matrix elements.

The \wallgoMatrix{} default format is
{\tt .json} and consists of two main keys pointing to arrays
\begin{lstlisting}
{
  "particles": array,
  "matrixElements": array
}
\end{lstlisting}
The
{\tt particles} array contains a list of particle objects, where each particle is defined by its index and name
\begin{lstlisting}
"particles":[
  {
    "index": number,
    "name": string
  }
]
\end{lstlisting}
While the naming does not need to match the naming in the \wallgo{} model file,
the indexing should.
For QCD this corresponds to
\begin{lstlisting}
"particles":[
  {
    "index":0,
    "name":"Top"
  },
  {
    "index":1,
    "name":"Gluon"
  },
  {
    "index":2,
    "name":"LightParticle"
  }
]
\end{lstlisting}
where the {\tt LightParticle} contains all light quarks.

The
{\tt matrixElements} array contains objects that define
the interactions between external particles, the matrix elements.
Each object has the following properties:
\begin{lstlisting}
"matrixElements":[
  {
    "externalParticles": array[number],
    "parameters": array[string]
    "expression": string
  }
]
\end{lstlisting}
The order in the
{\tt externalParticles} array corresponds to
the particles $a,c,d,e$ in eq.~\eqref{eq:collision-integral}.
Strictly speaking, only
the out-of-equilibrium particles need to appear on the first index.
Matrix elements with in-equilibrium particles on the first index will simply be ignored.
For the example of QCD,
{\tt [0,1,0,1]}, denotes scattering of a top quark with a gluon.
Concretely, the output corresponds to
\begin{lstlisting}
"matrixElements":[
  {
    "externalParticles":[0, 0, 0, 0],
    "parameters":["gs","mg2"],
    "expression":"(16*gs^4*(_s^2 + _t^2))\/(3*(mg2 - _u)^2) + (16*gs^4*(_s^2 + _u^2))\/(3*(mg2 - _t)^2)"
  },
  {
    "externalParticles":[0, 1, 0, 1],
    "parameters":["gs","mg2","mq2"],
    "expression":"(-64*gs^4*_s*_u)\/(9*(mq2 - _u)^2) + (16*gs^4*(_s^2 + _u^2))\/(mg2 - _t)^2"
  },
  {
  ...
  },
  {
    "externalParticles":[1, 1, 1, 1],
    "parameters":["gs","mg2"],
    "expression":"(18*gs^4*(_s - _t)^2)\/(mg2 - _u)^2 + (18*gs^4*(_s - _u)^2)\/(mg2 - _t)^2"
  }
]
\end{lstlisting}
The strong coupling constant, {\tt gs},
is the only relevant coupling and must match the corresponding name in \wallgo{}.
If other couplings or parameters are involved,
they must be specified for each matrix element.
The Mandelstam variables are represented by
{\verb!_s!},
{\verb!_t!},
{\verb!_u!}, while
{\tt mq2},
{\tt mg2}
denote the masses of quarks and gluons in the propagators.
In the leading-log approximation,
these masses can be treated as thermal masses.

\subsection{\wallgoCollision{}}
\label{sec:Collision}

Collision integrations are typically by far the most computationally intensive part of the \wallgo{} wall velocity pipeline. \wallgoCollision{} performs the integrations and related operations in native {\tt C++} for maximal performance and versatility. \wallgoCollision{} compiles to a {\tt Python} extension module for seamless interoperation with the main \wallgo{} package, but can also be used as a pure {\tt C++} library.

The purpose of \wallgoCollision{} is to compute elements of the collision tensor (\ref{eq:collisionTensorGeneral}). Integrals on different grid points are fully independent and parallelise trivially. \wallgoCollision{} supports OpenMP for parallel evaluation with shared memory, and has been tested on up to 96 cores. 

To compute collision integrals with \wallgoCollision{}, the user has to
\begin{itemize}
  \item[(i)]
    Create a {\tt PhysicsModel} object containing your particle and model parameter definitions. This should include all particles and parameters that appear in your collision matrix elements, including any light species that can be approximated as remaining in equilibrium but still appear as external particles in collision processes involving out-of-equilibrium particles.

  \item[(ii)]
    Load in matrix elements to the model in a symbolic form. The Mandelstam variables have to be denoted as {\verb!_s!}, {\verb!_t!}, {\verb!_u!}. The assumed physics conventions are as in (\ref{eq:matrixElementInput}). The output files of \wallgoMatrix{} can be used directly as long as the model definition is compatible.

  \item[(iii)]
    Use the model to create a {\tt CollisionTensor} object and pass it the size $N$ of your momentum grid. {\tt CollisionTensor} holds the integrals in an unevaluated but otherwise ready form.

  \item[(iv)]
    Call
    {\tt computeIntegralsAll()} from your
    {\tt CollisionTensor} object to perform the integrations. Once finished, the results can be stored in binary
    {\tt .hdf5} format and loaded into the \wallgo{} Boltzmann solver.
\end{itemize}
Unless otherwise specified, all classes described in this section can be found in
the \wallgoCollision{} {\tt Python} module, and in
the {\tt wallgo} namespace in native {\tt C++}
application programming interface (API).

Model definition in step (i) is done by filling in a {\tt ModelDefinition} helper object and passing it to the {\tt PhysicsModel} constructor. Parameters must be defined as (name, value) pairs; for example, to define symbol ``g" with initial value $0.42$:
{\tt modelDefinition.defineParameter("g", 0.42)}.
Any appearance of the symbol {\tt g} in matrix elements will then be replaced with value $0.42$ during collision integration. Dimensionful parameters must always be given in units of the temperature. In particular, particle masses appearing in propagators of matrix elements are also treated as ``model parameters" and must be given as dimensionless, floating point numbers.

A particle species is defined by passing a
{\tt ParticleDescription} object to your
{\tt ModelDefinition} instance.
A {\tt ParticleDescription} consists of a unique string name, unique integer identifier, matching the {\tt index} of the matrix element file, type specifier (boson or fermion) and a flag for keeping the particle species in thermal equilibrium. The last property can be used to reduce the number of collision integrations for models containing particle species for which deviations from equilibrium are negligible. The default behavior is to treat all particles as ultrarelativistic ($E_i = |p_i|$ during integration). This allows for heavy performance optimisations in collision integration. To go beyond the ultrarelativistic approximation, you can specify a mass function that computes the mass-square of this particle species from given model parameters, again in units of the temperature. 

\wallgoCollision{} performs the integrations using the Vegas algorithm~\cite{Lepage:1977sw}
which is an adaptive Monte Carlo integrator.%
\footnote{%
  Internally we use the GSL~\cite{gsl_vegas} implementation of Vegas.
}
Configuration settings for the integrator are available in
the class {\tt IntegrationOptions},
including the upper limit to use for momentum integration and error tolerances for the Monte Carlo method.
They can be passed to a
{\tt CollisionTensor} object by calling its
{\tt setIntegrationOptions()} function.

Examples of collision generation can be found in the \wallgo{} {\tt Models} folder.
In all these examples, the collision generation and the computation of $\vw$ are combined, though they can also be run independently.

\paragraph*{Limitations}
The release version of \wallgoCollision{} comes with the following limitations regarding the physics content:
\begin{itemize}
  \item
    Only $2 \to 2$ collision processes are supported.
  \item
    All momentum dependence in the matrix elements must be expressed in terms of the Mandelstam variables $s,t$ and $u$.
  \item
    All physics parameters that appear in the matrix elements must be
    constant, floating point valued numbers.
    Dimensionful parameters must be given in units of the temperature.
    Concretely, \wallgoCollision{} works in units of $T=1$.
    The limitation is thus that parameters that vary along the momentum or position grid are not supported.
\end{itemize}
The first two limitations are important for going beyond
the leading logarithm approximation.
The relevance of the third limitation is model dependent.

\subsection{\wallgo{}}
Now that the collision terms have been computed, \wallgo{} can be used to compute the wall velocity. In this subsection, we will discuss certain details of this computation. The computation is performed by a number of classes, such as
{\tt Thermodynamics},
{\tt EOM},
{\tt BoltzmannSolver}, etc.\
The user does not need to keep track of all these classes, as this is done by the {\tt WallGoManager} class, which initialises them in the appropriate order.
In what follows, we will refer to the instance of the
{\tt WallGoManager} as
{\tt manager}.

We will now give some more details about the computation, mainly focussing on the steps the user needs to perform for computing the wall velocity.
A more detailed documentation can be found in
\url{\wallgoDocsUrl},
which will also be up to date with the most recent release of the code. For illustration, we also refer the users to the examples in the {\tt Models} folder.
Note,
that most of the examples rely on the file {\tt Models/wallGoExampleBase.py}, which gives a template for the computation with a specific model.

\paragraph*{Configuration}
To compute $\vw$, \wallgo{} needs a number of model-independent configuration settings, such as the size of the grid on which the momentum and position are discretised, and error tolerances. These are stored in a {\tt Config} object in the {\tt WallGoManager} class, which is initialised with reasonable default values with {\tt WallGo.Config()}. This class contains an instance of the smaller dataclasses {\tt ConfigGrid}, {\tt ConfigEOM}, {\tt ConfigHydrodynamics}, {\tt ConfigThermodynamics} and {\tt ConfigBoltzmannSolver}, which contain settings for the corresponding class. Each individual setting can be accessed with, for example, {\tt config.configGrid.momentumGridSize}. Users can also read in a custom configuration file by running  {\tt config.loadConfigFromFile(<Path to configuration file>)}. See the {\tt Models/ManySinglets/manySingletsConfig.ini} file for a concrete example of a configuration file that can be loaded by {\tt Config}. It also contains a list of all the parameters stored in the {\tt Config} class. The user's custom configuration file should have the same format as this example (note that if some settings are absent from the file their value will not be updated compared to their default value), which can be found in the source code and the online documentation. \wallgo{} throws an error and crashes if the path is not correct.
Let us also stress that the number of Chebyshev polynomials {\tt momentumGridSize} \emph{must always be odd}.

\paragraph*{Setting up the {\tt WallGoManager}}
To prepare the {\tt manager} to compute the wall velocity,
the user must call
{\tt manager.setupThermodynamicsHydrodynamics()},%
\footnote{%
  The object {\tt manager.config} must contain the desired settings
  {\em before} calling this function.
  The settings can be changed individually or by loading a configuration file (see discussion in the previous paragraph).
}
which creates and initialises the objects
{\tt Thermodynamics} and
{\tt Hydrodynamics}
required for the calculation.
This function takes as parameter a {\tt PhaseInfo} object (to be described in the next paragraph) and a {\tt VeffDerivativeScales} object. This data class contains the two model-dependent quantities
\begin{itemize}
    \item {\tt temperatureVariationScale}:
      the temperature scale over which the potential changes by $\mathcal O(1)$ (often $\Tc - \Tn$ is a good estimate). This number is used to estimate the step size in the phase tracer and when taking the temperature derivatives of the potential. If it is chosen too large, \wallgo{} could crash as the phase tracer tries to probe a region where the phase no longer exists. It might also falsely return a run-away wall. If it is chosen too small, the phase tracing will take longer than necessary. In both cases, the temperature derivatives will become less accurate.

    \item {\tt fieldValueVariationScale}:
      the field scale over which the potential changes by $\mathcal O (1)$. This can be given as an array with the length of the number of fields, or as a single float. Usually, the value of the \vev{} is a good estimate. Choosing a too large or too small value will result in inaccuracies in the derivatives, though the effect should be minimal dependence if the order of mangitude is correct.
\end{itemize} 
These parameters are used by \wallgo{} to ensure that it uses an optimal step size when computing the derivatives of the effective potential by finite differences.

\paragraph*{Phase information}
\wallgo{} needs to know between which two phases the phase transition takes place, and at what temperature. This input is provided via a {\tt WallGo.PhaseInput} object, which is a data class, holding the nucleation temperature, and the (approximate) values of the \vev{}s of all fields in the two phases. The high-temperature (outside) phase is listed before the low-temperature (inside) phase. Once {\tt manager.setupThermodynamicsHydrodynamics()} has been called, the precise position of the phases is then determined. It is verified that the phases are not identical, and that the potential energy of the low-temperature phase is 
smaller than the potential energy of the high-temperature phase. If any of these conditions fails, an error will be thrown and the computation cannot continue.
Note,
that it is the responsibility of the user to choose an appropriate nucleation temperature,
as \wallgo{} does not perform additional checks.

\wallgo{} now traces the positions of the fields in the minima for the two phases as a function of the temperature. This serves two purposes. First, it determines for both phases the temperature range for which they exist, obtaining a maximum and minimum temperature for both. Second, it constructs an interpolated function of the value of the effective potential along the way, which speeds up the computation.
To limit the time spent on this step,
the phases are not traced over their entire range of existence. The minimum and maximum temperature are determined by the configuration file. If a too small maximum temperature is chosen, the wall velocity might not be found (\wallgo{} will print a warning if the choice of maximum temperature restricts the range of $\vw$). 

\paragraph{Registration of the model and the effective potential}
The user needs to define a {\tt model}, and register it with the {\tt WallGoManager} by running {\tt manager.registerModel(model)}. 
The model inherits from the abstract base class {\tt GenericModel}. It needs to have the
following properties:
\begin{itemize}
	\item {\tt outOfEquilibriumParticles}: a list of all the out-of-equilibrium particles that enter the friction and Boltzmann equations. See section~\ref{sec:simple} for a concrete example of an object of the {\tt Particle} class. The particles are labeled by a string identifier, e.g.\ {\tt "top"}. The {\tt index} should correspond to the {\tt index} in the corresponding matrix elements file. {\tt msqVacuum} denotes the mass squared without thermal corrections and {\tt msqDerivative} its field-derivative. {\tt statistics} should be equal to {\tt "Boson"} or {\tt "Fermion"}. {\tt totalDOFs} is the total number of degrees of freedom. E.g.\ tor a top quark with only SU(3) interactions this would be 12, but if we distinguish left-handed and right-handed top quarks they would both have 6 DOFs. The particles are added to  the model with the {\tt GenericModel} member function {\tt addParticle}.
	\item {\tt fieldCount}: property function that returns the number of (scalar) fields participating in the phase transition.
	\item{\tt getEffectivePotential}: a member function which returns the effective potential for the field(s) undergoing the phase transition (see below).
\end{itemize}
Typically,  users will add additional functions and properties to the model, such as {\tt modelParameters} and a function converting input parameters to model parameters. 

The {\tt effectivePotential} is also constructed by the user. It should have the same {\tt fieldCount} as the model and it should have a member function {\tt evaluate}, which gives the value of the potential as a function of the field(s) and temperature. Usually, the potential uses the same {\tt modelParameters} as the {\tt model}, but this is implemented by the user.
The {\tt effectivePotential} should also contain an estimate of the relative error of the potential in {\tt effectivePotentialError}.
Note,
that it is important to include field-independent contributions to the effective potential (e.g.\ the $T^4$ contribution). These terms are essential for a correct description of the hydrodynamics of the plasma. Initialisation of the potential class is done inside of the model.

Some functions that are commonly used in the construction of the effective potential are collected in {\tt src/PotentialTools/EffectivePotentialResum}, such as the Coleman-Weinberg potential and the one-loop thermal functions $J_{b,f}$. See e.g.\ the Inert Doublet and xSM model files for examples. The xSM model file also demonstrates how one can load a custom table for the $J_{b,f}$ (we do not advise this in principle, since \wallgo{} has its own interpolation tables, but we can imagine that this is useful for comparing to other implementations).

\paragraph*{Computation of the wall velocity}
The wall velocity and wall parameters can now be computed.
For deflagration or hybrid solutions, one calls
{\tt manager.solveWall(WallSolverSettings)}, where
{\tt WallSolverSettings} is a data class containing the parameters
\begin{itemize}
    \item {\tt bIncludeOffEquilibrium}: boolean which determines whether the out-of-equilibrium contributions should be included (True) or not (False),
    \item {\tt meanFreePathScale}: an estimate of the \textit{longest} mean free path of out-of-equilibrium plasma particles, given in units of $1/\Tn$. This number is used to estimate the extent of the solution for $\delta f$ (see section~\ref{sec:meanFreePathScale} for more details).
      The default value is $50$, which is roughly equal to
      $1/\alphas^2$, with
      $\alphas$ the strong coupling constant. In models having weaker interactions, this value likely has to be adjusted.
    \item  {\tt wallThicknessGuess}: an initial guess of the wall thickness, given in units of $1/\Tn$. This number is model-dependent, but should be larger than $1$ in order for a gradient expansion for the bubble wall to be appropriate. The default value is $5$.
\end{itemize} 

The function {\tt solveWall} relies on the fact that
the pressure is an increasing function of $\vw$ when $\vw<\vJ$.
This ensures that any deflagration or hybrid solution, if it exists,
will be unique and can be bracketed between
0 and $\vJ$.%
\footnote{%
  The allowed velocity range is determined by {\tt Hydrodynamics}. In principle it is given by $\vw \in [0,\vJ]$, but for strong phase transitions a minimum velocity might exist, and for a limited range of existence of the thermodynamic phases, the maximum velocity might be smaller than $\vJ$.
}
This property allows for the use of fast and robust root-finding algorithms
such as Brent's method.
It is first verified that the pressure at the minimum velocity is negative, and positive at the maximum velocity. If the minimum pressure is positive, this indicates a problem in the effective potential, as the low-temperature phase should have lower potential energy than the high-temperature phase. \wallgo{} will return a {\tt WallGoResults} object with ${\tt solutionType} = {\tt ESolutionType.ERROR}$. If the maximum pressure is negative, the friction is not large enough to stop the wall from accelerating and \wallgo{} will return ${\tt solutionType} = {\tt ESolutionType.RUNAWAY}$ as this typically means that the wall would continue accelerating until reaching runaway speeds. If the pressures do have the correct sign, 
the wall parameters are determined by finding the solution which yields zero pressure on the wall. For every attempted $\vw$, the pressure is computed by solving the Boltzmann equation and minimising the action~\eqref{eq:action} with a relative error tolerance for the pressure defined in {\tt WallGo.config}.
The maximum number of iterations for finding the pressure for a particular value of $\vw$ is also specified in
{\tt WallGo.config}.
 
The {\tt solveWall} function returns a {\tt WallGoResults} object,
containing parameters such as the wall velocity and width(s), but also the solution to the Boltzmann equation and the hydrodynamic boundary conditions.
The results for wall velocity, widths and offsets can be assessed via
{\tt results.wallVelocity},
{\tt results.wallWidths} and
{\tt results.wallOffsets}, respectively.
{\tt results.wallVelocityError} gives an estimate of the error of $\vw$, resulting from the finite grid size in the Boltzmann solver. \wallgo{} will throw a warning if the truncation error is estimated to be large. Whenever this happens, the user could increase the momentum or position grid size or adjust {\tt meanFreePathScale}.

To find the wall velocity for a detonation, $\vw > \vJ$, one calls {\tt manager.solveWallDetonation()}. For detonations, the pressure is generally not a monotonous function of $\vw$, and one can therefore find several solutions with zero pressure. In practice, only the roots with increasing pressure with respect to $\vw$ are stable, so the function will only return these roots in a list of {\tt WallGoResults} objects. Even if it is technically possible to have several stable solutions for the same model, most cases will either have zero or one solution. To avoid loosing time looking for a second root that usually does not exist, the user can pass the parameter {\tt onlySmallest=True}, which will force the solver to stop the calculation after finding the first root with the smallest $\vw$. 

Three situations can prevent
{\tt manager.solveWallDetonation()}
from finding a solution\te
each returning a list containing a single
{\tt WallGoResults} object with attribute
\begin{itemize}
  \item
    {\verb!solutionType = ESolutionType.RUNAWAY!}.
    If the pressure is consistently negative,
    the friction from the plasma is insufficient to stop the wall,
    leading to a runaway solution.
    Note, a deflagration or hybrid solution may exist but will not be returned here.
  \item
    {\verb!solutionType = SolutionType.DEFLAGRATION!}.
    If the pressure is consistently positive, friction is too strong to allow for a detonation or runaway solution.
    Here, the returned object
    indicates that only deflagration or hybrid solutions are possible.
  \item
    {\verb!solutionType = ESolutionType.DEFLAGRATION_OR_RUNAWAY!}.
    If the pressure is
    positive at $\vw = \vJ$ and
    negative at $\vw = 1$,
    and if there is no stable solution (there is however an unstable one),
    the outcome is uncertain and would require a time-dependent analysis.%
    \footnote{%
      The positive pressure at $\vw=\vJ$ should in principle stop the wall from accelerating and force it to be a deflagration or hybrid solution. But if for some reason the wall is able to overcome this pressure barrier (which can happen in a time-dependent analysis) and reach the region with negative pressure, the wall will become a runaway solution.
    }
\end{itemize}

A third option for the computation of $\vw$ is
{\tt WallGoManager.wallSpeedLTE}, which returns an approximation of the wall velocity, given by local thermal equilibrium, using hydrodynamics only. This function should give approximately the same result as
{\tt manager.solveWall(bIncludeOffEq = False)}, but is much faster. 
Note,
that this function will never return a detonation solution.

\paragraph*{The kinetic energy fraction}
One of the goals of \wallgo{} is to improve the prediction of GW signals generated in FOPTs. An essential parameter in this prediction is the energy budget
$K$~\cite{Hindmarsh:2015qta, Hindmarsh:2017gnf, Caprini:2019egz, Caprini:2024gyk} for GWs from sound waves. This quantity describes the ratio of kinetic energy in the fluid to the energy density of the symmetric phase. It can easily be obtained from the {\tt Hydrodynamics} module of \wallgo{}, which returns the related efficiency factor $\kappa$ via {\tt Hydrodynamics.efficiencyFactor(vw)}.
The energy budget is then obtained from
\begin{equation}
  K = \frac{3\alpha_n w_n}{4 e_n} \kappa
  \,,
\end{equation}
where the subscript $n$ denotes that the quantity is evaluated at the nucleation temperature. $\alpha_n$ denotes the strength of the phase transition, and we use the definition of~\cite{Giese:2020rtr, Giese:2020znk}.

\paragraph*{Units}
\wallgo{} works with natural units, but otherwise does not enforce units. The user can thus choose units that are appropriate for their physics application, by defining the effective potential and giving input such as the nucleation temperature, mass parameters and field expectation values in the units of choice (e.g.\ GeV, TeV). The only constraint is that one has to use the same units throughout the whole calculation. 

Some functions will ask for quantities given in particular units. If that is the case, it will be clearly stated in the documentation. For example, the {\tt WallSolverSettings} data class, which is used as input by {\tt manager.setupWallSolver()}, requires the {\tt meanFreePathScale} and {\tt wallThicknessGuess} to be given in units of $1/\Tn$.

\paragraph*{Multithreading}
To increase performance, \wallgo{} can be run in parallel to analyse multiple models at once. This can be especially helpful when doing a scan of the parameter space where many parameter points must be sampled. However, many {\tt numpy} and {\tt scipy} functions used by \wallgo{} use multiple threads by default, which saturates the CPU and makes running \wallgo{} in parallel very slow. Furthermore, these multithreaded functions must create and destroy threads frequently, which creates a lot of overhead and renders it quite inefficient. Therefore, we strongly suggest the user to turn off {\tt numpy}'s multithreading. This can be done by running the following before importing {\tt numpy}, {\tt scipy} or \wallgo{}:
\begin{lstlisting}[language=python]
import os
os.environ['OPENBLAS_NUM_THREADS'] = '1'
\end{lstlisting}
when using {\tt OpenBLAS}, or
\begin{lstlisting}[language=python]
os.environ['MKL_NUM_THREADS'] = '1'
\end{lstlisting}
for {\tt MKL}
(whether {\tt OpenBLAS} or {\tt MKL} is used can be found by running
{\verb!numpy.__config__.show()!}).
However, some integrated development environments (IDEs)
use their own multithreading settings,
in which case the previous command might not work.

\section{Convergence tests}
\label{sec:tests}

As with any numerical solver, usage of \wallgo{} depends on a number of meta-parameters, which determine the accuracy of the results.
These enter both in the calculation of the collision integrals, and in the solution of the coupled hydrodynamics, scalar field and Boltzmann equations.

In what follows, we test the dependence of the bubble wall speed on all the most relevant meta-parameters. Unless otherwise stated, we plot the results of these tests for one benchmark point in the Standard Model with singlet scalar, explained below in section~\ref{sec:xsm}.
Our default benchmark point \eqref{eq:BM1} has
\begin{align}
\label{eq:BM1}
  m_s&=120~{\rm GeV}
  \,,&
  \lambda_{hs}&=0.9
  \,,&
  \lambda_{s}&=1
  \,,
  \tag{BM1}
\end{align}
where
$m_s$ is the singlet mass,
$\lambda_{hs}$ the portal coupling and
$\lambda_{s}$ the self coupling.
The specific implementation can be found in
{\verb!Models/SingletStandardModel_Z2/singletStandardModelZ2.py!}.

\begin{figure}[t]
  \centering
  \includegraphics[width=1.0\textwidth]{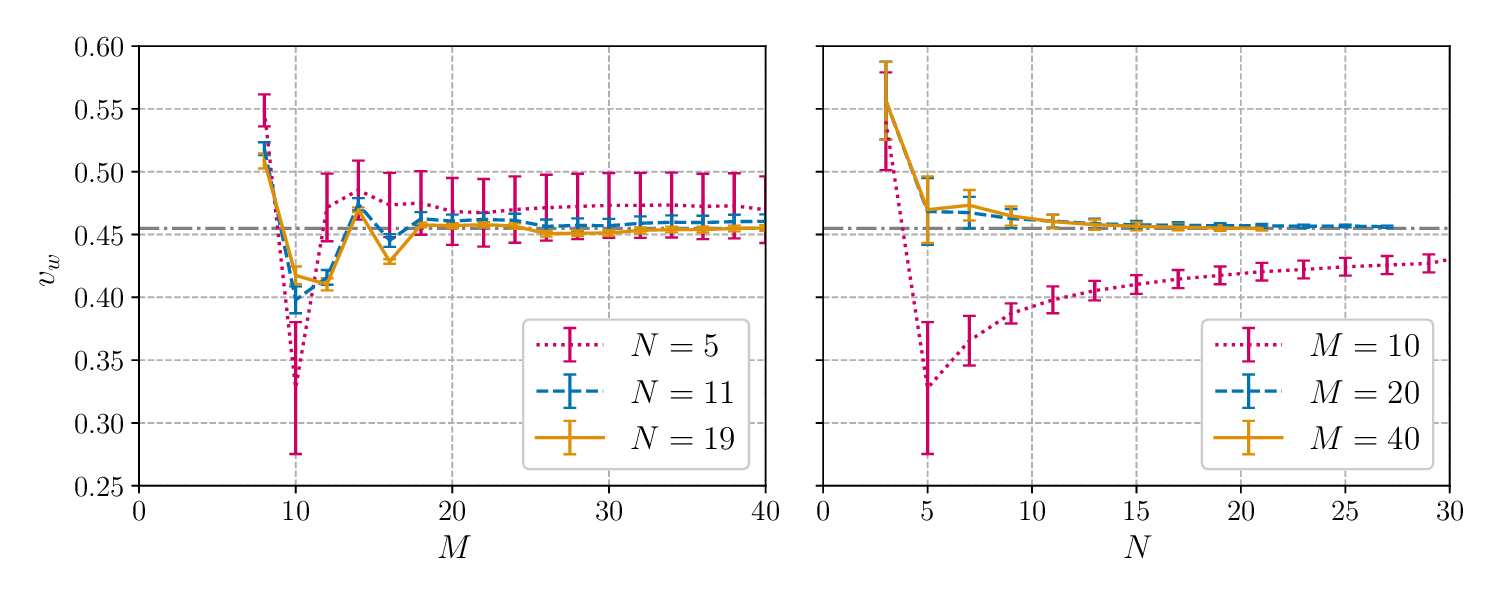}
  \caption{%
    Convergence for the bubble wall speed, at our parameter point~\eqref{eq:BM1} in the xSM, as the numbers of basis polynomials in the spatial ($M$) and momentum ($N$) dimensions are increased. The rate of convergence is expected to be exponential for sufficiently large $N$ and $M$. The dot-dashed line highlights the result for largest $N$ and $M$. All points with $N>20$ were computed on a cluster.
  }
  \label{fig:Convergence}
\end{figure}

\subsection{Testing convergence in the number of basis polynomials}
\label{sec:Convergence}

Perhaps the most crucial meta-parameters are the integer basis sizes $\Nchi$, the spatial basis size or {\tt spatialGridSize}, and $\Nrho$, the momentum basis size or {\tt momentumGridSize}.
These must be sufficiently large to resolve the length and momentum scales present in the system.
For sufficiently large $\Nchi$ and $\Nrho$, the spectral method that we adopt leads to exponential convergence to the continuum limit.

\begin{figure}[t]
  \centering
  \includegraphics[width=1.0\textwidth]{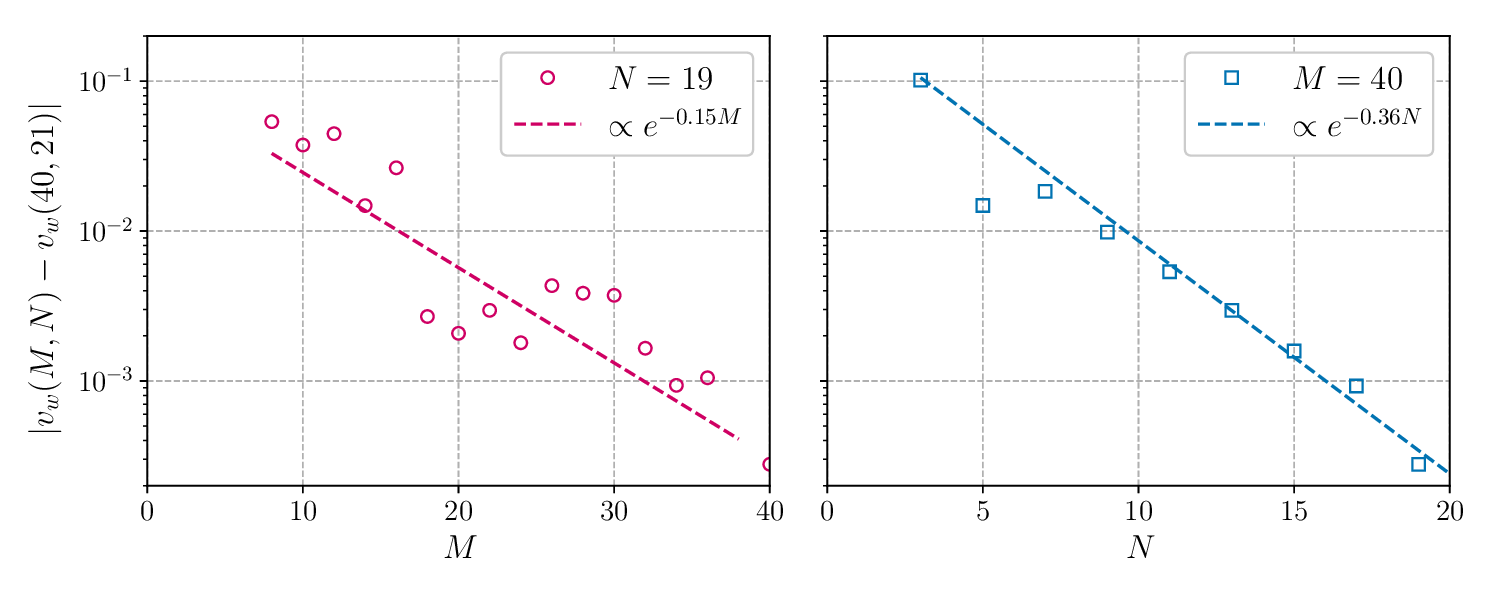}
  \caption{%
    Exponential fits to
    the rate of convergence for
    the bubble wall speed, at our parameter point~\eqref{eq:BM1} in the xSM, as
    the numbers of basis polynomials in
    the spatial ($M$) and momentum ($N$) dimensions increase.
    In computing these data,
    the error tolerance for the wall speed calculation was set to $10^{-3}$,
    establishing a floor for the exponential convergence of the data.
  }
  \label{fig:ConvergenceLog}
\end{figure}

We estimate truncation error of the spectral decomposition by the magnitude of the last coefficient in the Chebyshev basis.
This follows Boyd's Rule of Thumb~\cite{boyd2001chebyshev}, and should give the correct order of magnitude as long as the convergence in $\Nchi$ and $\Nrho$ is exponential rather than power-like.
For one benchmark point in the xSM, the deviation of $\vw$ as a function of $\Nrho$ and $\Nchi$ is shown in figure~\ref{fig:Convergence}.
At this benchmark point, the errors due to finite $\Nchi$ and $\Nrho$ are reduced to a few percent for $\Nchi\gtrsim 20$ and $\Nrho\gtrsim 10$.
Fits of the form $\delta \vw(\Nchi) = a\ e^{-b \Nchi}$ and $\delta \vw(\Nrho) = c\ e^{-d \Nrho}$ yield reasonably good agreement with $b\approx 0.15$ and $d\approx 0.36$ respectively, as can be seen in figure~\ref{fig:ConvergenceLog}. However, these values are not universal.

For additional insight into the approach to the continuum limit, in \wallgo{} we also solve the Boltzmann equation using a finite difference method, in addition to the spectral method.
The finite difference method is accurate up to $O(\Nchi^{-2})$ and $O(\Nrho^{-2})$ and hence its asymptotic convergence is much slower than the spectral method.
It is therefore expected that, for sufficiently large $\Nchi$ and $\Nrho$, differences between the two methods should be larger than the intrinsic truncation error of the spectral method.
\wallgo{} therefore throws a warning if this is not satisfied, and requests a larger basis set.

\subsection{Dependence on {\tt meanFreePathScale}}
\label{sec:meanFreePathScale}

\begin{figure}[t]
  \centering
  \includegraphics[width=1.0\textwidth]{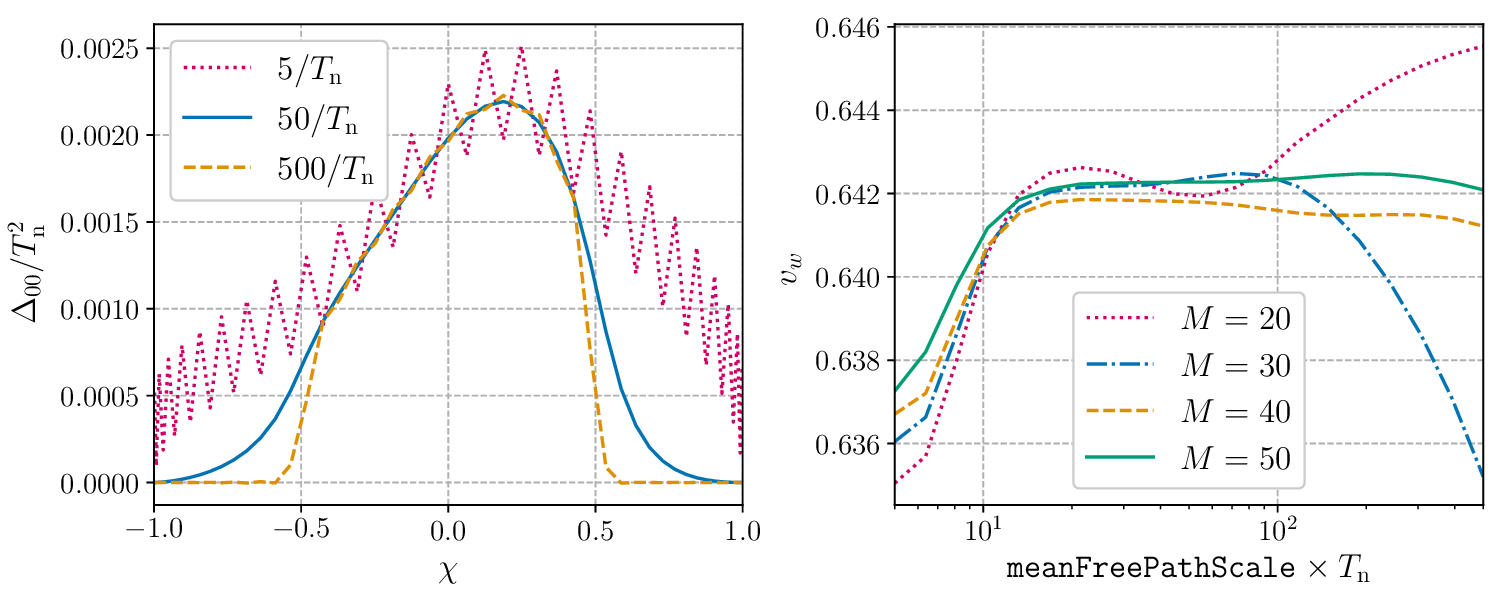}
  \caption{%
    Left: Value of the function $\Delta_{00}(\chi)$ for the top quark obtained by solving the Boltzmann equation with $M=50$ and $\vw=0.1$. Three different values of
    {\tt meanFreePathScale} were used (shown in the legend).
    Right: Variation of the wall velocity obtained by varying
    {\tt meanFreePathScale} with four different values of $M$.
  }
  \label{fig:boltzmannSolutions}
\end{figure}

An important parameter that the user has to select is {\tt meanFreePathScale}. It enters in the {\tt Grid} object via the definition of the mapping between the physical spatial coordinates $z$ and the compact ones $\chi(z)$, cf.\ section~\ref{sec:Boltzmann}. An optimal choice would be similar to the asymptotic decay length of the Boltzmann equation's solution when evaluated at a slow wall velocity. In other words, if the solution behaves like $\delta f(\vw\to 0, z\to\pm\infty)\sim \exp(\mp z/l)$, then the optimal choice for
{\tt meanFreePathScale} is $l$. 
With the chosen $z\to \chi(z)$ mapping and for the optimal choice of the
{\tt meenFreePathScale}, the solution expressed in the $\chi$ coordinates will be
a straight line close to the boundaries at $\chi=\pm1$.
This ensures that, in the $\chi$ coordinates in which the Boltzmann equation is solved, the solution is as smooth as possible which makes it easier to resolve.

In principle, the asymptotic decay length $l$ of the solution is completely determined by the collision operator $\mathcal{C}[f]$.
Therefore, the optimal
{\tt meanFreePathScale} should only depend on the matrix elements. Furthermore, if only strong interactions are included in the matrix elements, a simple estimate gives us $l\sim \frac{1}{\alphas^2 \Tn}\sim 70/\Tn$.%
\footnote{%
  If other interactions are included, $\alphas^2$ can simply be replaced by the relevant couplings. For having the Standard Model $W$s out of equilibrium, the relevant couplings would be $\alpha_w \alphas$.
  This corresponds to slower decay than decay from the strong interaction, and is thus relevant to $l$.
}

A simple way to test whether the chosen value of {\tt meanFreePathScale} is adequate is to plot the solution of the Boltzmann equation in the $\chi$ coordinates. This can be done from a {\tt WallGoResults} object with
\begin{lstlisting}[language=python]
WallGoResults.Deltas.Delta00.coefficients[a]
\end{lstlisting}
which returns the values of the function $\Delta_{00}^a(\chi^{(\alpha)})$ on the $\chi^{(\alpha)}$ grid coordinates. The latter can be accessed via
\begin{lstlisting}[language=python]
WallGoResults.Deltas.Delta00.grid.chiValues
\end{lstlisting}
The left side of figure~\ref{fig:boltzmannSolutions} shows an example of a solution for the xSM~\eqref{eq:BM1} where only the top quark was considered to be out-of-equilibrium, and only including the strong interaction in the matrix elements. Three solutions are shown, computed with three different values of {\tt meanFreePathScale}. Note that all three curves are meant to represent the same physical solution on $z$; only the mapping $z\to\chi(z)$ is different.

From this figure, it is quite apparent that the value
${\tt meanFreePathScale}=500/\Tn$ is too large to efficiently represent the solution. The problem is that the solver samples points too far from the wall where the solution is nearly vanishing. This means that all the points in the intervals
$\chi\in[-1,-0.6]$ and $[0.6,1]$ are essentially wasted. It would therefore be possible to get a similar accuracy with a smaller value of $M$ (and thus a faster calculation) with a better choice of
{\tt meanFreePathScale}.
On the other hand, the value
${\tt meanFreePathScale}=5/\Tn$ is clearly too small, which causes numerical instabilities, appearing as large oscillations in the solution. Here, the solver does not explore the space far away from the wall, so it cannot resolve the solution's tails. This causes the solution's derivative to become infinite at $\chi=\pm 1$, ultimately creating these numerical instabilities. Finally, the value of
${\tt meanFreePathScale}=50/\Tn$ seems much more adequate (note that it is quite close to our previous estimate of $70/\Tn$) as it leads to a smooth solution free from wild oscillations. Furthermore, the solution only vanishes at $\chi=\pm 1$ so each point sampled by the solver is actually increasing the solution's accuracy. 

To understand
to what extent the computed wall velocity depends on
{\tt meanFreePathScale},
we vary it over three orders of magnitude
resulting in less than a 2\% change in $\vw$;
see figure~\ref{fig:boltzmannSolutions} (right) for~\eqref{eq:BM1}.
Even in cases of significant numerical instabilities\te
such as in the
$5/\Tn$ curve of
figure~\ref{fig:boltzmannSolutions}\te
the oscillations tend to cancel out, keeping the effect on  $\vw$ minimal.
However, if these instabilities become excessive,
the solution would eventually diverge completely.
Therefore, we strongly recommend avoiding these instabilities by choosing an appropriate {\tt meanFreePathScale}.
Additionally, as expected,
increasing $M$ reduces the variation in $\vw$ since the larger number of points helps resolve the solution and makes up for a non-optimal value of {\tt meanFreePathScale}.
Also, all the curves roughly agree on the wall velocity when {\tt meanFreePathScale} is between $20/\Tn$ and $100/\Tn$, indicating that the optimal value for this model should be in that interval.

\subsection{Dependence on other {\tt Grid} parameters}
\label{sec:gridParameters}

\begin{figure}[t]
  \centering
  \includegraphics[width=1.0\textwidth]{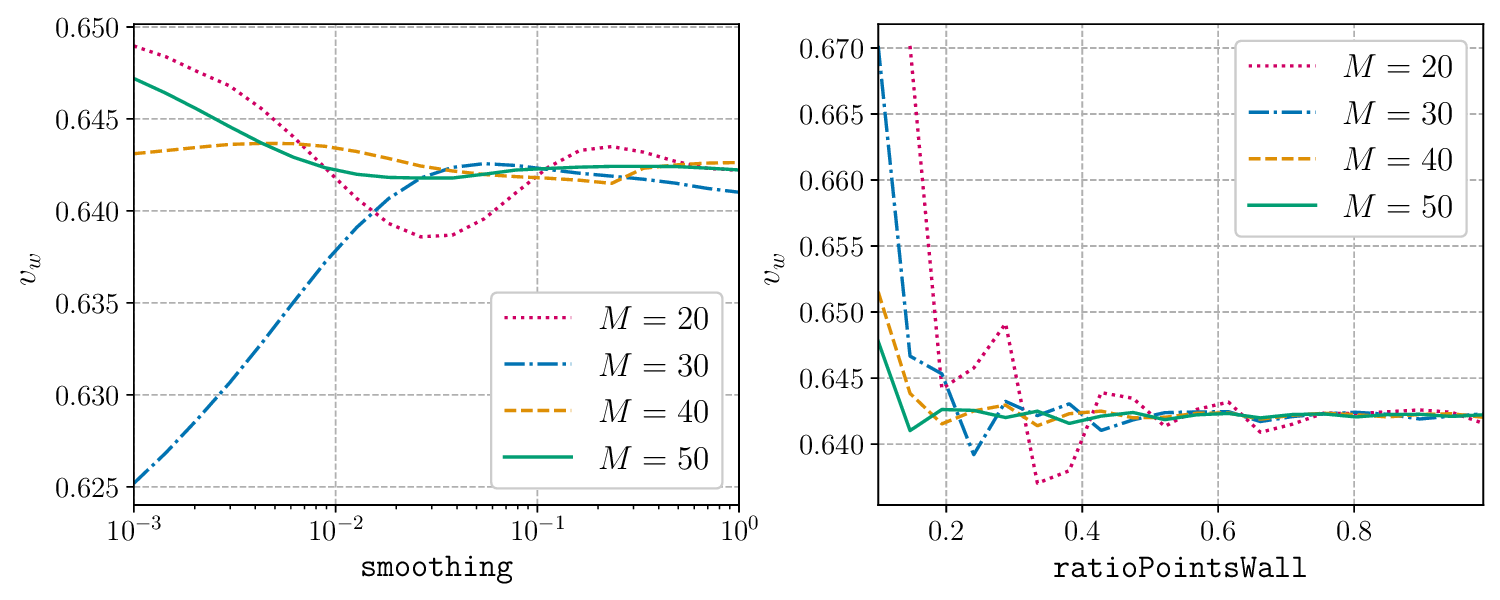}%
  \caption{%
    Dependence of $\vw$ on different values of
    {\tt smoothing} (left) and
    {\tt ratioPointsWall} (right) in the xSM. We used ${\tt ratioPointsWall}=0.5$ in the left plot, ${\tt smoothing}=0.1$ in the right plot, and ${\tt meanFreePathScale}=50/\Tn$ for both.
    }
  \label{fig:VaryingOtherGridParams}
\end{figure}

We investigate here the effect of the remaining {\tt Grid} parameters, {\tt smoothing} and {\tt ratioPointsWall}, on the wall velocity in~\eqref{eq:BM1}.
The current {\tt Grid3Scales} (which inherits from {\tt Grid}) implementation used by \wallgo{} divides the spatial direction into three distinct regions. The tails of the solution, which are located in the intervals
$\xi\in (-\infty,-L_{\tt Grid}]$ and
$\xi\in [L_{\tt Grid},\infty)$ are mapped to the compact intervals
$\chi\in[-1,-{\tt ratioPointWall}]$ and
$\chi\in[{\tt ratioPointsWall},1]$, respectively ($L_{\tt Grid}$ is the wall thickness in the {\tt Grid} object). The density of points in the tails is set to decay exponentially when $\xi\to\pm\infty$, with a decay length set by {\tt meanFreePathScale}.
The density of points in the wall (which is mapped from
$\xi\in[-L_{\tt Grid},L_{\tt Grid}]$ to
$\chi\in[-{\tt ratioPointsWall},{\tt ratioPointsWall}]$) is set to be roughly constant.
The {\tt smoothing} parameter controls the transition from one region to the other. In the ${\tt smoothing}\to 0$ limit, the mapping's first derivative becomes discontinuous at $\chi=\pm{\tt ratioPointsWall}$.
Increasing {\tt smoothing} makes the transitions smoother between the three regions, which removes the discontinuity introduced by the mapping, but makes the distinction between the regions less clear.

The effect of these two parameters on the wall velocity is shown in
figure~\ref{fig:VaryingOtherGridParams}.
Varying {\tt smoothing} by four orders of magnitude only changes the wall velocity by a maximum of 3\%, which shows that the results are not very sensitive to this parameter. One can still observe that best convergence is attained with ${\tt smoothing}\approx0.1$. However, we note that, when {\tt meanFreePathScale} is much larger than the wall width, it can be beneficial to increase {\tt smoothing} up to $\sim1$ to smooth out the large variation of scales between the different regions.
In the right panel of this figure, one can observe that the impact of {\tt ratioPointsWall} on the wall velocity is also small, although best convergence is attained when at least half of the points are used to resolve the wall.
 
\subsection{%
  Dependence on {\tt fieldValueVariationScale} and {\tt temperatureVariationScale}}

To numerically evaluate derivatives, \wallgo{} needs an estimate of the relevant temperature and field scale. A poor choice of these parameters might lead to inaccuracies in the interpolated {\tt FreeEnergy} object, which can result in problems while solving the wall velocity. The solver might e.g.\ mistakenly return a runaway solution, or not be able to find the local thermal equilibrium solution.
Typically,
{\tt temperatureVariationScale} ({\tt TVS}) can roughly be estimated by the difference of the critical and nucleation temperatures, and
{\tt fieldValueVariationScale} ({\tt FVVS}) can be estimated by the \vev{}s of the field.

\begin{figure}[t]
  \centering
  \includegraphics[width=1.0\textwidth]{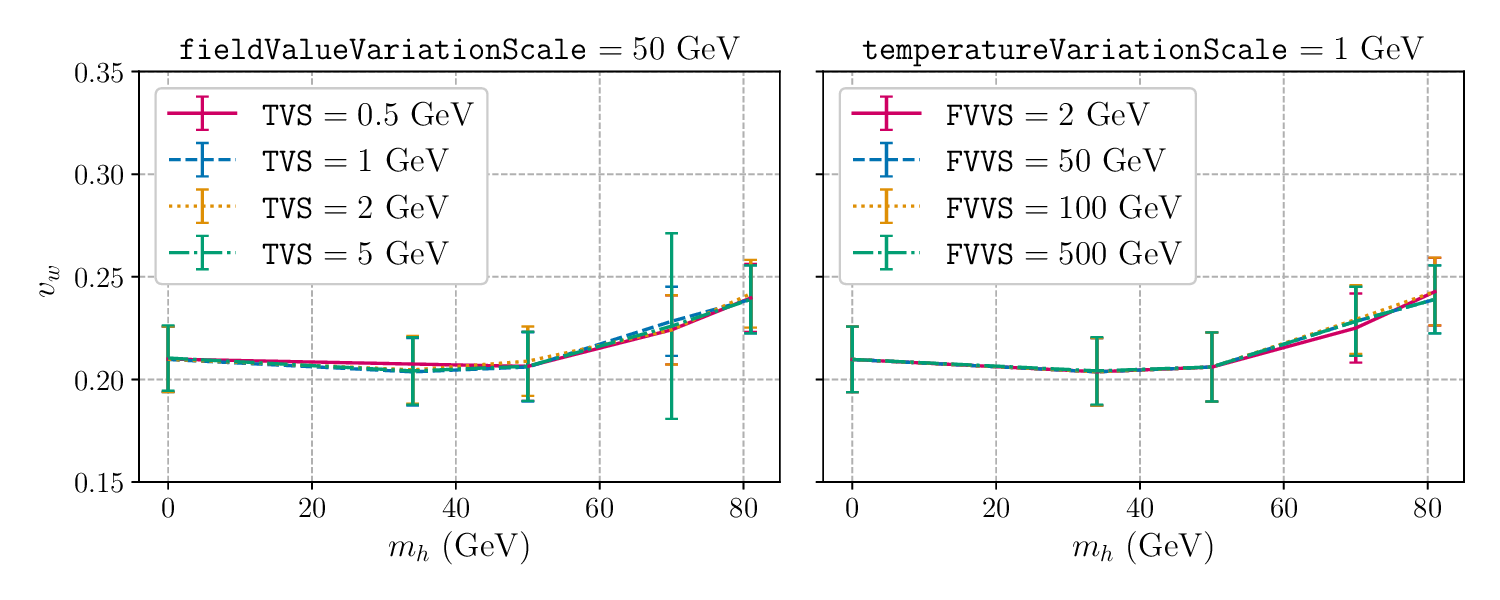}%
  \caption{%
      Dependence of $\vw$ on different values of
      {\tt temperatureVariationScale} ({\tt TVS}, left) and
      {\tt fieldValueVariationScale} ({\tt FVVS}, right)
      for the Standard Model with a light Higgs (cf.\ section~\ref{sec:SMLightHiggs}).
    }
  \label{fig:VaryingTempFieldScale}
\end{figure}
In figure~\ref{fig:VaryingTempFieldScale} we demonstrate the dependence of $\vw$ on these parameters for the Standard Model with a light Higgs
(see section~\ref{sec:SMLightHiggs}).
For the five different Higgs masses we consider, the difference between the critical temperature and nucleation temperature is of the
order of 0.2~GeV, and the \vev{} is of
order 60~GeV.
In the left panel, we keep
the {\tt FVVS} fixed at 50~GeV, and we vary
the {\tt TVS} from 0.5~GeV to 5.0~GeV.
We see that the value of $\vw$ depends only very weakly on the value of the
{\tt TVS}.
For a {\tt TVS} of $0.5$~GeV however, the wall velocity is not recovered for
$m_h = 34$~GeV.%
\footnote{%
  For this particular point, the correct wall velocity would be recovered by decreasing the tolerance of the phase tracer.
  We have however chosen to keep it fixed at $10^{-6}$ for the graph.
}
For a {\tt TVS} of $5.0$~GeV, \wallgo{} prints a warning:\\[2mm]
{\tt
    Warning: the temperature step size seems too large.\\
    \hphantom{Warning: }Try decreasing temperatureVariationScale.}\\[2mm]
Moreover, the obtained $\vw$ for $m_h = 70$~GeV has a large error bar. The latter is a result of \wallgo{} not recovering the LTE value of $\vw$, which enters in the truncation error estimate. From this graph we thus conclude that the optimal choice of {\tt TVS} is $\sim 1$~GeV, slightly larger than the difference between the critical and nucleation temperature for this model.

The right panel of figure~\ref{fig:VaryingTempFieldScale} shows the dependence on
the {\tt fieldValueVariationScale}, for a fixed {\tt TVS} of
$1.0$~GeV.
We see that we can vary the scale over a much larger range than the {\tt TVS}; the results between $2.0$ and $500$~GeV are almost identical, and the differences are much smaller than the truncation error. If we choose a smaller value for the {\tt FVVS}, the phase tracer throws an error.
We thus conclude that the \vev{} is an appropriate choice for the {\tt FVVS}, but that the solution is not very sensitive to it.

\section{Examples beyond the most simple one}
\label{sec:BSM}
\subsection{Standard model with singlet scalar}
\label{sec:xsm}

A simple model that renders the electroweak phase transition first-order and is still allowed by experimental data is the singlet scalar extension (xSM). It is obtained by augmenting the SM with a new scalar field $s$, which is not charged under the SM gauge group. To simplify the analysis, one can impose the singlet field to have an additional $\mathbb Z_2$ symmetry,
in which case the effective potential reads
\begin{align}
  V^{\rm eff}(\Phi,s, T) &=
      \mu_h^2 \Phi^\dagger \Phi
    + \lambda_h (\Phi^\dagger \Phi)^2
    + \frac{1}{2}\mu_s^2 s^2
    + \frac{1}{4}\lambda_s s^4
    + \frac{1}{2}\lambda_{hs} (\Phi^\dagger \Phi)s^2
    \nn &
    + V_\rmii{CW}(\Phi,s)
    + V_\T(\Phi,s, T)
    \,,
\end{align}
where
$V_\rmii{CW}$ is the Coleman-Weinberg potential,
$V_\T$ the thermal potential, and
$\Phi$ denotes the Higgs doublet.
This model only depends on three new parameters:
the singlet mass in the broken phase
$m_s^2=-\lambda_{hs}\mu_h^2/(2\lambda_h)+\mu_s^2$,
its self-coupling
$\lambda_s$, and
its coupling with the Higgs
$\lambda_{hs}$.

We follow here the methodology of~\cite{Laurent:2022jrs} and choose a renormalisation scheme where
$V_\rmii{CW}$ does not change the scalar fields' \vev{}s and masses.
We then have
\begin{align}
    V_\rmii{CW} = \sum_a \pm\frac{n_a}{64\pi^2}\left\lbrace m_a^4(\Phi,s)\left[\log\left(\frac{m_a^2(\Phi,s)}{{\bar m}_a^2}\right)-\frac{3}{2}\right] + 2m_a^2(\Phi,s){\bar m}_a^2\right\rbrace
    \,,
\end{align}
where the upper sign is for bosons and lower one for fermions, the sum is over all the massive particles (which we take to be the top quark, the $W$ and $Z$ bosons, the Higgs, the singlet and the Goldstone bosons), $n_a$ is their number of degrees of freedom, $m_a(\Phi,s)$ their field-dependent mass and ${\bar m}_a$ their mass in the broken phase at $T=0$.
Finally, the thermal potential is given by the one-loop contribution
\begin{align}
  V_\T=\frac{T^4}{2\pi^2}\sum_a \pm n_a \int_0^\infty\! {\rm d}y\, y^2\log\left[1\mp \exp\left(-\sqrt{y^2+m_a^2(H,s)/T^2}\right)\right]
    -\frac{{\tilde g}\pi^2T^4}{90}
  \,,
\end{align}
where the second term is the thermal contribution from the massless degrees of freedom, with ${\tilde g}=83.25$.

To study the stability of \wallgo{} against a wide range of models, we performed a parameter scan of the xSM. We used the same data points and nucleation temperatures that were computed in~\cite{Laurent:2022jrs}, which were obtained with
$\lambda_s=1$,
$m_s\in [62.5,160]$~GeV and
$\lambda_{hs}\in[0.5,1.4]$.
To compute the wall velocity, we only considered the top quark to be out-of-equilibrium.

\begin{figure}[t]
  \centering
  \hspace{-0.09\textwidth}\textbf{Deflagration/hybrid}%
  \hspace{0.33\textwidth}\textbf{Detonation}
  \includegraphics[width=0.5\textwidth]{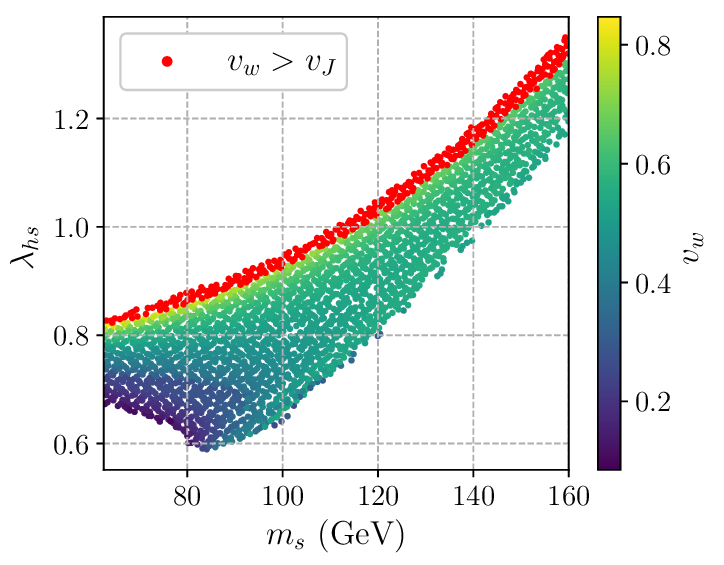}%
  \includegraphics[width=0.5\textwidth]{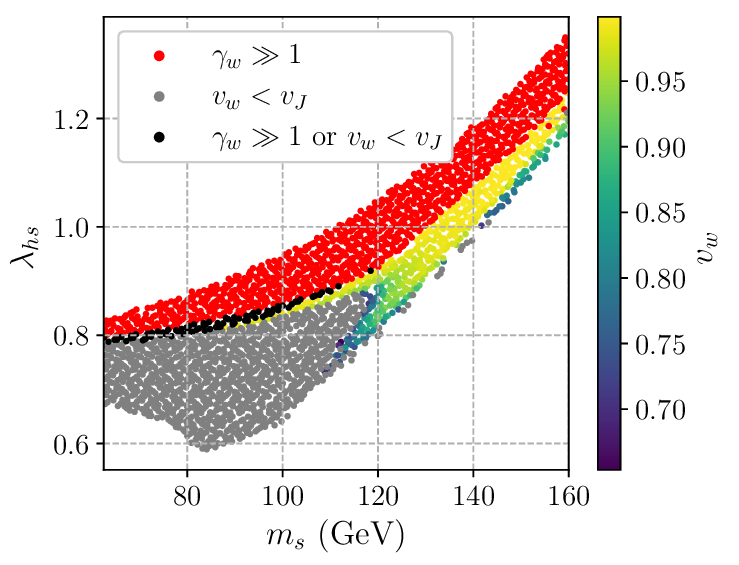}
  \caption{%
    Scan of the xSM parameter space where the singlet mass $m_s$ and the coupling $\lambda_{hs}$ are varied with constant $\lambda_s=1$. The
    left plot shows deflagration and hybrid solutions and the right plot shows detonations.
    The red, black and grey points do not have any solution of the corresponding type. The legend shows what their wall velocity will be based on the sign of the pressure on the wall. 
  }
  \label{fig:scans}
\end{figure}

We show the result of this scan in figure~\ref{fig:scans}, which shows the wall velocity computed by
{\tt WallGoManager} {\tt .solveWall()} on the left side (deflagration and hybrid solutions)
and
{\tt WallGoManger.solveWall} {\tt Detonation()} on the right side (detonation solutions).
In both cases, the corresponding solution did not necessarily exist for every model. If the net pressure on the wall is always negative in the relevant velocity interval (shown in red in the figure), the friction with the plasma would not be strong enough to stop the wall from accelerating and the wall velocity would end up exceeding $\vJ$ for deflagration/hybrid solutions and becoming ultrarelativistic for detonations.
In detonation solutions, there is also a possibility of having a positive pressure everywhere%
\footnote{%
  This cannot happen for deflagration solutions because the pressure is always negative at $\vw=0$, otherwise tunneling would be impossible and the phase transition would not happen.
}
(shown in grey) which indicates that the friction is too strong to allow solutions with $\vw>\vJ$ and that these models can only have deflagration or hybrid solutions. Detonations can also have models with positive pressure at $\vw=\vJ$ and negative pressure at $\vw=1$ with no stable solution (shown in black). A pure static analysis would seem to indicate that these models would become deflagration/hybrid solutions since they would not be able to accelerate over the positive pressure at $\vw=\vJ$, but time-dependent effects could allow them to overcome that positive pressure barrier and end up in the region with negative pressure, in which case it would accelerate to an ultrarelativistic speed. It is unfortunately not possible to conclude with certainty which one of these outcomes would turn out to be true with the static analysis made by \wallgo{}.

It can also be seen that certain models exhibit two solutions:
one deflagration or hybrid and
one detonation solution.
In fact, every model with a detonation solution also has a deflagration or hybrid one.
Again, \wallgo{} is not able to determine which one would be realised in nature because it relies on a static analysis. On one hand, one could argue that because the wall starts still and then accelerates, it would be stopped at the first static solution it encounters, which would be the deflagration/hybrid. On the other hand, these subjouguet solutions sometimes rely on the presence of a shock wave to increase the friction on the wall. If the shock wave does not have enough time to form before reaching the corresponding velocity, the wall could overshoot this first solution and reach the detonation solution (see~\cite{Krajewski:2024gma}, but note that no out-of-equilibrium effects are included in this study). 

Finally, note that, although it is in principle possible to have several detonation solutions for the same model, we found no instance of that happening in this scan.

\subsection{Standard Model with light Higgs}\label{sec:SMLightHiggs}
In~\cite{Moore:1995si,Huber:2013kj, Konstandin:2014zta}, the wall velocity was computed for the Standard Model with a light Higgs mass. The computations all rely on a moment expansion of $\delta f$, and use three moments. We will compare our results with those obtained by Moore and Prokopec in~\cite{Moore:1995si} and Konstandin, Nardini and Rues in~\cite{Konstandin:2014zta}, and from now we will refer to these results as MP and KNR respectively. 
For a sufficiently light Higgs, the electroweak phase transition becomes first order without addition of new particles. The wall velocity becomes a function of the Higgs mass only. Lattice computations show that the phase transition becomes a cross-over for
$m_h \gtrsim 72$~GeV~\cite{Kajantie:1996mn, Karsch:1996yh, Gurtler:1997hr, Rummukainen:1998as}, but in the perturbative description of~\cite{Moore:1995si,Huber:2013kj, Konstandin:2014zta}, the phase transition remains (weakly) first order even for larger Higgs masses as well, and the maximum value we consider is $m_h = 80$~GeV.

The temperature-dependent effective potential used by MP (and also~\cite{Huber:2013kj}) is given in eq.~(7) of~\cite{Moore:1995si}%
\footnote{%
  Eq.~(7) of MP contains one additional term, proportional to $A_F$. This term is set to zero in the benchmark points that we compare to. It is also set to zero by KNR.
}
\begin{equation}
  V^{\rm eff}(h, T) = D(T^2 - T_0^2)h^2 - C T^2 h^2 \log\left(\frac{h}{T} \right) - E T h^3 + \frac{\lambda_T}{4} h^4
  \,,
\end{equation}
where $h$ denotes the (background) Higgs field.
Note that the term proportional to $C$ is absent in the potential of KNR. The coefficients are given by (note that MP's expression for $E$ contains a typo)
\begin{align}
  \lambda_T & = \frac{m_h^2}{2 v_0^2} - \frac{3}{16\pi^2 v_0^4}\left(2 m_w^4 \log{\left(\frac{m_w^2}{a_b T^2} \right)} + m_z^4 \log{\left(\frac{m_z^2}{a_b T^2} \right)} - 4m_t^4 \log{\left(\frac{m_t^2}{a_f T^2} \right)} \right)
  \,, \\
  D & = \frac{1}{ 8 v_0^2} \left(2 m_w^2 + m_z^2 + 2m_t^2 \right)
  \,, \\
  C & = \frac{1}{ 16 \pi^2} (1.42 g_w^4 + 4.8 g_w^2 \lambda_T - 6 \lambda_T^2)
  \,, \\
  E & = \frac{1}{12 \pi} \left(4 \frac{m_w^3}{v_0^3} + 2\frac{m_z^3}{v_0^3} + (3 + 3^{3/2}) \lambda_T^{3/2} \right)
  \,, \\
  B & = \frac{3}{64 \pi^2 v_0^4}\left( 2 m_w^4 + m_z^4  - 4 m_t^4\right)
  \,, \\
  T_0 & = \sqrt{\frac{1}{4D} \left(m_h^2 - 8 B v_0^2 \right)}
  \,.
\end{align}
The term proportional to $\lambda_T^{3/2}$ in $E$ is absent in KNR's potential, and
$E$ gets multiplied by an overall factor $3/2$.
The input parameters can be found in~\cite{Huber:2013kj}
\begin{align}
  g_w &= \frac{2 M_w}{v_0}
  \,,&
  m_w  &= 80.4 \, {\rm GeV}
  \,,&
  m_z  &= 91.2 \, {\rm GeV}
  \,,&
  m_t &= 174.0 \, {\rm GeV}
  \,,
  \nn
  v_0 & = 246.0 \, {\rm GeV}
  \,,&
  a_b &= 49.78019250
  \,,&
  a_f &= 3.111262032
  \,.
  & &
\end{align}
We follow the procedure described by MP and KNR to find the nucleation temperature (this happens outside of \wallgo, we make use of~\cite{Guada:2020xnz}) as a function of $m_h$. Note that MP and~\cite{Huber:2013kj} use $S_3/T = 97$ as nucleation criterion, whereas KNR use $S_3/T = 140$. 

\begin{figure}[t]
  \centering
  \includegraphics[width=0.5\textwidth]{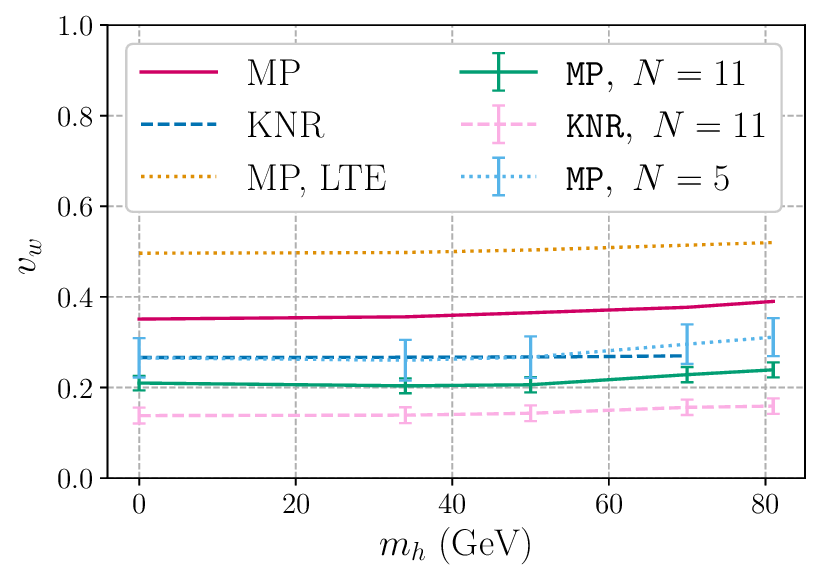}
  \caption{%
      Wall velocity as function of a light the Higgs mass, $m_h$, in the Standard model.
      The results are obtained by \wallgo{} using
      the potentials of MP and KNR.
      The results from MP and KNR are displayed for comparison.
    }
  \label{fig:ComparevwMP}
\end{figure}

We compute $\vw$ with out-of-equilibrium contributions from tops and $W$- and $Z$-bosons (the latter two are treated as a single species, with the mass of the $W$-boson). We do not allow for out-of-equilibrium Higgs particles. This approximation is also made in MP and KNR. We obtain the matrix elements from \wallgoMatrix{}, where we consider the same couplings as MP and KNR.  We perform the computation at $N = 11$, $M = 20$ for the two potentials of MP and KNR. Our results and the results from MP, KNR (we take the results with Tanh ansatz from~\cite{Moore:1995si}) are summarised in figure~\ref{fig:ComparevwMP}.

Before comparing to our results, let us discuss the differences between the results of MP and KNR. The wall velocity reported by KNR (blue dashed line) is significantly smaller than the values obtained by MP (magenta solid line). KNR gives two reasons for the difference. First, the potentials and the nucleation criterion are not identical; we find that the value of the phase transition strength $\alpha_n$ of KNR is a factor $ 0.61 - 0.95 $ smaller than the $\alpha_n$ of MP, which could partially explain the smaller velocity. Second, the treatment of the shock front is different: in MP the fluid profile in front of the wall is approximated by a linearisation of the fluid equations. KNR solve the full fluid equation (as does \wallgo{}) and argues that they therefore correctly capture the backreaction effect from heating~\cite{Konstandin:2010dm, Balaji:2020yrx, Ai:2021kak}, whereas~\cite{Moore:1995si} do not. 
 
The results obtained with \wallgo{}, for $N = 11$, $M = 20$ are smaller than the ones obtained with the moment expansion, for both choices of potential (solid green for the potential of MP, pink dashed for the one of KNR). Part of the difference with MP could be explained by the different treatment of the hydrodynamics. In addition, MP and KNR use the same matrix elements, but some errors in those were pointed out in~\cite{Arnold:2000dr}, and our matrix elements agree with~\cite{Arnold:2000dr}. We include mixing between the different out-of-equilibrium species in the Boltzmann equations, which MP and KNR do not. The method to extract the wall velocity is also slightly different: MP and KNR minimise moments of the equation of motion, whereas we minimise the action to obtain the wall parameters. 

A last prominent difference between our approach and the MP and KNR approach is the different method used to solve the Boltzmann equations - we expand $\delta f$ in Chebyshev polynomials rather than taking a moment expansion. For comparison, we demonstrate our result for $N = 5$, $M = 20$ for the MP potential in dotted light blue, and we see that our value for $\vw$ becomes significantly larger (but still smaller than the result of MP). This begs the question whether the computation of $\vw$ obtained with just two moments has really converged. In~\cite{Dorsch:2021nje}, the corrections from higher moments to the friction was computed. The distribution from particles out-of-equilibrium increased with the number of moments, suggesting that $\vw$ decreases with an increasing number of moments. Unfortunately, the dependence of $\vw$ on the number of moments was not computed. It is therefore yet an open question whether the two methods converge towards the same value of $\vw$, which we leave to future work.

For comparison, figure~\ref{fig:ComparevwMP} also shows the local thermal equilibrium result for the potential of MP in dotted yellow. As expected, $\vw$ is larger than with out-of-equilibrium effects included. Unlike for the xSM, where it was demonstrated in~\cite{Laurent:2022jrs} that LTE often gave a reasonable estimate for the full $\vw$, the LTE result does not give a very good prediction of $\vw$, as the two values are different by about a factor 3.

\subsection{Inert doublet model}
\label{sec:IDM}

The wall velocity was computed for three benchmark points of the Inert Doublet Model in~\cite{Jiang:2022btc}, and we will now compare to the results obtained with \wallgo{}. The corresponding model file can be found in {\tt Models/InertDoubletModel/inertDoubletModel.py}. 
The Inert Doublet Model~\cite{Deshpande:1977rw, Ma:2006km, Barbieri:2006dq} is a special case of the Two Higgs Doublet Model, where the new doublet has the same SM gauge charges as the SM Higgs field. In addition, it has a global $\mathbb Z_2$-symmetry, which prevents tree-level couplings to the Standard Model fermions.
The model can provide
a dark matter candidate~\cite{Barbieri:2006dq, LopezHonorez:2006gr, LopezHonorez:2010eeh},
and features a first-order phase transition in part of its
parameter space~\cite{Ginzburg:2010wa, Chowdhury:2011ga,Borah:2012pu, Gil:2012ya, Cline:2013bln, Blinov:2015sna, Blinov:2015vma, Basler:2016obg, Laine:2017hdk}. 

To make a faithful comparison of $\vw$, we will follow closely the implementation of the effective potential used in~\cite{Jiang:2022btc}. Note that the computation of~\cite{Laine:2017hdk} suggests that an accurate computation might require the inclusion of higher-order corrections in the effective potential, but we leave these to future work. The scalar potential at zero-temperature is given by (note that our normalisation of $\lambda_{1,2}$ differs from~\cite{Jiang:2022btc})
\begin{align}
  V_0 =
      \mu_1^2 \Phi^\dagger \Phi
    + \mu_2^2 \chi^\dagger \chi
    &
    + \lambda_1 (\Phi^\dagger \Phi)^2
    + \lambda_2 (\chi^\dagger \chi)^2
    \nn &
    + \lambda_3 \Phi^\dagger \Phi \chi^\dagger \chi
    + \lambda_4 \Phi^\dagger \chi \chi^\dagger \Phi
    + \biggl[ \frac{\lambda_5}{2} (\Phi^\dagger \chi)^2
        + \mbox{H.c.}
    \biggr]
  \,,
\end{align}
where $\Phi$ denotes the SM Higgs boson, and $\chi$ the inert doublet
\begin{equation}
	\chi = \begin{pmatrix}H^+ \\ \frac{1}{\sqrt 2} (H + iA) \end{pmatrix}.
\end{equation}
We have four new degrees of freedom: the CP-even and CP-odd scalars $H$ and $A$ respectively, and the charged scalars $H^\pm$. Their zero-temperature masses correspond to
\begin{align}
  \bar m_{\rmii{$H$}}^2 &= \mu_2^2 + \frac 1 2 (\lambda_3 + \lambda_4 + \lambda_5) v^2
  \,, \\
  \bar m_{\rmii{$A$}}^2 &= \mu_2^2 + \frac 1 2 (\lambda_3 + \lambda_4 - \lambda_5) v^2
  \,, \\
  \bar m_{\rmii{$H^\pm$}}^2&= \mu_2^2 + \frac 1 2 \lambda_3 v^2
  \,,
\end{align}
where $v$ denotes the zero-temperature Higgs \vev{}. In the following, we will choose $\lambda_4 = \lambda_5$, such that
$\bar m_{\rmii{$A$}}^2 = \bar m_{\rmii{$H^\pm$}}^2$.
The combination appearing in the mass
$\bar m_{\rmii{$H$}}^2$ is denoted as
$\frac{1}{2} (\lambda_3 + \lambda_4 + \lambda_5) \equiv \lambda_{\rmii{$L$}}$.

We consider a one-step phase transition where the Higgs-field obtains a \vev{},
$\Phi = 1/\sqrt 2(0, h)^T$. The effective potential for $h$ is given by
\begin{equation}
  V^{\rm eff}(\phi, T) =
      V_0(\phi)
    + V_\rmii{CW}(h)
    + V_\T(h, T)
    \,.
\end{equation}
For further details on the implementation, see~\cite{Jiang:2022btc} or the {\tt inertDoubletModel.py} file. We have confirmed that we obtain the same critical temperature as~\cite{Jiang:2022btc} with our implemented potential. We take the value of the nucleation temperature from~\cite{Jiang:2022btc}. The parameters of the three benchmark points are given in table~\ref{tab:IDM}.
For all benchmark points, $\lambda_2 = 0.1$.
Note that the running of the couplings is not considered.
\begin{table}
\centering
\begin{tabular}{ |c|c|c|c|c|c|c|c|c|}
 \hline
  BM & $\bar m_{\rmii{$H$}}$ [GeV] &
  $\bar m_{\rmii{$A$}}, \bar m_{\rmii{$H^\pm$}}$ [GeV] &
  $\lambda_{\rmii{$L$}}$ &
  $\Tc$ [GeV] &
  $\Tn$ [GeV] &
  $\vw$~\cite{Jiang:2022btc} &\
  $\vw$ [\wallgo{}]\\
  \hline\hline
  A & 62.66 & 300 & 0.0015 & 118.3& 117.1 & 0.165 & $0.191 \pm 0.024$ \\
  B & 65.00 & 300 & 0.0015 & 118.6& 117.5 & 0.164 & $0.180 \pm 0.025$ \\
  C & 63.00 & 295 & 0.0015 & 119.4& 118.4 & 0.164 & $0.182 \pm 0.024$ \\
 \hline
 \end{tabular}
 \caption{%
    Benchmark (BM) input parameters used in~\cite{Jiang:2022btc} and results for $\vw$ in~\cite{Jiang:2022btc} and \wallgo{}.}
 \label{tab:IDM}
\end{table}

\cite{Jiang:2022btc} considers the out-of-equilibrium contributions from the top quark, the $W$- and $Z$-boson and the $A$ and $H^\pm$ scalars.
The $W$- and $Z$-boson are treated as a single species (with the mass of the $W$), and similarly the $H^\pm$ and $A$ are treated as a single species with the mass of the $A$. The Boltzmann equations are solved with a moment expansion as in~\cite{Moore:1995si}. The wall is approximated with a tanh-ansatz. The hydrodynamics equations are solved without linearisation, but the fluid equation of motion is approximated by the pure radiation-contribution and temperature-independent vacuum energy difference only (bag equation of state). 

We use \wallgo{} to compute the wall velocity for the same out-of-equilibrium particles at a \\
{\tt momentumGridSize = 11} and
{\tt spatialGridSize = 30}.
The matrix elements are determined with \wallgoMatrix{}. We include the same external particles and couplings as in~\cite{Jiang:2022btc},
(i.e.\ we set $\lambda_{1,2,4,5}, \gY = 0$).
We end up with a larger set of diagrams than~\cite{Jiang:2022btc}, since certain diagrams were neglected there, for example interactions between $A$ and $W$.
As position-dependent \vev{}s are currently not supported in \wallgo{},
we fix the \vev{} to be equal to half of
$v_\rmii{LT}(\Tn)$ of benchmark point A.

The last two columns of table~\ref{tab:IDM} compare the obtained values of $\vw$. In all cases, the result of \wallgo{} is slightly larger than the result of~\cite{Jiang:2022btc}, but overall there is a reasonable agreement, with deviations smaller than $20 \%$. The observed differences are likely caused by the different approaches used in the Boltzmann equations (taking moments versus the spectral method), the slightly different treatment of the hydrodynamics of the plasma, and the differences in the matrix elements and corresponding collision terms. As we included a larger set of matrix elements, this could also explain why we observe a larger value of $\vw$, as the particles can get slightly closer to their equilibrium distribution, reducing the corresponding friction.

\section{Conclusions}
\label{sec:conclusion}

We present \wallgo{},
a {\tt Python} package for calculating the wall velocity, $\vw$,
in first-order cosmological phase transitions.
\wallgo{} includes two auxiliary packages,
\wallgoMatrix{} and
\wallgoCollision{}.
Together, they perform the entire computation from
calculating the matrix elements for in- and out-of-equilibrium particles, to
computing the collision integrals for the Boltzmann equations, to
solving the coupled hydrodynamics, scalar-field equation of motion and
Boltzmann equations for the out-of-equilibrium particles.
The model, along with the set of out-of-equilibrium particles, is fully user-defined.
Since the wall velocity significantly affects the predictions for GW spectra and the baryon or dark matter abundance,
\wallgo{} has the potential to enhance future predictions of observables related to cosmological first-order phase transitions.

In this article,
we have presented a quick start guide, theoretical background and a succinct documentation of the code.
A more extensive documentation of the code can be found in
\url{\wallgoDocsUrl}.
The online documentation will also stay up-to-date with future versions of \wallgo{}.
By varying the number of basis polynomials 
we demonstrate that the spectral expansion indeed converges exponentially in section~\ref{sec:tests}. 
In addition, we show that the wall velocity has only a mild sensitivity to the other configuration parameters.
We also provide examples of tests that the user can perform to determine whether their obtained value of $\vw$ has converged.
In section~\ref{sec:BSM},
we demonstrate computations of $\vw$ for several BSM models.
In section~\ref{sec:xsm},
we use \wallgo{} to study the xSM, demonstrating that \wallgo{} can be used to scan over a parameter space.
The results agree with those obtained earlier in~\cite{Laurent:2022jrs}.

For the Standard Model with a light Higgs in section~\ref{sec:SMLightHiggs}, and the Inert Doublet Model (IDM) in section~\ref{sec:IDM}, we compare our results to the results presented in~\cite{Moore:1995si, Konstandin:2014zta} and~\cite{Jiang:2022btc} respectively. In the case of the Standard Model, we observe that the obtained wall velocities differ significantly from the values obtained in~\cite{Moore:1995si, Konstandin:2014zta} with our results being almost a factor 2 smaller.
For the IDM,
we only deviate from the results of~\cite{Jiang:2022btc} by less than $20\%$.

A number of factors could explain the significant differences with~\cite{Moore:1995si, Konstandin:2014zta} and the milder differences with~\cite{Jiang:2022btc}:
all three references expand the $\delta f_a$ in three moments, which is known to yield a singular solution at $\cs$~\cite{Laurent:2020gpg, Dorsch:2021nje}, due to the linearisation of the background solution. We use the spectral expansion, which allows us to converge exponentially quickly towards the exact solution for $\delta f_a$. Moreover, we do not need to linearise the background solution. Unlike~\cite{Moore:1995si, Konstandin:2014zta, Jiang:2022btc}, we have included mixing in the collision terms in the Boltzmann equations. There are also some errors in the matrix elements used in~\cite{Moore:1995si, Konstandin:2014zta}, pointed out by~\cite{Arnold:2000dr}. We also use a slightly larger set of matrix elements than~\cite{Jiang:2022btc}. Lastly, there are differences with~\cite{Moore:1995si} in the treatment of the hydrodynamics outside of the bubble wall. As pointed out by~\cite{Konstandin:2014zta}, that treatment incorrectly captures the backreaction force from heating.
Finding the main source of the discrepancy will require a more in-depth comparison, which will be the topic of future work.

For future versions of \wallgo{},
we foresee several improvements, such as dropping the Tanh-ansatz,
including matrix elements and collision integrals beyond leading logarithmic order,
allowing for field-dependent couplings and masses, and the implementation of  a consistent treatment of soft gauge modes.
These improvements involve both technical and theoretical challenges.

Having automated the full computation of the wall velocity,
from matrix elements to Boltzmann equations,
\wallgo{} has the potential to greatly improve predictions of FOPT observables.
Rather than being the final verdict in the computation of $\vw$,
\wallgo{} opens a new chapter of precision computations for FOPTs.
The automation of the computation will allow us to investigate and address
the main sources of uncertainty
which will be a topic of upcoming work
after \href{\wallgoPypiUrl}{{\tt pip} {\tt install}ing \wallgo}.

\clearpage

\section*{Acknowledgements}
We thank
Tuomas Tenkanen for his impact and contributions
during the initial stages of this project.
We are also grateful to
Mark Hindmarsh,
Maciej Kierkla,
Thomas Konstandin,
Johan L\"ofgren,
and Kari Rummukainen
for discussions as this project was carried out.
We thank Fa Peng Huang for communication on the implementation of the Inert Doublet Model.
We thank Wen-Yuan Ai, Jonas Matuszak, Jasmine Thomson-Cooke and Miguel Vanvlasselaer for testing \wallgo{}.

AE was supported by the Swedish Research Council, under project number VR:2021-00363.
OG was supported by a Royal Society Dorothy Hodgkin Fellowship, and is grateful for access to the University of Nottingham's Ada HPC service.
JH was supported by the same Royal Society Dorothy Hodgkin Fellowship and by Research Council of Finland grants 1345070 and 1354533.
BL was supported by the Fonds de recherche du Québec Nature et technologies (FRQNT).
LN was supported by Academy of Finland grant 354572.
PS was supported by
the Swiss National Science Foundation (SNF) under grant PZ00P2-215997
and
the Deutsche Forschungsgemeinschaft (DFG, German Research Foundation) through
the CRC-TR 211 `Strong-interaction matter under extreme conditions' --
project no.~315477589 --
TRR 211.
JvdV was supported by the Dutch Research Council (NWO), under project number VI.Veni.212.133.

We are very grateful to the Quantum Universe Postdoc Council of Hamburg University and DESY,
for funding the workshop ``How fast does the bubble grow?", where this work was initiated and acknowledge
the support by the Deutsche Forschungsgemeinschaft under Germany’s Excellence Strategy - EXC 2121 “Quantum Universe” - 390833306.
We also acknowledge the hospitality of the Helsinki Institute of Physics during
the second \wallgo{} workshop.

\appendix
\renewcommand{\thesection}{\Alph{section}}
\renewcommand{\thesubsection}{\Alph{section}.\arabic{subsection}}
\renewcommand{\theequation}{\Alph{section}.\arabic{equation}}

\section{Using the \wallgoMatrix{} matrix-element generator}
\label{sec:Matrix}

In this section,
we detail the usage of matrix-element generator \wallgoMatrix{}.
Calculations of out-of-equilibrium contributions require a multitude of matrix elements.
These can either be supplied to \wallgo{} directly, or
alternatively generated automatically with
\wallgoMatrix{}
which is available via the repositories:\\
\url{\wallgoMatrixUrl},\\
\url{\wallgoMatrixPacletUrl}.

\subsection{Loading the required packages}

The main routines for matrix element generation
are collected in the files:
{\tt WallGoMatrix.m} and
{\tt matrixElements.m}.
The model generation of
\wallgoMatrix{} utilises
\dralgo{}~\cite{Ekstedt:2022bff} and
{\tt GroupMath}~\cite{Fonseca:2020vke}.
We ask the user to cite these references when making use of \wallgoMatrix{}.
Before loading the package it either needs to be installed as
in listing~\ref{lst:wallgomatrix:pacletInstall} or by placing
the \wallgoMatrix{} directory inside
the {\tt Wolfram} user base directory\begin{lstlisting}[language=Mathematica]
<UserBaseDirectory>/Applications/WallGoMatrix
\end{lstlisting}
where the latter can be found under
{\verb!$UserBaseDirectory!} in
{\tt Mathematica}.
Thereafter, \wallgoMatrix{} is loaded via
\begin{lstlisting}[
  style=backtickavailable,
  language=Mathematica]
WallGo@\backtick@WallGoMatrix@\backtick@$InstallGroupMath=True;
<<WallGo@\backtick@WallGoMatrix@\backtick
\end{lstlisting}
where the flag
{\tt WallGo\backtick WallGoMatrix\backtick\$InstallGroupMath}
automatically installs {\tt GroupMath}
in the user base directory.

\subsection{Defining the model}
\label{sec:matrix:model}

We first need to define the model, in the same way as in {\tt DRalgo}.
For the Standard Model
without hypercharge group%
\footnote{
  We follow the common practice in the wall velocity literature,
  of ignoring the $\mathrm{U}(1)$ contributions.
  However, this assumption could also be dropped.
}
this is done by writing
\begin{lstlisting}[language=Mathematica]
Group={"SU3","SU2"};
CouplingName={gs,gw};
\end{lstlisting}
The user needs to specify all particles that should be included
in the computation of $\vw$;
cf.\ discussion in sections~\ref{sec:install} and~\ref{sec:code}.
Just as in \dralgo{}, this is done by giving the Dynkin coefficients for each representation.
These coefficients allow for handling a general model, but they might be unfamiliar.
{\tt GroupMath} provides a map to a more familiar notation;
for $\mathrm{SU}(3)$ one can obtain all representations up to size 30~\cite{Fonseca:2020vke}
\begin{lstlisting}[language=Mathematica]
su3Reps = RepsUpToDimN[SU3,30];
Grid[Prepend[{#,RepName[SU3,#]}&/@ su3Reps,{"Dynkin coefficients","Name"}],
    Frame->All,FrameStyle->LightGray]
\end{lstlisting}
where a typical representation is the adjoint one with
Dynkin index
{\verb!{1,1}!} and
representation $R = {\bm 8}$.
As a consequence,
the (adjoint) gauge representation of ${\rm SU}(3)\times{\rm SU}(2)$ is
\begin{lstlisting}[language=Mathematica]
RepAdjoint={{1,1},{2}};
\end{lstlisting}
where the notation is
${\tt RepAdjoint }=\{
\{{\tt SU3}_\text{dynkin}\},
\{{\tt SU2}_\text{dynkin}\}\}$.
For fermions and scalars the same notation as
for the vector particles is used.

As an example,
we focus on a simplified fermion sector.
One generation of quarks consists of
one left-handed
{\tt SU2}-doublet quark and
two right-handed
{\tt SU2}-singlet quarks.
These are specified as 
\begin{lstlisting}[
  label={lst:fermion:rep},
  language=Mathematica]
Rep1={{{1,0},{1}},"L"};
Rep2={{{1,0},{0}},"R"};
Rep3={{{1,0},{0}},"R"};
RepFermion1Gen={Rep1,Rep2,Rep3};
\end{lstlisting}
All three quark generations can now be created by stacking
together three copies of {\tt RepFermion1Gen},
\begin{lstlisting}[language=Mathematica]
RepFermion3Gen={RepFermion1Gen,RepFermion1Gen,RepFermion1Gen}//Flatten[#,1]&;
\end{lstlisting}

The Higgs field is similarly specified by 
\begin{lstlisting}[
  label={lst:higgs:su2},
  language=Mathematica]
HiggsDoublet={{{0,0},{1}},"C"};
RepScalar={HiggsDoublet};
\end{lstlisting}
The last entry
indicates whether the scalar representation is
complex
({\tt"C"}) or
real
({\tt"R"}).
In the example of the
$\mathrm{SU}(2)$ triplet,
choosing a
real ({\tt "R"}) or
complex ({\tt "C"}) scalar representation would result
for the scalar to have
three or six
degrees of freedom, respectively.

\wallgoMatrix{} takes ordered lists of representations as an input.
Generically a fermionic or scalar {\tt Rep}
will take the form
\begin{lstlisting}[language=Mathematica,mathescape=true]
Rep={Rep[r$_1$],Rep[r$_2$],Rep[r$_3$],...,Rep[r$_n$]};
\end{lstlisting}
The ordering of the indices
$
\{{\tt r}_{1},\dots,{\tt r}_{n}\} \in
\overline{{\tt s}}\lor\overline{{\tt f}}\lor\overline{{\tt v}}
$
in which representations are stacked
for scalars
($\overline{{\tt s}}=\{{\tt s}_{1},\dots,{\tt s}_{m}\}$),
fermions
($\overline{{\tt f}}=\{{\tt f}_{1},\dots,{\tt f}_{n}\}$),
and vectors
($\overline{{\tt v}}=\{{\tt v}_{1},\dots,{\tt v}_{o}\}$)
respectively,
will be fixed throughout the computation and generation of the matrix elements
in the following section; see e.g.\ section~\ref{sec:outOfEq:particles}.
In our example of three quark families this corresponds to
\begin{lstlisting}[language=Mathematica,mathescape=true]
RepFermion3Gen={Rep1,Rep2,Rep3,Rep1,Rep2,Rep3,Rep1,Rep2,Rep3};
\end{lstlisting}

Once all particles have been defined,
the model can be loaded by writing
\begin{lstlisting}[language=Mathematica,mathescape=true]
{gvvv,gvff,gvss,$\lambda$1,$\lambda$3,$\lambda$4,$\mu$ij,$\mu$IJ,$\mu$IJC,Ysff,YsffC}=
    AllocateTensors[Group,RepAdjoint,CouplingName,RepFermion3Gen,RepScalar];
\end{lstlisting}

To distinguish fermions by their masses
one has to define Yukawa couplings.
An example is the left-handed quark representation that splits into
one left-handed top and
bottom quark, respectively.
The corresponding Yukawa couplings are definable as
\begin{lstlisting}[language=Mathematica]
InputInv={{1,1,2},{False,False,True}};
YukawaDoublet=CreateInvariantYukawa[Group,RepScalar,RepFermion3Gen,InputInv]//Simplify;
Ysff=-GradYukawa[yt*YukawaDoublet[[1]]];
\end{lstlisting}
The corresponding input invariant is of the form
\begin{lstlisting}[language=Mathematica]
InputInv={{s1,f1,f2},{s1c,f1c,f2c}};
\end{lstlisting}
where
${\tt s1} \in \overline{{\tt s}}$
and
${\tt f1}, {\tt f2} \in \overline{{\tt f}}$ are
indices that refer to the definition of scalar or fermionic representation
in {\tt RepScalar} and {\tt RepFermion3Gen}, respectively.
Here,
${\verb!{s1,f1,f2}!}={\verb!{1,1,2}!}$ corresponds to
\begin{align}
\text{Higgs doublet}\times
\text{Left-handed top-quark}\times
\text{Right-handed top-quark}\sim
\overline{Q}_{\rmii{L}} \Phi^\dagger t_{\rmii{R}}
\,,
\end{align}
and the attribute
${\verb!{s1c,f1c,f2c}!}={\verb!{False,False,True}!}$ indicates
which of the representations are
unconjugated ({\tt True}) and
conjugated ({\tt False}).
By instead setting ${\tt f2} = {\tt 3}$,
the right-handed top quark is swapped for
the right-handed bottom quark.
For the definition of scalar couplings see~\cite{Ekstedt:2022bff}.

Finally,
the model is initiated via
\begin{lstlisting}[language=Mathematica,mathescape=true]
ImportModel[Group,gvvv,gvff,gvss,$\lambda$1,$\lambda$3,$\lambda$4,$\mu$ij,$\mu$IJ,$\mu$IJC,Ysff,YsffC];
\end{lstlisting}

\subsection{Implementing spontaneous symmetry-breaking}

After symmetry-breaking,
interaction eigenstates and
mass eigenstates are generally not the same.
While one can still use the interaction eigenstates from
the particle definitions of the previous section,
one can also use the mass eigenstates directly.
To this end,
\wallgoMatrix{} groups each degree of freedom of a particle depending
on the corresponding mass.
We now discuss how to make use of this feature.

To group particles depending on the symmetry-breaking induced masses,
first a vacuum-expectation value must be defined.
In the case of
the doublet in the fundamental representation of the ${\rm SU}(2)$ of
listing~\ref{lst:higgs:su2},
this amounts to
\begin{lstlisting}[language=Mathematica]
vev={0,v,0,0};
\end{lstlisting}
The
{\tt vev} here has size
{\tt Length[gvss[[1]]]},
where {\tt gvss} is the \dralgo{} internal
vector-scalar trilinear coupling tensor.
The components belonging to each representation can be found by
\begin{lstlisting}[
  label={lst:PrintFieldRepPositions},
  language=Mathematica]
PrintFieldRepPositions["Vector"]
PrintFieldRepPositions["Fermion"]
PrintFieldRepPositions["Scalar"]
\end{lstlisting}
which gives
{\verb!{1;;4}!}
for a fundamental ${\rm SU}(2)$ doublet.
For a complex representation (like ours),
the first two indices specify the real-components, and
the last two the complex components.
Schematically
\begin{align}
\Phi=\begin{pmatrix}
    {\tt vev1}+i {\tt vev3}\\
    {\tt vev2}+i {\tt vev4}\\
\end{pmatrix}
\,,
\end{align}
or correspondingly
{\verb!vev={vev1,vev2,vev3,vev4}!}.
By writing
{\verb!vev={0,v,0,0}!},
we specify that the Higgs \vev{} is of the form
$\Phi=
  (0, {\tt v})^T
$.

After specifying the {\tt vev},
the representations can be separated via their masses by writing
\begin{lstlisting}[language=Mathematica]
SymmetryBreaking[vev];
\end{lstlisting}
and subsequently referenced by the {\tt CreateParticle} command.
The information from the output of {\tt SymmetryBreaking} can be used
to deal field-dependent masses to the particles during the matrix element
generation in section~\ref{sec:matrix:generation}.
However, in the context of the \wallgoMatrix{} examples,
only thermal masses are used and therefore calling
{\tt SymmetryBreaking} does not affect the matrix elements.

The command becomes effective,
when turning on \vev-dependent couplings.
By default,
\vev-dependence is turned off but can be invoked by
setting the following option inside
{\tt SymmetryBreaking}
\begin{lstlisting}[language=Mathematica]
VevDependentCouplings->True
\end{lstlisting}

\subsection{Specifying out-of-equilibrium particles}
\label{sec:outOfEq:particles}

We now specify representations of particles which,
during collision integration,
can be taken in- or out-of-equilibrium.
For these particles,
a complete set of matrix elements with them appearing on
the external legs will be provided.

As an example,
we create particles for
the left-handed top quark,
the right-handed top-quark,
the gluons,
the $W$-bosons, and
the scalar doublet,
by creating a separate distribution for each of these.
The to-be equilibrium-particles can be grouped
in individual distributions or,
if they are considered to be of the same light mass,
into a single collective distribution.
One such example is grouping together all quarks except for the top.
If also leptons would be included, it is sensible to create a
separate distribution for all left-handed leptons and all right-handed leptons.

By calling {\tt SymmetryBreaking[vev]},
a list of
representations and their corresponding masses is generated
only for representations with at least one massive degree of freedom
\begin{lstlisting}[
  language=Mathematica,
  mathescape=true]
SymmetryBreaking[vev];
Gauge rep 2 splits into particles with mass squared:
{2,1}:    $({\tt gw}^2 {\tt v}^2)/{\tt 4}$
Fermion rep 1 splits into particles with mass squared:
{1,1}:    $({\tt v}~{\tt yt})/\sqrt{{\tt 2}}$
{1,2}:    0
Fermion rep 2 splits into particles with mass squared:
{2,1}:    $({\tt v}~{\tt yt})/\sqrt{{\tt 2}}$
\end{lstlisting}
where, for the quarks,
we chose to put
the left-handed quark doublet on position 1
of {\tt RepFermion3Gen} and
the right-handed top quark on position 2;
cf.\ listing~\ref{lst:fermion:rep}.
In the above listing, the bracketed integer indices,
{\verb!{r,m}!},
indicate
the particle with mass
$M_{\tt r}[{\tt m}]$ of
the corresponding representation
${\tt r} \in \{\overline{{\tt s}},\overline{{\tt f}},\overline{{\tt v}}\}$.
An example is
$M_{\tt r =1}[{\tt m = 1}] = m_\rmi{\tt ReptL}=({\tt v}~{\tt yt})/\sqrt{{\tt 2}}$.
An entire representation,
without distinguishing mass states, would correspond to writing
{\verb!{r}!}.

Using this notation,
several particles can be grouped together into one distribution using
the explicit list of indices from listing~\ref{lst:PrintFieldRepPositions}
\begin{lstlisting}[
    label={lst:no:createparticle},
    language=Mathematica,mathescape=true]
{{i$_1$,...,i$_l$},"R",M,"particleName"};
\end{lstlisting}
where
${\tt i}_i \in {\tt PrintFieldRepPositions["Field"]}$,
with ${\tt Field} \in \{{\tt Vector},{\tt Fermion},{\tt Scalar}\}$.
Alternatively, after having called {\tt SymmetryBreaking},
one can also use the command
\begin{lstlisting}[
    label={lst:createparticle:list},
    language=Mathematica,mathescape=true]
CreateParticle[{{r$_1$,m$_1$},...,{r$_n$,m$_n$},"R",M,"particleName"];
\end{lstlisting}
Above,
$n$ is the number of particles grouped into the distribution,
${\tt R} = \{{\tt F},{\tt V},{\tt S}\}$ is the corresponding particle species,
namely {\tt F}ermion, {\tt V}ectors, and {\tt S}calars,
{\tt M} the particle mass squared, and
{\tt "particleName"} the particle name.
Specifically,
the distribution for
the massive
left-handed and
right-handed quarks, namely
the top quark,
can be specified by writing
\begin{lstlisting}[language=Mathematica]
(*left-handed top-quark*)
ReptL=CreateParticle[{{1,1}},"F",mt2,"TopLeft"];
(*right-handed top-quark*)
ReptR=CreateParticle[{{2,1}},"F",mt2,"TopRight"]
\end{lstlisting}
where
{\tt mt2} is the squared top-quark mass.
Distributions for the gauge-bosons and scalars can similarly be created by writing
\begin{lstlisting}[language=Mathematica]
(*Vector bosons*)
RepGluon=CreateParticle[{1},"V",mg2,"Gluon"];
RepW=CreateParticle[{2},"V",mW2,"W"];

(*Higgs*)
RepH=CreateParticle[{1},"S",mH2,"Higgs"];
\end{lstlisting}
As seen above, in the definition of
{\tt CreateParticle},
several particles can be grouped together and dealt with in the same distribution.

To generate the matrix elements in section~\ref{sec:matrix:generation},
all particles of the theory defined in section~\ref{sec:matrix:model}
need to be declared as with {\tt CreateParticle[]}.
Therefore, we group the remaining light quarks into a single representation
\begin{lstlisting}[language=Mathematica]
(*Light quarks*)
LightQuarks=CreateParticle[{{1,2},3,4,5,6,7,8,9},"F",mq2,"LightQuarks"];
\end{lstlisting}
with mass squared {\tt mq2}.
By separately stacking together
out-of-equilibrium representations via
{\tt ReptL},
{\tt ReptR},
{\tt RepGluon},
{\tt RepW}, and
{\tt RepH} as well as
equilibrium representations
via {\tt LightQuarks},
we create the two lists of particles
that are relevant for collisions in the model,
{\em viz.}
\begin{lstlisting}[language=Mathematica]
ParticleList={ReptL,ReptR,RepGluon,RepW,RepH};
LightParticleList={LightQuarks};
\end{lstlisting}
The ordering of the particles can be arbitrary but
this numeric ordering (starting from 0)
will be subsequently used in all output of the matrix elements.
This is the same {\tt index} as in
the collision model from section~\ref{sec:Collision}.
Particles of the
{\tt LightParticleList} can never be ingoing in collisions
and they will thus never appear on the first index of
the generated matrix element file.
The representations listed in {\tt ParticleList} can still be assumed
to be out-of- or in-equilibrium when using them during the collision integration
in \wallgoCollision{}.
If a particle appears on the first index of the matrix element but it is in-equilibrium, 
this matrix element will simply be skipped in collision generation.

\subsection{Generating the matrix elements}
\label{sec:matrix:generation}

In calculations of the collision operators, we will encounter logarithmic divergences for small-angle scattering.
These are softened by thermal masses.
In this tutorial,
the top quarks was given
the mass {\tt mt2},
the remaining fermions
the mass {\tt mq2},
the gluons the mass {\tt mg2},
the $W$-bosons {\tt mW2}, and
the scalars {\tt ms2}.
The matrix elements are then calculated by writing
\begin{lstlisting}[language=Mathematica]
MatrixElements=ExportMatrixElements[<filePath>,ParticleList,LightParticleList,OptionPattern];
\end{lstlisting}
where its default options are given as
\begin{lstlisting}[language=Mathematica]
OptionPattern={
    Replacements->{},
    NormalizeWithDOF->True,
    TruncateAtLeadingLog->True,
    Verbose->False,
    Format->"none"};
\end{lstlisting}
By default,
the generated matrix elements are truncated at
leading logarithmic order.

The indices in
the elements {\tt M[a,c,d,e]} of
{\tt MatrixElements} (cf.\ eq.~\eqref{eq:collision-integral})
correspond to the position indices of
the respective particles inside the total list
{\verb!{ParticleList,LightParticleList}!}.
Each element contains all (squared) matrix elements that contribute with those specific particles.
The matrix-elements are also normalised by
the degrees~of~freedom of the incoming particle, that is by
the degrees~of~freedom of
{\tt particle1}.
By modifying {\tt OptionPattern},
it is possible to omit this normalisation factor by setting
\begin{lstlisting}[language=Mathematica]
NormalizeWithDOF->False
\end{lstlisting}
While there is no default output format of \wallgoMatrix{},
the default input format of \wallgo{} is
{\tt .json};
see section~\ref{sec:wallgomatrix}.
The following formats are implemented:
\begin{lstlisting}[language=Mathematica]
Format->{"json","txt","hdf5"}
\end{lstlisting}

In some cases, one might generate certain matrix elements that one does not want to include in the collisions. For example, to include the lepton exchange diagram in $W$-scattering, the leptons have to be included in the (light) particle representations, and this will generate matrix elements with leptons on the second, third and fourth index.
If we do not want to include these in the collision computation (e.g.\ because of computational costs), we simply do not include the corresponding particles in
the \wallgoCollision{} model.
We then only need to provide an input for the lepton asymptotic thermal mass.

\section{General matrix elements}
\label{sec:GenMat}

\wallgo{} and specifically \wallgoMatrix{},
employ the leading-logarithmic approximation.
To this end,
we take a given (vacuum) matrix-element, and
regulate terms that diverge logarithmically via adding
the corresponding (thermal, or rather asymptotic, cf.\ section~\ref{sec:Boltzmann:llog}) mass.
Since we only need vacuum matrix-elements, we can generate these efficiently.
A useful way to do this for general models, is to use general coupling tensors (see \cite{Machacek:1983fi, Machacek:1983tz, Machacek:1984zw, Martin:2017lqn,Martin:2018emo}) to handle the group algebra; and spinor-helicity methods to handle
the Lorentz algebra.%
\footnote{
  Using Feynman diagrams for expressions with chiral particles is not very practical and the spinor-helicity formalism has the added advantage that 
  gauge-invariance is manifest~\cite{Mangano:1990by}.
}
This appendix illustrates the approach by focussing on three
specific scattering processes.

Along these calculations,
we first treat all particles as outgoing.
All possible particle flows then follow upon changing
the sign of the relevant momenta.
Henceforth, we employ the notation (cf.~\cite{Srednicki:2007qs})
\begin{align}
  s_{ij}=-(p_i+p_j)^2
  \,,
\end{align}
and label particles as in section~\ref{sec:Matrix} as
$V$ for vector bosons,
$F$ for fermions, and
$S$ for scalars.
The coupling tensors for the different interactions appearing in
the calculations below are given by
\begin{align}
  S_{i} S_{j} V^a &:
  \quad
  g_{ij}^{a}
  \;,\\
  F_{\rmii{$I$}} F_{\rmii{$J$}} V^a &:
  \quad
  g_{\rmii{$I$}\rmii{$J$}}^{a}
  \;,\\
  F_{\rmii{$I$}} F_{\rmii{$J$}} S_{i} &:
  \quad
  Y_{i\rmii{$I$}\rmii{$J$}}^{ }
  \;,\\
  S_{i} S_{j} S_{k} &:
  \quad
  \lambda_{ijk}^{ }
  \;,
\end{align}
where
$I,J,L,M$ describe spinor indices,
$a,b,c,d$ vector indices, and
$i,j,k,l$ scalar indices.

\subsection{Matrix elements for $F_1 F_2 \to F_3 F_4$}
We first focus on fermion scattering with a vector or scalar boson as propagator.
The contributing diagrams
to the $s$-, $t$-, and $u$-channels of
the vector- and scalar-exchange are
\begin{align}
\label{eq:M:FFFF:diags}
  -\mathcal T_{\lambda_1\lambda_2\lambda_3\lambda_4} &\supset
        \VtxvSn(\Lqu,\Lqu,\Luq,\Luq,\Lglx,
            {\lambda_3,\rmii{$L$}},
            {\lambda_1,\rmii{$I$}},
            {\lambda_2,\rmii{$J$}},
            {\lambda_4,\rmii{$M$}})
      + \VtxvTn(\Luq,\Lqu,\Lqu,\Luq,\Lglx,
            {\lambda_3,\rmii{$L$}},
            {\lambda_1,\rmii{$I$}},
            {\lambda_2,\rmii{$J$}},
            {\lambda_4,\rmii{$M$}})
      + \VtxvUn(\Luq,\Lqu,\Lqu,\Luq,\Lglx,
            {\lambda_3,\rmii{$L$}},
            {\lambda_1,\rmii{$I$}},
            {\lambda_2,\rmii{$J$}},
            {\lambda_4,\rmii{$M$}})
    \\[6mm] &=
\label{eq:M:FFFF}
      \underbrace{
        g^{a\,\rmii{$I$}}_{\rmii{$J$}}
        g^{a\,\rmii{$M$}}_{\rmii{$L$}}}_{\equiv G_{s}}
      A^s_{\lambda_1\lambda_2\lambda_3\lambda_4}
    + \underbrace{
        g^{a\,\rmii{$I$}}_{\rmii{$L$}}
        g^{a\,\rmii{$J$}}_{\rmii{$M$}}}_{\equiv G_{t}}
      A^t_{\lambda_1\lambda_2\lambda_3\lambda_4}
    + \underbrace{
        g^{a\,\rmii{$I$}}_{\rmii{$M$}}
        g^{a\,\rmii{$J$}}_{\rmii{$L$}}}_{\equiv G_{u}}
      A^u_{\lambda_1\lambda_2\lambda_3\lambda_4}
    \,,
  \\[6mm]
\label{eq:M:FFFF:scalar}
  -\mathcal T_{\lambda_1\lambda_2\lambda_3\lambda_4} &\supset
        \VtxvSn(\Lqu,\Lqu,\Luq,\Luq,\Lxx,
            {\lambda_3,\rmii{$L$}},
            {\lambda_1,\rmii{$I$}},
            {\lambda_2,\rmii{$J$}},
            {\lambda_4,\rmii{$M$}})
      + \VtxvTn(\Luq,\Lqu,\Lqu,\Luq,\Lxx,
            {\lambda_3,\rmii{$L$}},
            {\lambda_1,\rmii{$I$}},
            {\lambda_2,\rmii{$J$}},
            {\lambda_4,\rmii{$M$}})
      + \VtxvUn(\Luq,\Lqu,\Lqu,\Luq,\Lxx,
            {\lambda_3,\rmii{$L$}},
            {\lambda_1,\rmii{$I$}},
            {\lambda_2,\rmii{$J$}},
            {\lambda_4,\rmii{$M$}})
      \,,
      \\[-2mm] \notag
\end{align}
where
wiggly lines denote vector bosons,
directed lines fermions, and
dashed lines scalars.
We now focus only on
the $u$- and $t$-channels of the vector exchange.
The contributing scalar-exchange diagrams
of this scattering process
are implemented in \wallgoMatrix{}.

Here, the $\lambda_1,\dots,\lambda_4$
specify the helicities of the particles.
The coupling-tensor
$
g^{a\,\rmii{$I$}}_{\rmii{$L$}}=
g^{a\,\rmii{$I$}}_{13,\rmii{$L$}}$
describes the interactions between
two fermions and a vector boson, namely
fermion 1 with spinor index $I$,
fermion 3 with spinor index $L$, and
the vector boson with label $a$.
We treat all particles as Weyl; which means that an upper spinor index always must contract with a lower one in the end\te
the hermitian conjugate flips upper and lower signs.

Furthermore,
$A^t_{\lambda_1 \cdots \lambda_4}$ denotes the Lorentz-structure of the $t$-channel diagram, and
$A^u_{\lambda_1 \cdots \lambda_4}$ of the $u$-channel one.
Since
$\BraKet{p}{\gamma^\mu}{p}=\braket{p}{\gamma^\mu}{p}=0$,
the amplitudes are only non-zero if two helicities at a fermion vertex
are of opposite sign,
leaving six helicity signatures, namely
\begin{align}
  A^{t,u}_{++--}\,,&&
  A^{t,u}_{+-+-}\,,&&
  A^{t,u}_{+--+}\,,&&
  \Bigl[
  A^{t,u}_{--++}\,,&&
  A^{t,u}_{-+-+}\,,&&
  A^{t,u}_{-++-}
  \Bigr]\,.
\end{align}
Via complex-conjugation,
the last three bracketed amplitudes can be related to
the first three ones which reduces
the total number of helicity signatures to three.

For the signature $A_{+--+}$,
the $u$-channel diagram vanishes as two positive-helicity lines meet,
hence we are left with the $t$-channel diagram.
The signature $A_{+-+-}$ only has a $u$-channel, and
$A_{++--}$ has both $t$ and $u$-channel contributions.
The contributions are,
\begin{align}
  \label{eq:A:t:+--+}
  A^t_{+--+} &=
  + \braKet{3}{\gamma^\mu}{1}
    \Braket{4}{\gamma_\mu}{2}/s_{13}
  =
  - 2\BraKetT{1}{4}\braketT{2}{3}/s_{13}
  \,,\\
  A^u_{+-+-} &=
  - \braKet{4}{\gamma^\mu}{1}
    \Braket{3}{\gamma_\mu}{2}/s_{14}
  =
  + 2\BraKetT{1}{3}\braketT{2}{4}/s_{14}
  \,,\\
  A^t_{++--} &=
  + \braKet{3}{\gamma^\mu}{1}
    \Braket{2}{\gamma_\mu}{4}/s_{13}
  =
  + 2\BraKetT{1}{2}\braketT{3}{4}/s_{13}
  \,, \\
  A^u_{++--} &=
  - \braKet{4}{\gamma^\mu}{1}
    \Braket{2}{\gamma_\mu}{3}/s_{14}
  =
  + 2\BraKetT{1}{2}\braketT{3}{4}/s_{14}
  \,,
\end{align}
where in the second step of eq.~\eqref{eq:A:t:+--+},
we used the Fierz identity
$
\braKet{1}{\gamma^\mu}{2}
\braKet{3}{\gamma_\mu}{4}
=
2\braketT{1}{3}\BraKetT{2}{4}
$,
$\Braket{k}{\gamma^\mu}{p}=
\braKet{p}{\gamma^\mu}{k}$,
and
the anti-symmetry
$\braketT{p}{q}=-\braketT{q}{p}$.
Using that
$\braketT{n}{m}^* = \BraKetT{m}{n}$ and
$s_{mn}=\braketT{m}{n}\BraKetT{n}{m}$,
we find
\begin{align}
  \abs{A^t_{+--+}}^2 &=
  4s_{14}^2/s_{13}^2
  \,,&
  \abs{A^u_{+-+-}}^2 &=
  4s_{13}^2/s_{14}^2
  \,,&
  A^u_{++--} \left[A^t_{++--}\right]^{\dagger} &=
  4s_{12}^2/s_{13}/s_{14}
  \,,\nn
  \abs{A^t_{++--}}^2 &=
  4s_{12}^2/s_{13}^2
  \,,&
  \abs{A^u_{++--}}^2 &=
  4s_{12}^2/s_{14}^2
  \,,
\end{align}
including the contributions from complex conjugation.

The remaining step is to
multiply with the complex conjugate and sum over
helicities, particles and anti-particles,
such that
\begin{align}
  \sum \abs{\mathcal T}^2 &=
    8 (G_t^{ } G_t^\dagger)\frac{s_{14}^2+s_{12}^2}{s_{13}^2}
  + 8 (G_u^{ } G_u^\dagger)\frac{s_{13}^2+s_{12}^2}{s_{14}^2}
  + 8 (G_t^{ } G_u^\dagger+G_u^{ } G_t^\dagger)\frac{s_{12}^2}{s_{13}s_{14}}
  \,,
\end{align}
where
the $G_t$ and $G_u$ are defined in eq.~\eqref{eq:M:FFFF}.

We can now get any amplitude of this type by adapting the signs
of some particles as incoming,
e.g.\
\begin{align}
  F_1 F_2 &\to F_3 F_4
  :&
  p_1, p_2 &\to -p_1, -p_2
  \,,&
  s_{13},s_{14},s_{12} &\to t,u,s
  \,, \\
  F_1 \bar{F}_2 &\to F_3 \bar{F}_4
  :&
  p_1, p_4 &\to -p_1, -p_4
  \,,&
  s_{13},s_{14},s_{12} &\to t,s,u
  \,.
\end{align}

\subsection{Matrix elements for $S_1 S_2 \to F_1 F_2$}

As before,
we initially take all particles as outgoing.
The momenta
$p_1$, $p_2$ correspond to the scalars
$S_1$, $S_2$, and
$p_3$, $p_4$ to the fermions
$F_1$, $F_2$.
The corresponding diagrams constitute
a $s$-, $t$-, and $u$-channel,
{\em viz.}
\begin{align}
\label{eq:M:SSFF:diagrams}
  -\mathcal T_{00\lambda_3\lambda_4} &=
        \VtxvSn(\Luq,\Lxx,\Lxx,\Lqu,\Lxx,
            {\lambda_3,\rmii{$L$}},
            {0,i},
            {0,j},
            {\lambda_4,\rmii{$M$}})
      + \VtxvSn(\Luq,\Lxx,\Lxx,\Lqu,\Lglx,
            {\lambda_3,\rmii{$L$}},
            {0,i},
            {0,j},
            {\lambda_4,\rmii{$M$}})
      + \VtxvTn(\Luq,\Lxx,\Lxx,\Lqu,\Luq,
            {\lambda_3,\rmii{$L$}},
            {0,i},
            {0,j},
            {\lambda_4,\rmii{$M$}})
      + \VtxvUn(\Luq,\Lxx,\Lxx,\Lqu,\Lqu,
            {\lambda_3,\rmii{$L$}},
            {0,i},
            {0,j},
            {\lambda_4,\rmii{$M$}})
    \\[6mm]
  \label{eq:M:SSFF}
    &=
      \lambda_{ijk}^{ }\,Y_{\rmii{$M$}}^{k\,\rmii{$L$}}
      \mathcal{A}^{s}_{00\lambda_3\lambda_4}
    + g_{ij}^{a}\,
      g_{\rmii{$M$}}^{a\,\rmii{$L$}}
      A^{s}_{00\lambda_3\lambda_4}
    + Y_{i}^{\rmii{$I$}\rmii{$M$}}
      Y_{j\rmii{$I$}\rmii{$L$}}^{ }
      A^{t}_{00\lambda_3\lambda_4}
    + Y_{i}^{\rmii{$I$}\rmii{$M$}}
      Y_{j\rmii{$I$}\rmii{$L$}}^{ }
      A^{u}_{00\lambda_3\lambda_4}
    \,,
\end{align}
where
$Y_{i\rmii{$I$}\rmii{$J$}}$ are the Yukawa coupling tensors and
$\lambda_{ijk}$ the cubic scalar coupling tensor.
We henceforth only focus on
the vector- and fermion-exchange contributions
from eq.~\eqref{eq:M:SSFF}.
Since the helicities of the fermions have to be opposing,
we consider
$\mathcal{T}_{00+-}$.
We can write down the corresponding channels
using $\braKet{p}{\bsl{P}}{q}=\braketT{p}{P}\BraKetT{P}{q}$ and
$p_n\Ket{n}=0$,
such that
\begin{align}
  A^{s}_{00+-}
  &=
  - \frac{\braKet{4}{1-2}{3}}{s_{12}}
  =
  + 2\frac{
    \braketT{1}{4}
    \BraKetT{1}{3}}{s_{12}}
  \,,\nn
  A^{t}_{00+-}
  &=
  - \frac{\braKet{4}{1-3}{3}}{s_{13}}
  =
  + \frac{
    \braketT{1}{4}
    \BraKetT{1}{3}}{s_{13}}
  \,,\nn
  A^{u}_{00+-}
  &=
  + \frac{\braKet{4}{1-4}{3}}{s_{14}}
  =
  - \frac{
    \braketT{1}{4}
    \BraKetT{1}{3}}{s_{14}}
  \,.
\end{align}
For simplicity,
we only display quadratic contributions at leading logarithmic order
and suppress mixed contributions.
The sum of the (possibly) logarithmically divergent matrix elements is
\begin{align}
  \bigl| \mathcal{T}_{00+-}\bigr|^2 \supset
     4\tr\Bigl[\bigl(
      g_{ij}^{a}\,
      g_{\rmii{$M$}}^{a\,\rmii{$L$}}
    \bigr)^2\Bigr]
    \frac{s_{13} s_{14}}{s_{12}^2}
    + \tr\Bigl[\bigl(
      Y_{i}^{\rmii{$I$}\rmii{$M$}}
      Y_{j\rmii{$I$}\rmii{$L$}}^{ }
    \bigr)^2\Bigr]
    \frac{s_{13} s_{14}}{s_{13}^2}
    + \tr\Bigl[\bigl(
      Y_{i}^{\rmii{$I$}\rmii{$M$}}
      Y_{j\rmii{$I$}\rmii{$L$}}^{ }
    \bigr)^2\Bigr]
    \frac{s_{13} s_{14}}{s_{14}^2}
  \,.
\end{align}
The $\mathcal{T}_{00-+}$ process is the same as the above.

\subsection{Matrix elements for $F_1 F_2 \to S V$}

The corresponding diagrams that contribute to
this process are\\[1mm]
\begin{align}
\label{eq:M:FFVS:diagrams}
  -\mathcal T_{\lambda_1 \lambda_2 0 \lambda_4} &=
        \VtxvSn(\Lxx,\Lqu,\Luq,\Lglx,\Lxx,
            {\quad0,k},
            {\lambda_1,\rmii{$I$}},
            {\lambda_2,\rmii{$J$}},
            {\lambda_4,d})
      + \VtxvTn(\Lxx,\Lqu,\Luq,\Lglx,\Lqu,
            {0,k},
            {\lambda_1,\rmii{$I$}},
            {\lambda_2,\rmii{$J$}},
            {\;\lambda_4,d})
      + \VtxvUn(\Lxx,\Lqu,\Luq,\Lglx,\Lqu,
            {\;\;0,k},
            {\lambda_1,\rmii{$I$}},
            {\lambda_2,\rmii{$J$}},
            {\;\lambda_4,d})
    \\[6mm]
\label{eq:M:FFVS:diagrams}
    &=
      Y_{i\rmii{$J$}}^{\rmii{$I$}}\,
      g_{ik}^{d}
      A^{s}_{\lambda_1\lambda_2 0\lambda_4}
    + Y_{k}^{\rmii{$M$}\rmii{$I$}}
      g_{\rmii{$M$}\rmii{$J$}}^{d}
      A^{t}_{\lambda_1\lambda_2 0\lambda_4}
    + Y_{k}^{\rmii{$M$}\rmii{$J$}}
      g_{\rmii{$M$}\rmii{$I$}}^{d}
      A^{u}_{\lambda_1\lambda_2 0\lambda_4}
  \,.
\end{align}
The only non-zero contribution to this process originates
from fermions with
identical helicity opposed to the helicity of the vector spin-1 particle.
By first considering $\mathcal{T}_{++0-}$,
we list the corresponding channel contributions
\begin{align}
  A^{t}_{++0-}
  &=
  \frac{\sqrt{2}}{\BraKetT{3}{4}}
    \frac{\BraKetT{2}{3}\braKet{4}{3}{1}}{s_{13}}
  \,,\\
  A^{u}_{++0-}
  &=
  \frac{\sqrt{2}}{\BraKetT{4}{3}}
    \frac{\BraKetT{2}{4}\braKet{3}{4}{1}}{s_{14}}
  \,,
\end{align}
where we
used the external line rule for outgoing
spin-1 massless vectors by inserting their
polarization vectors
in the spinor helicity notation~\cite{Srednicki:2007qs}, namely
$\msl{\epsilon}_{-}(p;q) = \frac{\sqrt{2}}{\BraKetT{q}{p}}
\left(
    \ket{p}\Bra{q}
  + \Ket{q}\bra{p}\right)$
and
$\msl{\epsilon}_{+}(p;q) = \frac{\sqrt{2}}{\braketT{q}{p}}
\left(
    \Ket{p}\bra{q}
  + \ket{q}\Bra{p}\right)$.
The matrix element squared gives
the following terms
\begin{align}
  \bigl| \mathcal{T}_{++0-}\bigr|^2 &=
      2\tr\Bigl[\bigl(
        Y_{k}^{\rmii{$M$}\rmii{$I$}}
        g_{\rmii{$M$}\rmii{$J$}}^{d}
      \bigr)^2\Bigr]
      \frac{s_{14}s_{13}}{s_{13}^2}
    + 2\tr\Bigl[\bigl(
        Y_{k}^{\rmii{$M$}\rmii{$J$}}
        g_{\rmii{$M$}\rmii{$I$}}^{d}
      \bigr)^2\Bigr]
      \frac{s_{13}s_{14}}{s_{14}^2}
    \nn &
    - 2\tr\Bigl[
      \bigl(
        Y_{k}^{\rmii{$M$}\rmii{$I$}}
        g_{\rmii{$M$}\rmii{$J$}}^{d}
      \bigr)
      \bigl(
        Y_{k}^{\rmii{$M$}\rmii{$J$}}
        g_{\rmii{$M$}\rmii{$I$}}^{d}
      \bigr)^\dagger
      +
      \bigl(
        Y_{k}^{\rmii{$M$}\rmii{$J$}}
        g_{\rmii{$M$}\rmii{$I$}}^{d}
      \bigr)
      \bigl(
        Y_{k}^{\rmii{$M$}\rmii{$I$}}
        g_{\rmii{$M$}\rmii{$J$}}^{d}
      \bigr)^\dagger
      \Bigr]
  \,,
\end{align}
where the last term contains the contributions
from mixed $u$- and $t$-channels.

\section{Reduction of the collision integral}
\label{sec:CollisionReduction}

The integral (\ref{eq:collision-integral}) is nine dimensional, but four of the integrals are trivial due to the momentum-conserving $\delta$-function. This leaves 5 integrals for numerical evaluation. 
Here we discuss a convenient parametrisation for use with numerical integrators.%
\footnote{%
  As discussed
  in e.g.~\cite{Hannestad:1995rs,Ala-Mattinen:2022nuj}, it is possible to perform one more integral analytically, leaving 4 numerical integrals.
  However, this requires a nontrivial
  rotation of the coordinate system in our case, as we work in a frame where $P_1$ is not aligned along any of the $x,y,z$ axes.
  We will thus not discuss the $9\rightarrow 4$ reduction here.
}
For generality,
we will keep particle masses explicit, so that
$E_i = \sqrt{p_i^2 + m_i^2}$. The momentum $P_1$ is a known input, fixed e.g.\
by giving the $\rho_z, \rho_{||}$ momentum components on the grid.

We begin by doing the integral over $p_4$.
For this, write
\begin{align}
  \int \frac{{\rm d}^3p_4}{2E_4} (\cdots) = \int {\rm d}^4 P_4 \; \delta(P_4^2 - m_4^2) \theta(E_4) (\cdots)
\,,
\end{align}
and to make use of the newly introduced $\delta$ we define a function $g(p_3) \equiv P_4^2 - m_4^2$.  
Then
\begin{align}
    \mathcal{C}_{ab}^\mathrm{lin}[\delta f^b] &= \frac{1}{4} \sum_{cde}
    \int \frac{
      {\rm d}^3 p_2\,
      {\rm d}^3 p_3\,
      {\rm d}^3 p_4}{(2\pi)^5 2E_2 2E_3 2E_4}
    \delta^4(P_1 + P_2 - P_3 - P_4) |M|^2 \mathcal{P}^\mathrm{lin}[\delta f^b] \nn
    &= \frac{1}{4} \sum_{cde}
    \int \frac{
        {\rm d}^3p_2\,
        {\rm d}^3p_3}{(2\pi)^5 2E_2 2E_3} \delta(g(p_3))
      \theta(E_4)|M|^2 \mathcal{P}^\mathrm{lin}[\delta f^b]
      \Big|_{P_4 = P_1 + P_2 - P_3}\label{eq:Cactingondeltaf}
    \,,
\end{align}
where $\mathcal{P}^\mathrm{lin}[\delta f^b]$ is given by
\begin{equation}
	\mathcal{P}^\mathrm{lin}[\delta f^b] = f^a f^c f^d f^e \left(
          \delta_{ab}^{ } F^c_a
        + \delta_{cb}^{ } F^a_c
        - \delta_{db}^{ } F^e_d
        - \delta_{eb}^{ } F^d_e
      \right) \delta f^b
        \,,
\end{equation}
and we have defined $F^a_b \equiv e^{E_a / T} / f_b^2$.

Next we go to spherical coordinates: 
\begin{align}
    \mathbf{p}_2 &= p_2(\sin\theta_2 \cos\phi_2, \sin\theta_2 \sin\phi_2, \cos\theta_2)
    \,,\nn
    \mathbf{p}_3 &= p_3(\sin\theta_3 \cos\phi_3, \sin\theta_3 \sin\phi_3, \cos\theta_3)
    \,,
\end{align}
where $\theta_i \in [0, \pi]$ and $\phi_i \in [0, 2\pi]$.
The integral over $p_3$ is done with the remaining $\delta$ function:
\begin{align}
\label{eq:collision-final}
    \mathcal{C}_{ab}^\mathrm{lin}[\delta f^b] &=
      \frac{1}{4} \sum_{cde}
      \int_0^\infty\!
      \frac{p_2^2 p_3^2
        {\rm d}p_2
        {\rm d}p_3}{(2\pi)^5 2E_2 2E_3}
      \int_0^{2\pi}\!
        {\rm d}\phi_2
        {\rm d}\phi_3
      \int_{-1}^1
        {\rm d}\cos\theta_2\,
        {\rm d}\cos\theta_3 \,
        \delta(g(p_3)) \theta(E_4) |M|^2  \mathcal{P}^\mathrm{lin}[\delta f^b]
    \nn &=
    \frac{1}{4} \sum_{cde} \sum_i
    \int_0^\infty\!\frac{p_2^2 p_3^2 {\rm d}p_2}{(2\pi)^5 2E_2 2E_3}
    \left| \frac{1}{g'(p_3^{(i)})} \right|
    \nn &\hphantom{{}\frac{1}{4} \sum_{cde} \sum_i}
    \times
    \int_0^{2\pi}\!
      {\rm d}\phi_2
      {\rm d}\phi_3
    \int_{-1}^1\!
      {\rm d}\cos\theta_2\,
      {\rm d}\cos\theta_3\,
      \theta(E_4) |M|^2 \mathcal{P}^\mathrm{lin}[\delta f^b] \Big|_{p_3 = p_3^{(i)}}
    \,.
\end{align}
Here the $i$-sum is over non-negative roots $p_3^{(i)}$ of $g(p_3) = 0$ and $p_3$ has been fixed by the $\delta$-function. 

To find these roots, note that $P_4$ is already fixed in terms of the other momenta:
\begin{align}
    g(p_3) &= P_4^2 - m_4^2 = (P_1 + P_2 - P_3)^2 - m_4^2 \nn 
    &= Q + 2(P_1 \cdot P_2) + 2\mathbf{p}_3 \cdot (\mathbf{p}_1 + \mathbf{p}_2) - 2\sqrt{p_3^2 + m_3^2} (E_1 + E_2)
    \,,
\end{align}
where $Q = m_1^2 + m_2^2 + m_3^2 - m_4^2$. The square root comes from energy $E_3(p_3)$. We now define $\hat{\mathbf{p}}_3 = \mathbf{p}_3 /p_3$ and after some algebra,
we get a quadratic equation for $p_3$: 
\begin{align}
    \left( \delta^2 - \epsilon^2 \right) p_3^2 + 2\kappa\delta p_3 + \kappa^2 = 0
    \,,
\end{align}
with
\begin{align}
\delta &= 2 \hat{\mathbf{p}}_3 \cdot (\mathbf{p}_1 + \mathbf{p}_2)
  \,, \nn
\epsilon &= 2(E_1 + E_2)
  \,, \nn
\kappa &= 2 P_1 \cdot P_2 + Q
  \,.
\end{align}
Solving this gives two roots $p_3^{(i)}$ of which we only need the non-negative ones. The derivative is given by
\begin{align}
g'(p_3) = \delta - \epsilon \frac{p_3}{E_3}
  \,.
\end{align}
Using this parameterisation it is now straightforward to integrate
eq.~(\ref{eq:collision-final}) using e.g.\
a numerical Monte Carlo method.

This expression for the root simplifies in the ultrarelativistic limit. Setting $E_i = p_i$ one gets that the only non-negative solution to $g(p_3) = 0$ is
\begin{align}
    p_3 = \frac{p_1 p_2 - \mathbf{p}_1 \cdot \mathbf{p}_2}{p_1 + p_2 - \hat{\mathbf{p}}_3 \cdot (\mathbf{p}_1 + \mathbf{p}_2)} \quad \text{(ultrarelativistic limit)}
\end{align}
and $g'(p_3) = -2 (p_1 + p_2 - \hat{\mathbf{p}}_3 \cdot (\mathbf{p}_1 + \mathbf{p}_2))$.

{\small
\bibliographystyle{utphys}
\bibliography{ref.bib}
}
\end{document}